\documentclass[floatfix,a4paper,aps,prd,twocolumn,preprintnumbers]{revtex4} 

\usepackage{epsfig}
\usepackage{axodraw}
\usepackage{amsmath}
\usepackage{amssymb}
\usepackage{amsfonts}
\usepackage{graphicx}
\usepackage{epic}
\usepackage{overpic}
\usepackage{graphics,subfigure}
\usepackage{axodraw}
\usepackage{float}
\usepackage{booktabs}
\usepackage{color}
\usepackage{geometry}
\usepackage{rotating}
\usepackage{psfrag}

\newbox\charbox
\newbox\slabox
\def\s#1{{      
 \setbox\charbox=\hbox{$#1$}
 \setbox\slabox=\hbox{$/$}
 \dimen\charbox=\ht\slabox
 \advance\dimen\charbox by -\dp\slabox
 \advance\dimen\charbox by -\ht\charbox
 \advance\dimen\charbox by \dp\charbox
 \divide\dimen\charbox by 2
 \raise-\dimen\charbox\hbox to \wd\charbox{\hss/\hss}
 \llap{$#1$}
}}

\graphicspath{%
{./}%
{figs/}%
}

\newcommand{\newc}{\newcommand}
\newc{\wt}{\widetilde}
\newc{\cL}{{\cal L}}
\newc{\cM}{{\cal M}}
\newc{\ra}{\rightarrow}
\newc{\eps}{\epsilon}
\newc{\bino}{\widetilde{\cal B}}
\newc{\wino}{\widetilde{\cal W}}
\newc{\gluino}{\widetilde{\cal G}}
\newc{\half}{\frac{1}{2}}
\newc{\third}{\frac{1}{3}}
\newc{\fourth}{\frac{1}{4}}
\newc{\eighth}{\frac{1}{8}}
\newc{\gev}{\mbox{~GeV}}
\newc{\lra}{\leftrightarrow}
\newc{\Dslash}{\not\!\! D}
\newc{\sg}{{\cal G}}
\newc{\ovl}{\overline}
\newc{\ok}{$\surd$}
\newc{\etal}{{\it et al.}\ }
\newc{\Hbar}{{\bar H}}
\newc{\hhbar}{{\overline h}}
\newc{\Ubar}{{\bar U}}
\newc{\Dbar}{{\bar D}}
\newc{\Ebar}{{\bar E}}
\newc{\eg}{{\it e.g.}\ }
\newc{\ie}{{\it i.e.}\ }
\newc{\nonum}{\nonumber}
\newc{\kap}{\kappa}
\newc{\Dt}{\frac{d}{dt}}
\newc{\rpv}{{\mbox{${\not\!\!R_p}$}}}
\newc{\bpv}{$\not\!\!B_p$}
\newc{\mpl}{$M_{Pl}$\ }
\newc{\mx}{$M_X$\ }
\newc{\mgut}{$M_{\rm GUT}$}
\newc{\tev}{\mbox{~TeV}}
\newc{\sect}[1]{\ref{sec:#1}}
\newc{\nonr}{\nonum}
\newc{\vev}[1]{\langle{#1}\rangle}
\newc{\eq}[1]{(\ref{eq:#1})}
\newc{\eqs}[2]{(\ref{eq:#1},\ref{eq:#2})}
\newc{\lab}[1]{\label{eq:#1}}
\newc{\Lam}{{\bf \Lambda}}
\newc{\ltau}{\lambda_\tau}
\newc{\lt}{\lambda_t}
\newc{\lb}{\lambda_b}
\newc{\lae}{{\Lam}_E}
\newc{\lad}{{\Lam}_D}
\newc{\lau}{{\Lam}_U}
\newc{\lame}[1]{{\Lam}_{E^{#1}}}
\newc{\lamhe}[1]{{\h}_{E^{#1}}}
\newc{\lamhed}[1]{{\h}_{E^{#1}}^\dagger}
\newc{\lamhd}[1]{{\h}_{D^{#1}}}
\newc{\lamhdd}[1]{{\h}_{D^{#1}}^\dagger}
\newc{\lamhu}[1]{{\h}_{U^{#1}}}
\newc{\lamhud}[1]{{\h}_{U^{#1}}^\dagger}
\newc{\lamd}[1]{{\Lam}_{D^{#1}}}
\newc{\lamu}[1]{{\Lam}_{U^{#1}}}
\newc{\lamet}[1]{{\Lam}_{E^{#1}}^T}
\newc{\lamdt}[1]{{\Lam}_{D^{#1}}^T}
\newc{\lamut}[1]{{\Lam}_{U^{#1}}^T}
\newc{\lames}[1]{{\Lam}_{E^{#1}}^*}
\newc{\lamds}[1]{{\Lam}_{D^{#1}}^*}
\newc{\lamus}[1]{{\Lam}_{U^{#1}}^*}
\newc{\lamed}[1]{{\Lam}_{E^{#1}}^\dagg}
\newc{\lamdd}[1]{{\Lam}_{D^{#1}}^\dagg}
\newc{\lamud}[1]{{\Lam}_{U^{#1}}^\dagg}
\newc{\lam}{{\bf \lambda}}
\newc{\lamp}{{\bf \lambda}^{\prime}}
\newc{\lampp}{{\bf \lambda}^{\prime\prime}}
\newc{\Y}{{\bf Y}}
\newc{\h}{{\bf h}}
\newc{\meee}{{{\rm {\bf  m}}_e}}
\newc{\mdee}{{{\rm {\bf  m}}_d}}
\newc{\myew}{{{\rm {\bf m}}_u}}
\newc{\ye}{{\Y}_E}
\newc{\he}{{\h}_E}
\newc{\hed}{{\h}_E^\dagger}
\newc{\yd}{{\Y}_D}
\newc{\hd}{{\h}_D}
\newc{\hdd}{{\h}_D^\dagger}
\newc{\yu}{{\Y}_U}
\newc{\hu}{{\h}_U}
\newc{\hud}{{\h}_U^\dagger}
\newc{\yes}{{\Y}_E^*}
\newc{\yds}{{\Y}_D^*}
\newc{\yus}{{\Y}_U^*}
\newc{\yet}{{\Y}_E^T}
\newc{\ydt}{{\Y}_D^T}
\newc{\yut}{{\Y}_U^T}
\newc{\yed}{{\Y}_E^\dagg}
\newc{\ydd}{{\Y}_D^\dagg}
\newc{\yud}{{\Y}_U^\dagg}
\newc{\dagg}{\dagger}
\newc{\lp}{\left(}
\newc{\rp}{\right)}
\newc{\inv}{\frac{1}{16\pi^2}}
\newc{\invsq}{\frac{1}{(16\pi^2)^2}}
\newc{\ggam}[2]{\gamma_{#2}^{#1}}
\newc{\yukgam}[2]{\inv \gamma_{#1}^{(1){#2}}+\invsq\gamma_{{#1}}^{(2){#2}}}
\newc{\susyunif}{ohman,nirpaul,marcelacarlos,susyunif}
\newc{\lsim}{\stackrel{<}{\sim}}
\newc{\gsim}{\stackrel{>}{\sim}}
\newc{\Tr}{{~\rm Tr}}
\newc{\me}{{(\bf m_{\tilde{E}}}^2)}
\newc{\mh}[1]{m_{H_{#1}}^2}
\newc{\ml}{{\bf m_{\tilde{L}}}^2}
\newc{\md}{{(\bf m_{\tilde{D}}}^2)}
\newc{\mup}{{(\bf m_{\tilde{U}}}^2)}
\newc{\mq}{{(\bf m_{\tilde{Q}}}^2)}
\newc{\mlh}[1]{{\bf m}_{ \tilde{L}_{#1} H_1}^2}
\newc{\mhl}[1]{{\bf m}_{ H_d \tilde{L}_{#1}}^2}
\newc{\del}{\partial}
\newc{\beq}{\begin{equation}}
\newc{\eeq}{\end{equation}}
\newc{\barr}{\begin{eqnarray}}
\newc{\earr}{\end{eqnarray}}
\newc{\dspl}{\displaystyle}
\newc{\phmin}{\phantom{-}}
\newc{\stau}{{\tilde\tau}}
\newc{\mnu}{$m_{\nu}$ }
\newc{\AO}{$A_0$ }
\newc{\vd}{$v_d$ }
\newc{\MGUT}{$M_{\text{GUT}}$}

\newc{\stext}[1]{{\color{green}  #1}}
\newc{\mtext}[1]{{\color{blue}  #1}}
\newc{\htext}[1]{{\color{red}  #1}}

\begin{document}


\title{Bounds on R--parity Violating Couplings at the Grand Unification 
Scale from Neutrino Masses}
\author{H.~K.~Dreiner}
\email[]{dreiner@th.physik.uni-bonn.de}
\affiliation{Bethe Center for Theoretical Physics and Physikalisches 
Institut, Universit\"at Bonn, Bonn, Germany}
\affiliation{SCIPP, University of California Santa Cruz, Santa Cruz, 
CA 95064, USA}

\author{M.~Hanussek}
\email[]{hanussek@th.physik.uni-bonn.de}
\affiliation{Bethe Center for Theoretical Physics and Physikalisches 
Institut, Universit\"at Bonn, Bonn, Germany}

\author{S.~Grab}
\email[]{sgrab@scipp.ucsc.edu}
\affiliation{SCIPP, University of California Santa Cruz, Santa Cruz, 
CA 95064, USA}

\begin{abstract}
We consider the embedding of the supersymmetric Standard Model with
broken R--parity in the minimal supergravity (mSUGRA) model. We
restrict ourselves to the case of broken lepton number, the B$_3$
mSUGRA model. We first study in detail how the tree--level neutrino
mass depends on the mSUGRA parameters. We find in particular a strong
dependence on the trilinear supersymmetry breaking $A$--parameter,
even in the vicinity of the mSUGRA SPS1a point. We then reinvestigate
the bounds on the trilinear R-parity violating couplings at the
unification scale from the low--energy neutrino masses including
dominant one--loop contributions. These bounds were previously shown to
be very strict, as low as $\mathcal{O}(10^{-6})$ for SPS1a. We show
that these bounds are significantly weakened when considering the full
mSUGRA parameter space. In particular the ratio between the
tree--level and 1--loop neutrino masses is reduced such that it may
agree with the observed neutrino mass hierarchy. We discuss in detail
how and in which parameter regions this effect arises.

\end{abstract}

\preprint{BONN--TH--2010-02, SCIPP 10/02}

\maketitle

\section{Introduction}

The experimental observation of neutrino oscillations, and thus of
neutrino masses, is an experimental indication that the Standard
Model of particle physics (SM) is incomplete
\cite{Cleveland:1998nv,Fukuda:1998ah,Fukuda:1998fd,Ahmad:2002jz,Aharmim:2005gt,Apollonio:2002gd,GonzalezGarcia:2007ib}.

Experimentally, neutrinos must be relatively light. Direct laboratory
measurements restrict their masses to be below $\mathcal{O}(10
\,\text{MeV} - 1\, \text{eV})$
\cite{GonzalezGarcia:2007ib,Amsler:2008zzb,
Barate:1997zg,Assamagan:1995wb,Bonn:2001tw,Lobashev:2001uu}, depending
on the flavor. Cosmological observations even give upper bounds of
$\mathcal{O}(0.1\,\text{eV})$ on the sum of the neutrino masses
\cite{GonzalezGarcia:2007ib,Amsler:2008zzb,Cirelli:2006kt,Goobar:2006xz}.
Furthermore, the atmospheric and solar neutrino oscillation data are
best fit if the squared neutrino mass differences are
$\mathcal{O}(10^{-3} \text{eV}^2)$ and $\mathcal{O}(10^{-5}
\text{eV}^2)$, respectively
\cite{GonzalezGarcia:2007ib,Wendell:2010md}.  This allows for one
massless neutrino.

In principle, it is easy to extend the SM Lagrangian by a Dirac
neutrino mass term \cite{GonzalezGarcia:2007ib}.  However,
right-handed neutrinos and new Yukawa couplings of $\mathcal{O}( \lsim
10^{-12})$ are in this case needed.  Such tiny couplings seem to be
very unnatural and might point towards a dynamical mechanism, that
explains the small neutrino masses. Furthermore, the right--handed
neutrinos can have an unspecified Majorana neutrino mass.

Most prominently discussed are extensions of the SM involving the
see--saw mechanism, by introducing right-handed neutrinos and
fixing the new Majorana neutrino mass scale to be large, \textit{cf.}
Refs.~\cite{GonzalezGarcia:2007ib,Minkowski:1977sc,Yanagida,Mohapatra:1979ia,GellMann:1980vs,Mohapatra:1980yp,Jezabek:1998du}.
The see--saw mechanism is also naturally incorporated into
supersymmetry (SUSY) \cite{Haber:1984rc,Martin:1997ns}.

Supersymmetry is one of the most promising extensions of the SM.  It
is the unique extension of the Lorentz spacetime symmetry, when
allowing for graded Lie algebras
\cite{Coleman:1967ad,Haag:1974qh}. Furthermore, it provides a
solution to the hierarchy problem of the SM
\cite{Drees:1996ca,Gildener:1976ai,Veltman:1980mj,Sakai:1981gr,Witten:1981nf}.
More importantly here: neutrino masses can be generated without
introducing right-handed neutrinos if
lepton number is violated, {\it cf.} for example
Refs.~\cite{Hall:1983id,Joshipura:1994ib,Nowakowski:1995dx,Nardi:1996iy,Davidson:2000ne,Dedes:2006ni,Dreiner:2007uj,Grossman:1997is,Grossman:1999hc,Grossman:2000ex,Allanach:2007qc}.

The most general gauge invariant and renormalizable superpotential of
the supersymmetric extension of the SM with minimal particle content
(SSM) possesses lepton number conserving (LNC) terms
\cite{Sakai:1981pk,Weinberg:1981wj}
\begin{eqnarray}
W_{\text{LNC}} &=& \eps_{ab} [ (\ye)_{ij} L_i^a
H_d^b {\bar E}_j + (\yd)_{ij} Q_i^{ax} H_d^b {\bar D}_{jx} \nonumber \\
& &+(\yu)_{ij} Q_i^{ax} H_u^b {\bar U}_{jx} - \mu H_d^a H_u^b ]
\label{LNC_superpot}
\end{eqnarray}
and also lepton number violating (LNV) terms
\begin{eqnarray}
W_{\text{LNV}} &=& \eps_{ab} \left[ \frac{1}{2} \lam_{ijk} L_i^a L_j^b{\bar E}_k +
\lamp_{ijk} L_i^a Q_j^{xb} {\bar D}_{kx} \right] \nonumber \\
& & - \eps_{ab} \kap_i L_i^a H_u^b  \, ,
\label{LNV_superpot}
\end{eqnarray} 
where $i,j,k=1,2,3$ are generation indices. We have employed the standard notation 
of Ref.~\cite{Allanach:1999ic}. 

The LNV interactions violate the discrete symmetries R--parity and
proton--hexality ($\text{P}_6$), however, they conserve baryon triality
($\text{B}_3$)
\cite{Ibanez:1991hv,Dreiner:2005rd,Ibanez:1991pr,Banks:1991xj}.  Note
that $\text{B}_3$ stabilizes the proton because it suppresses the
baryon number violating interactions. R--parity, $\text{P}_6$ and
$\text{B}_3$ are the only discrete symmetries, which can be written as
a remnant of a broken anomaly free gauge symmetry
\cite{Dreiner:2005rd,Ibanez:1991hv,Ibanez:1991pr,Banks:1991xj}.  In
the following, we assume that $\text{B}_3$ is conserved and thus
R--parity and $\text{P}_6$ are violated. 
Eq.~(\ref{LNC_superpot}) and Eq.~(\ref{LNV_superpot}) constitute the
full renormalizable superpotential allowed by this symmetry.
For reviews of such theories see for example
Refs.~\cite{Bhattacharyya:1996nj,Dreiner:1997uz,Barbier:2004ez}.

Beside the superpotential, also the soft-breaking Lagrangian of the
$\text{B}_3$ conserving SSM exhibits lepton number violating operators
\cite{Allanach:2003eb}
\begin{eqnarray}
-{\cal L}_{\rm soft}^{\rm LNV} 
&=& \eps_{ab} \left[  \frac{1}{2} h_{ij k}\tilde{L}_i^a \tilde{L}_j^b \tilde{\bar{E}}_k 
+h^\prime_{i jk} \tilde{L}_i^a \tilde{Q}_j^b \tilde{\bar{D}}_k ~+~{\rm h.c.}  \right] \nonum \\
& & - \, \eps_{ab} \tilde{D}_i \tilde{L}_i^a h_u^b ~+~{\rm h.c.}
+  \, ({h^*_d})^a  \mathbf{m}^2_{h_d \tilde{L}_i} {\tilde{L}^a_i} \, , \nonum \\
\label{LNV_Lsoft}
\end{eqnarray}
where again $i,j,k=1,2,3$ are generation indices. $\tilde{L}$,
$\tilde{\bar{E}}$, $\tilde{Q}$ and $\tilde{\bar{D}}$ are the scalar
components of the lepton doublet, lepton singlet, quark doublet and
down quark singlet superfield, respectively. Furthermore, $h_u$
($h_d$) denotes the up--type (down--type) scalar Higgs
field. Beside the term proportional to $\mathbf{m}^2_{h_d
\tilde{L}_i}$, the operators in Eq.~(\ref{LNV_Lsoft}) are the
soft-breaking analog of the terms in Eq.~(\ref{LNV_superpot}).

The LNV terms in Eq.~(\ref{LNV_superpot}) and Eq.~(\ref{LNV_Lsoft})
lead to the dynamical generation of neutrino masses. For example, the
bilinear terms in Eq.~(\ref{LNV_superpot}) mix the Higgsinos, the
supersymmetric partners of the Higgs bosons, with the neutrino fields
and thus generate one non--vanishing neutrino mass at tree--level
\cite{Hall:1983id,Joshipura:1994ib,Nowakowski:1995dx,Nardi:1996iy,
Davidson:2000ne,Dedes:2006ni,Dreiner:2007uj}. 

In this paper, we derive bounds on the trilinear LNV couplings of the
superpotential, Eq.~(\ref{LNV_superpot}), from the upper cosmological
bound on the sum of neutrino masses
\cite{Cirelli:2006kt,Goobar:2006xz}, {\it i.e.}
\begin{equation}
\sum {m_{\nu_i}} < 0.40 \, \text{eV} \, ,
\label{bound_numass}
\end{equation}
at $99.9\%$ confidence level. The bound was determined by a
combination of the Wilkinson Microwave Anisotropy Probe (WMAP) and
Large Scale Structure (LSS) data.

In order to perform a systematic study, we restrict ourselves to the
well motivated framework of the $\text{B}_3$ minimal supergravity
model (mSUGRA) \cite{Allanach:2003eb}, which provides simple boundary
conditions for the SSM parameters at the grand unification scale
($M_{\rm GUT}$).  We describe the model in the next section in
detail. We employ the full set of renormalization group equations
(RGEs) at one loop
\cite{Carlos:1996du,Besmer:2000rj,Allanach:2003eb,Martin:1993zk}
in order to obtain the $\text{B}_{3}$ SSM spectrum and the neutrino
masses at the electroweak scale ($M_{\rm EW}$).  We then derive bounds
on the LNV trilinear couplings at $M_{\rm GUT}$.

Bounds on trilinear LNV couplings within this model were also derived
in Ref.~\cite{Allanach:2003eb} from the generation of neutrino masses
at tree--level. It was claimed that neutrino masses put an upper bound
of $\mathcal{O}(10^{-3}-10^{-6})$ on most of the trilinear couplings
in Eq.~(\ref{LNV_superpot}).  However, it was shown in
Ref.~\cite{Carlos:1996du} that the tree--level neutrino mass can
vanish in certain regions of the $\text{B}_3$ mSUGRA parameter space.
In our analysis, we especially focus on these regions of parameter
space. We show that the bounds on the trilinear couplings can be
weakened up to $\mathcal{O}(10^{-1})$, depending on the boundary
conditions.

We go beyond the former work in several aspects. Beside the
tree--level neutrino mass, we also include the dominant contributions
to the neutrino mass matrix at one--loop. These contributions were
neither included in the calculation of the bounds in
Ref.~\cite{Allanach:2003eb} nor in Ref.~\cite{Carlos:1996du}. However,
as we show in Sec.~\ref{mSUGRA}, the loops dominate in the regions of
parameter space where the tree--level mass vanishes.  They must thus
be included when determining the bounds.

In Ref.~\cite{Carlos:1996du} there is only a brief explanation of the
dominant effect that leads to a vanishing tree--level mass in
$\text{B}_3$ mSUGRA.  We give for the first time a detailed and
complete explanation of how different configurations of the
$\text{B}_3$ mSUGRA parameters at $M_{\rm GUT}$ can affect the
tree--level {\it and} loop contributions to the neutrino masses at
$M_{\rm EW}$. Although we restrict ourself to the framework of
$\text{B}_3$ mSUGRA, the mechanisms described in this publication also
work in more general models. Furthermore, we calculate bounds for {\it
all} trilinear LNV couplings, whereas Ref.~\cite{Carlos:1996du}
focused only on the couplings $\lambda_{i33}$ and $\lambda'_{i33}$.
We also update the bounds given in Ref.~\cite{Allanach:2003eb}
according to the more recent and stronger bound on the sum of neutrino
masses, {\it cf.}  Eq.~(\ref{bound_numass}).

Going beyond the work presented here, we believe our results can help
find LNV SUSY scenarios that explain the observed neutrino masses and
mixing angles.  Within the framework of $\text{B}_3$ mSUGRA,
Ref.~\cite{Allanach:2007qc} searched for a minimal set of LNV
parameters which can explain the measured neutrino parameters. They
found sets of five parameters [two trilinear LNV couplings together
with the three mixing angles that describe the lepton Yukawa matrix,
{\it cf.}  Eq.~(\ref{LNC_superpot})] that give the right masses and
mixing angles.  Ref.~\cite{Allanach:2007qc} claimed that the
tree--level mass is always much larger than the loop induced
masses. But we show in the following, that the loops can exceed the
tree--level masses in $\text{B}_3$ mSUGRA.  Therefore, it should be
possible to find a smaller set of LNV parameters that lies in this
region of parameter space and thus posses much larger LNV couplings
than those found in Ref.~\cite{Allanach:2007qc}. However, an
investigation of the complete neutrino sector is beyond the scope of
this paper and will be postponed to a future publication.

We finally note that (large) trilinear LNV couplings
can lead to distinct collider signatures at the Large Hadron Collider
(LHC), {\it e.g.} 
\begin{itemize}
\item Supersymmetric particles (sparticles) can be produced singly at
a collider, possibly on resonance
\cite{Barbier:2004ez,Dreiner:1991pe,Dreiner:2000vf,Dreiner:2006sv,
Dreiner:2008rv,Bernhardt:2008mz, singleslep, Allanach:1997sa, Arai:2010ci}.
For example, single resonant slepton production at the LHC via
$\lambda'_{ijk}$, Eq.~(\ref{LNV_superpot})
\cite{Dreiner:2000vf,Dreiner:1991pe,Dreiner:2006sv,Dreiner:2008rv,singleslep}.
An excess over the SM backgrounds is visible if $\lambda'_{ijk}
\gtrsim \, \mathcal{O}(10^{-3})$, depending also on the sparticle masses
\cite{Dreiner:2000vf,Dreiner:2006sv,Dreiner:2008rv,Allanach:2009iv}.
\item A LNV coupling $\lambda_{ijk}$ ($\lambda'_{ijk}$) of
  $\gtrsim \,
\mathcal{O}(10^{-2})$ at $M_{\rm GUT}$ can significantly change the
running of the sparticle masses, such that the scalar electron or muon
(sneutrino) is the LSP \cite{Allanach:2003eb,Bernhardt:2008jz,Dreiner:2008ca,
Allanach:2006st,Jack:2005id}.  This
can dramatically change the SUSY collider signatures, because (heavy)
sparticles normally cascade decay down to the LSP
\cite{Bernhardt:2008jz,Dreiner:2009fi}.
\end{itemize}

Note that the generation of neutrino masses via the bilinear terms in
Eq.~(\ref{LNV_superpot}) and the corresponding collider signatures
have also been investigated; see for example
Refs.~\cite{Hempfling:1995wj,Kaplan:1999ds,Mira:2000gg,Hirsch:2000ef,Hirsch:2002ys,Diaz:2003as,
Bartl:2003uq,Hirsch:2003fe,Hirsch:2004he,Hirsch:2005ag,deCampos:2007bn,deCampos:2008av}
and references therein.

This paper is organized as follows. In Sec.~\ref{model}, we review the
parts of the $\text{B}_3$ mSUGRA model that are relevant for this
work. Sec.~\ref{chapM} then shows the different contributions to the
neutrino mass matrix that we employ to derive the bounds.  We explain
the main mechanism leading to a vanishing tree--level neutrino
mass in $\text{B}_3$ mSUGRA in Sec.~\ref{mSUGRA} and derive the bounds
on the LNV trilinear couplings in Sec.~\ref{bounds}. These sections
are the central part of our paper. We
conclude in Sec.~\ref{conclusion}.

App.~\ref{further_msugra} explains the additional subleading
dependence of the neutrino masses on the $\text{B}_3$ mSUGRA
parameter not described in Sec.~\ref{mSUGRA}.

\section{The $\text{B}_3$ mSUGRA Model}
\label{model}

The general B$_3$ SSM has more than 200 free parameters
\cite{Haber:1997if}. This large number is intractable for detailed
phenomenological studies.  For that purpose the simplifying
$\text{B}_3$ mSUGRA model was proposed in Ref.~\cite{Allanach:2003eb},
which we now discuss.

\subsection{Free Parameters}

In the $\text{B}_3$ mSUGRA model the boundary conditions at $M_{\rm
GUT}$ are described by the six parameters
\begin{equation}
M_0, \; M_{1/2}, \; A_0,  \;\tan\beta,  \;\textrm{sgn}(\mu),  \;\mathbf{\Lambda}\, ,
\end{equation}
with 
\begin{equation}
\mathbf{\Lambda} \in \{ \lambda_{ijk}, \lambda'_{ijk} \} \, .
\end{equation}
Here $M_0, \; M_{1/2}$ and $A_0$ are the universal scalar mass, the
universal gaugino mass and the universal trilinear scalar coupling at
the grand unification scale ($M_{\rm GUT}$), respectively.
$\tan\beta$ denotes the ratio of the Higgs vacuum expectation values
(vevs) $v_u$ and $v_d$, and sgn($\mu$) fixes the sign of the bilinear
Higgs mixing parameter $\mu$.  The magnitude of $\mu$ is
determined dynamically by radiative electroweak symmetry
breaking (REWSB) \cite{Ibanez:1982fr}.  These five parameters are the
conventional free parameters of the R-parity or proton-hexality
conserving mSUGRA model \cite{msugramodel}.

In order to incorporate the effects of the LNV interactions in
Eq.~(\ref{LNV_superpot}) and Eq.~(\ref{LNV_Lsoft})
\textit{exactly} one additional non--vanishing trilinear coupling
$\mathbf{\Lam}\in\{ \lam_{ijk}, \lam'_{ijk} \} $ is assumed at
$M_{\rm GUT}$. Further LNV couplings are generated via the RGEs
at the lower scale. Note, that the bilinear couplings $\kappa_i$ and
$\tilde{D}_i$ are both set to zero at $M_{\rm GUT}$ via a basis
transformation of the lepton and Higgs superfields \cite{Hall:1983id}.
(For the most general case of a complex rotation see
Ref.~\cite{Dreiner:2003hw}.) This is natural for universal SUSY
breaking \cite{Allanach:2003eb}.  However, at lower scales $\kappa_i$
and $\tilde{D}_i$ are generated via the RGEs \cite{Nardi:1996iy}; see
Sec.~\ref{REWSB}.

The complete low energy spectrum is obtained by running the RGEs down
from $M_{\rm GUT}$ to $M_{\rm EW}$.  For that purpose we employ the
program {\tt SOFTSUSY-3.0.12} \cite{Allanach:2001kg,Allanach:2009bv}.
We calculate the neutrino masses with our own program. Note that we
work in the CP-conserving limit throughout this paper.

\subsection{Benchmark Scenarios for Parameter Scans}
\label{example_points}

We center our analysis around the following $\text{B}_3$ mSUGRA
parameter points\\

\textbf{Point I: } $M_{1/2} = 500$ GeV, 
$M_0 = 100$ GeV,
tan$\beta= 20$, 
sgn$(\mu) = +1$,
$A_0 = 900$ GeV,
$\mathbf{\Lambda} = \lambda'_{233}$\\

\textbf{Point II: } $M_{1/2} = 500$ GeV, 
$M_0 = 100$ GeV,
tan$\beta= 20$, 
sgn$(\mu) = +1$,
$A_0 = 200$ GeV, 
$\mathbf{\Lambda} = \lambda_{233}$\\

Point II differs from Point I only by the choice of the LNV coupling
and the size of $A_0$.  We have chosen these points as examples
because the tree--level contribution to the neutrino mass is small
around Point I and II and therefore one--loop contributions are
important. Both points lead to squark masses of $\mathcal{O}$(1 TeV)
and slepton masses of around 300 GeV, with a scalar tau (stau) as the
LSP.

Note that in the LNV SSM a stau LSP is as well motivated as a
neutralino LSP
\cite{Allanach:2003eb,Allanach:2006st,Allanach:2007vi,Dreiner:2008rv,Akeroyd:1997iq,Akeroyd:2001pm}.
Either will decay via the LNV interactions and cosmological
constraints do not apply \cite{Ellis:1983ew}.

In addition, we ensured that both points lie in regions of parameter
space where various other experimental constraints are fulfilled, such
as the lower bound on the lightest Higgs mass from LEP2
\cite{Schael:2006cr,Barate:2003sz} and constraints from the anomalous
magnetic moment of the muon \cite{Bennett:2006fi}, from $b \rightarrow
s \gamma$ \cite{Barberio:2008fa}, and from $B_s \rightarrow \mu^+
\mu^-$ \cite{Barberio:2008fa}; see Sec.~\ref{bounds} for
details.

\subsection{Renormalization Group Equations and Radiative Electroweak Symmetry 
Breaking}
\label{REWSB}
An important feature of the $\text{B}_3$ mSUGRA model is that lepton
number violation leads to mixing between the lepton superfields $L_i$
and the Higgs superfield $H_d$. Furthermore, sneutrinos, the
superpartners of the neutrinos, can acquire vevs $v_i$
($i=1,2,3$). Note that it is possible to rotate away the $\kappa_i$
terms in the superpotential at any given energy scale by an orthogonal
rotation of the fields $\cL_\alpha \equiv (H_d, L_i)$
\cite{Hall:1983id,Dreiner:2003hw,Allanach:2003eb}.

The corresponding bilinear soft-breaking terms proportional to $\tilde
D_i$, Eq.~(\ref{LNV_Lsoft}), can be rotated away in conjunction with
$\kappa_i$ if $\tilde D_i$ and $\kappa_i$ are aligned. This condition
is fulfilled at $M_{\rm GUT}$ in the $\text{B}_{3}$ mSUGRA model if
the underlying supergravity superpotential satisfies the quite natural
condition~\cite{Allanach:2003eb}
\beq f(z_i;y_{\alpha}) = f_1(z_i) +
f_2(y_{\alpha}) \, , 
\eeq 
where the superfields $z_i$ belong to the observable sector and the
superfields $y_{\alpha}$ to the hidden sector.

However, when evolving the parameters down to the weak scale, $\kap
_i,\,\tilde{D}_i\not=0$ are generated via the RGEs. The leading terms
for $\Lam\in\{\lam'_{ijk}\}$ are given
by~\cite{Allanach:2003eb}
\begin{eqnarray}
16 \pi^2 \frac{d\kap_i}{dt} &=& - 3 \kap_i \left[ \frac{g_1^2}{5} + 
g_2^2 - (\textbf{Y}_U)^2_{33} - \frac{(\textbf{Y}_E)^2_{33}}{3} 
\delta_{3i} \right] \nonumber \\
& & -  3 \mu  \lam'_{ijk} (\textbf{Y}_D)_{jk} + ... 
\label{RGEkappa}
\end{eqnarray}
and 
\begin{eqnarray}
16 \pi^2 \frac{d\tilde{D}_i}{dt} &=&  - 3 \tilde{D}_i 
\left[ \frac{g_1^2}{5} + g_2^2 - (\textbf{Y}_U)^2_{33} - 
\frac{(\textbf{Y}_E)^2_{33}}{3} \delta_{3i} \right] \nonumber \\
& & + 6 \kap_i \left[ \frac{g_1^2}{5} M_1 + g_2^2 M_2 \right] \nonumber \\
& & + 6 \kap_i \left[ (\textbf{Y}_U)_{33} (\textbf{h}_U)_{33} + 
\frac{(\textbf{Y}_E)_{33}}{3}
(\textbf{h}_E)_{33} \, \delta_{3i} \right] \nonumber \\
& &-3 (\textbf{Y}_D)_{jk} (2 \mu \: h'_{ijk} \: + \tilde{B} \lam'_{ijk} )  + ... \,.
\label{RGEDi}
\end{eqnarray}
Here $t\equiv\ln(Q/\mu_0)$ with $Q$ the renormalization scale and
$\mu_0$ an arbitrary reference scale. $h'_{ijk}\equiv A_0 \times
\lam'_{ijk}$ at $M_{\text{GUT}}$, {\it cf.}
Eq.~(\ref{LNV_Lsoft}). $\tilde{B}$ is the soft supersymmetry breaking
analog of the Higgs mixing parameter $\mu$ and $(\textbf{h}_U)_{33}$
$[(\textbf{h}_E)_{33}]$ is the soft-breaking analog of the Yukawa
coupling $(\textbf{Y}_U)_{33}$ $[(\textbf{Y}_E)_{33}]$
\cite{Allanach:2003eb}.  $g_1$ and $g_2$ ($M_1$ and $M_2$) are the
$\text{U}(1)_Y$ and $\text{SU}(2)$ gauge couplings (gaugino masses),
respectively.  We see in Eqs.~(\ref{RGEkappa}) and (\ref{RGEDi}) that
the RGEs differ, and therefore $\kap_i$ and $\tilde{D}_i$ will no
longer be aligned at the weak scale \cite{Nardi:1996iy}. The case
$\mathbf{\Lam} \in \{\lam_{ijk}\}$ is analogous up to
the color factor 3.

The sneutrino vevs $v_i$, the bilinear Higgs parameter $|\mu|$ and the
corresponding soft breaking term $\tilde{B}$ are determined by REWSB,
which has been discussed in detail in~Ref.~\cite{Allanach:2003eb} for
the LNV case.

Neglecting higher order corrections
\cite{Chun:1999bq,Davidson:1999mc,Dedes:2002dy}, which are not
important for the following qualitative discussion {\footnote{We will briefly
turn to a discussion of the NLO corrections in
Sect.~\ref{msugra_abh_loops}. }}, the sneutrino vevs can be
written as \cite{Allanach:2003eb}
\begin{eqnarray}
(M_{\tilde{\nu}}^2)_{ij} v_j &=& -\biggl [ \mathbf{m}^2_{h_d
  \tilde{L}_i} +\mu\kap_i \biggr ]
v_d   + \tilde{D}_i v_u \;, 
\label{minVI}
\end{eqnarray}
with
\begin{eqnarray}
(M_{\tilde{\nu}}^2)_{ij} &=& (\ml)_{{ij}}+\kap_i\kap_j +\frac{1}{2}
M_Z^2 \cos2\beta \: \delta_{ij} \nonumber \\[1mm] 
&+& \frac{(g^2+g_2^2)}{2} \sin^2\beta\, \sum_l v_l^2 \: \delta_{ij} \;,
\label{sneutrino_mass}
\end{eqnarray}
where $(\ml)$ is the squared soft-breaking lepton doublet mass matrix
and $g=\sqrt{3/5} \, g_1$. $\mathbf{m}^2_{h_d \tilde{L}_i}$ originates
from the LNV soft-breaking Lagrangian, Eq.~(\ref{LNV_Lsoft}).  It mixes
the down--type Higgs fields, $h_d$, with the lepton doublet scalars,
$\tilde{L}_i$, and is zero at $M_{\rm GUT}$. That is, because we take
within mSUGRA the mass matrix of the fields
$\tilde{\mathcal{L}}_\alpha = (h_d, \tilde{L}_i)$ to be diagonal and
proportional to $M_0$ at $M_{\rm GUT}$. However, $\mathbf{m}^2_{h_d
\tilde{L}_i}\not=0$ is subsequently generated via the
RGEs, {\it cf.} Eq.~(\ref{running_mHL}).

As we will see in Sec.~\ref{tree_level}, sneutrino vevs and non--zero
bilinears $\kappa_i$ lead to neutrino masses at tree--level because
they mix neutrinos and neutralinos.

\subsection{Quark Mixing}
\label{quark_mixing}
The RGE evolution of the parameters in the $\text{B}_3$ mSUGRA model
from $M_{\rm GUT}$ to $M_{\rm EW}$ depends on the Higgs--Yukawa
coupling matrices $\mathbf{Y}_E$, $\mathbf{Y}_D$ and $\mathbf{Y}_U$,
\textit{cf.}  Eqs.~(\ref{RGEkappa}) and (\ref{RGEDi}). In particular,
the RGEs of the LNV violating parameters are coupled via the
non--diagonal matrix elements of the Higgs--Yukawa couplings.
Therefore a knowledge of the latter is crucial for the analysis of
bounds on the LNV parameters.

The initial parameter set of the $\text{B}_3$ mSUGRA model at $M_{\rm
GUT}$ is given in the electroweak basis so that for the RGE evolution
the Higgs--Yukawa couplings (or the quark-- and lepton--mass matrices)
are also needed in the electroweak basis. However, from experiment we
only know the masses and the CKM matrix
\begin{equation}
{\bf V}_{CKM}={\bf U}_{\bf L}^\dagger \, {\bf D}_{\bf L}
\end{equation}
at $M_{\rm EW}$. Here ${\bf U}_{\bf L}^\dagger$ (${\bf D}_{\bf
L}^\dagger$) rotate the left--handed up-- (down--) quark fields from the
mass eigenstate basis to the electroweak basis.  For simplicity, we
take ${\bf Y}_D$ and ${\bf Y}_U$ to be real and symmetric and thus the
rotation matrices for the right--handed quark fields are identical to
the ones for left--handed quark fields, ${\bf U}_{\bf R}={\bf U}_{\bf
L}$ and ${\bf D}_{\bf R}={\bf D}_{\bf L}$.  Because of the uncertainty
about the neutrino masses and mixings we will assume a diagonal
$\mathbf{Y}_E$ in the weak basis.

When determining the neutrino masses, we will consider two limiting
cases at $M_{\rm EW}$, following
Ref.~\cite{Agashe:1995qm,Allanach:2003eb,Dreiner:2008rv}:
\begin{itemize}
\item ``\textbf{up--type mixing}" the quark mixing is only in the
up--quark sector, 
\barr & & {\bf U}_{\bf L,R}= {\bf V}_{CKM},
\hspace{0.5cm} {\bf D}_{\bf L,R}={\bf 1}\,, \nonumber\\[0.2cm] & &
\mathbf{Y}_D \times v_d = \text{diag}(m_d,m_s,m_b)\,,
\label{up-mixing} \\ & & \mathbf{Y}_U \times v_u = {\bf
V}_{CKM}\cdot \text{diag}(m_u,m_c,m_t) \cdot {\bf V}_{CKM}^T
\nonumber \,.  
\earr
\item ``\textbf{down--type mixing}" the mixing is only in the
down--quark sector,
\barr 
& & {\bf D}_{\bf L,R}={\bf V}_{CKM}, \hspace{0.5cm} {\bf U}_{\bf L,R}=
{\bf 1}\,, \nonumber \\[0.2cm]
& & \mathbf{Y}_D \times v_d = {\bf V}_{CKM}\cdot \text{diag}(m_d,m_s,m_b) 
\cdot {\bf V}_{CKM}^T\,, \nonumber \\ 
& & \mathbf{Y}_U \times v_u = \text{diag}(m_u,m_c,m_t)  \label{down-mixing}\,.
\earr
\end{itemize}
Here $m_d,m_s,m_b$ ($m_u,m_c,m_t$) denote the masses of the down--type
(up--type) quarks.

The choice between up-- and down--type mixing has a strong effect on
the final results for the LNV couplings $\mathbf\Lam\in\{
\lam'_ {ijk}\}$ with $j\neq k$, as we will show in
Sec.~\ref{bounds} (see Tab.~\ref{tab:bounds}). The reason is that the
generated tree level neutrino mass is proportional to the
off--diagonal matrix element $(\mathbf{Y}_D)^2_{jk}$, {\it
cf.} the discussion in Sec.~\ref{chapM} and Sec.~\ref{mSUGRA}.  Our
results (for the tree--level neutrino mass) in Sec.~\ref{bounds} can
be easily translated to scenarios which lie between the limiting cases
of Eqs.~(\ref{up-mixing}) and (\ref{down-mixing}). One only needs to
know the respective Yukawa matrix elements $(\mathbf{Y}_D)_{jk}$.

\section{Neutrino Masses}
\label{chapM}

In this paper, we investigate bounds on lepton--number violating
couplings at $M_{\rm GUT}$ within the $\text{B}_3$ mSUGRA model, which
arise from the generation of too large neutrino masses at $M_{\rm
EW}$.  We therefore need to identify the dominant contributions to the
neutrino masses.

It was stated in Ref.~\cite{Allanach:2003eb} that the main
contribution stems from mixing between neutralinos and neutrinos,
which leads to one non-vanishing neutrino mass at tree--level, {\it
cf.}  Sec.~\ref{tree_level}. However, as we will show in the next two
sections, this is only true in parts of the $\text{B}_3$ mSUGRA
parameter space. It is possible that the different terms in the
tree--level mass formula cancel each other. We then need to identify
the dominant contributions, which arise at one--loop.

A complete list of all one--loop contributions is given in
Ref.~\cite{Davidson:2000ne}, where they are formulated in a
basis--independent manner. Most of the one--loop contributions are
proportional to the mass insertions that mix the neutrinos with the
neutralinos. They thus also vanish when the tree--level
neutrino mass vanishes and are negligible in the region we are
interested in.

\begin{figure}[t!]
\begin{center}
\epsfig{figure=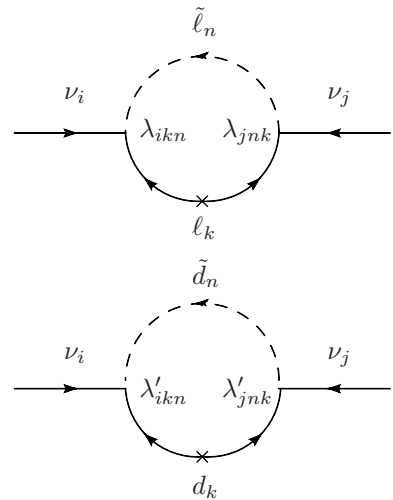}
\caption{Loop contributions to the neutrino mass matrix via a
 non-vanishing product of $\text{B}_3$ couplings
 $\lam_{ikn}\times \lam_{jnk}$ (upper figure) and
 $\lam'_{ikn}\times \lam'_{jnk}$ (lower figure).  See
 Sec.~\ref{lamlam_loop} for more details.}
\label{lamlam}
\end{center}
\end{figure}

\begin{figure}[t!]
\begin{center}
\epsfig{figure=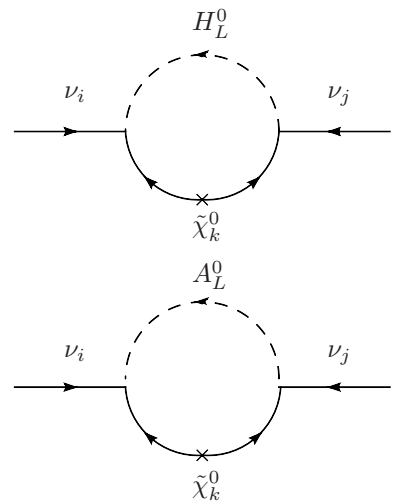}
\caption{Loop contributions to the neutrino mass matrix via a
  non--exact cancellation of loops with CP-even and CP-odd neutral
  scalars. Note, that there is a relative minus sign between the two
  diagrams. See Sec.~\ref{snusnu_loop} for more
  details.\label{snusnu}}
\end{center}
\end{figure}

The remaining dominant one--loop contributions are on the one hand due
to loops involving two R--parity violating vertices and are
thus either proportional to $\lam^2$ or to $\lam'^2$, {\it cf.}
Fig.~\ref{lamlam} \footnote{Mixed terms are suppressed due to
our assumption of a single dominant coupling at $M_{GUT}$. }. We will
review these contributions in Sec.~\ref{lamlam_loop}. On the other
hand, loops with virtual neutral scalars ({\it i.e.} Higgses and
sneutrinos) and neutralinos, which are shown in Fig.~\ref{snusnu}, can
also give large contributions to neutrino masses.  These loops are
proportional to the mass difference between CP-even and CP-odd
sneutrinos, {\it cf.} Sec.~\ref{snusnu_loop}. 

According to Ref.~\cite{Davidson:2000ne}, there is in principle also a
contribution which is proportional to $\lam \times
\tilde{D}_i$. However, this contribution is suppressed by two ore more
orders of magnitude in the regions of parameter space where the loops
dominate over the tree--level mass. Note that $\tilde{D}_i$ vanishes
near the tree--level mass minimum as we will show in
Sec.~\ref{mSUGRA}. We therefore neglect it in the following.

Further one--loop contributions are only present in a lepton- and
Higgs-superfield basis with non-vanishing sneutrino
vevs. This is the case in the $\text{B}_3$ mSUGRA model.  However, we
have checked that in our parameter scans, these contributions are at
least one order of magnitude smaller than the dominant one and thus
negligible for calculating the bounds.  Note, that they are also
aligned with the tree--level mass, because the sneutrino vevs vanish
near the tree--level mass minimum, {\it cf.}  Sec.~\ref{A0}.

We conclude, that the contributions to the neutrino masses which we
review in the following, are sufficient to calculate the correct
bounds on the LNV couplings $\lam$ and $\lam'$. However, in
order to calculate the correct neutrino mass spectrum and mixing
angles, all one--loop contributions given in
Ref.~\cite{Davidson:2000ne} must be taken into account. This lies
beyond the scope of this paper.

\subsubsection{Tree--Level Contributions}
\label{tree_level}

In the context of the $\text{B}_3$ mSUGRA Model, neutrino masses are
generated at tree--level through mixing between neutrinos and
neutralinos. Analogously to the standard see-saw mechanism
\cite{GonzalezGarcia:2007ib,Minkowski:1977sc,Yanagida,Mohapatra:1979ia,GellMann:1980vs,Mohapatra:1980yp,Jezabek:1998du}
(with the neutralinos taking over the role of the right-handed
neutrinos), an effective $3\times3$ neutrino mass matrix is
generated~\cite{Joshipura:1994ib,Nowakowski:1995dx},

\begin{widetext}
\begin{eqnarray}
\cM^{\nu}_{\textrm{eff}} &=& \frac{\mu (M_1g^2_2 + M_2g^2)}{2 v_u v_d(M_1g^2_2 + M_2g^2)- 2 \mu M_1M_2}
\left( \begin{array}{ccc}
\Delta_1\Delta_1 & \Delta_1\Delta_2 &\Delta_1\Delta_3\\
\Delta_2\Delta_1 & \Delta_2\Delta_2 &\Delta_2\Delta_3\\
\Delta_3\Delta_1 & \Delta_3\Delta_2 &\Delta_3\Delta_3 \end{array} \right)\, ,
\label{mnutree_matrix}
\end{eqnarray}
\end{widetext}
where $M_1$ ($M_2$) is the bino (wino) soft-breaking mass and
\begin{eqnarray}
\Delta_i &\equiv&  v_i - v_d \frac{\kappa_i}{\mu}, \qquad i=1,2,3 \, .
\label{Lambda}
\end{eqnarray}
This matrix has one non--zero eigenvalue which can at
$M_{\textrm{EW}}$ be simplified to \cite{Allanach:2003eb}
\begin{eqnarray}
m^{\textrm{tree}}_{\nu} \approx - \frac{16 \pi \alpha_{\textrm{GUT}}}{5} 
\frac{\sum_{i=1}^3 \Delta_i^2}{M_{1/2}}\,, 
\label{mnutree}
\end{eqnarray}
if we take into account the gaugino universality assumption at $M_{\rm
GUT}$, leading to $M_2 = \frac{3}{5} \frac{\alpha_2^2}{\alpha_1^2}
M_1 = \frac{\alpha_2^2}{\alpha_{GUT}^2} M_{1/2}$ at
$M_{\textrm{EW}}$~\cite{Allanach:2003eb}.  Here $\alpha_{\rm GUT} =
g_{\rm GUT}^2/4\pi \approx 0.041$ is the grand unified gauge coupling
constant~\cite{Allanach:2003eb}.

\subsubsection{Contributions from $\lam\lam$- and $\lam'\lam'$-Loops}
\label{lamlam_loop}
In the region of parameter space where the tree-level
neutrino mass, Eq.~(\ref{mnutree}), vanishes, loop induced neutrino masses 
give the dominant contributions. As we will show in Sec.~\ref{bounds},
the most important loops are those proportional to the product of 
two LNV trilinear couplings. 
The corresponding squark-quark and slepton-lepton loops are shown 
in Fig.~\ref{lamlam}. The resulting neutrino mass contributions 
are~\cite{Grossman:1999hc}
\barr 
(m_{\nu}^{\mathbf{\lam\lam}} )_{ij} &
=& \frac{1}{32 \pi^2} \sum_{k,n}\lam_{ikn}\lam_{jnk} m_{\ell_k} \sin
2\tilde{\phi}_n^{\ell}
\ln \left( \frac{m^2_{\tilde{\ell}_{1 n}}}{m^2_{\tilde{\ell}_{2 n}}} 
\right) \nonumber \\
&+& \frac{3}{32\pi^2} \sum_{k,n}\lam'_{ikn}\lam'_{jnk} m_{d_k} \sin
2\tilde{\phi}_n^{d} \ln\left( \frac{m^2_{\tilde{d}_{1
     n}}}{m^2_{\tilde{d}_{2 n}}} \right)
\, ,\nonumber \\
\label{llloop}
\earr
where $m_{\ell_k}$ ($m_{d_k}$) are the lepton (down-quark) masses of
generation $k$, and $\tilde{\phi}^\ell_n$ ($\tilde{\phi}^d_n$) the mixing
angles that describe the rotation of the left-- and right--handed
slepton (down-squark) current eigenstates of generation $n$ to the two
mass eigenstates, $m_{\tilde{\ell}_{1n}}$ and $m_{\tilde{\ell}_{2n}}$ 
($m_{\tilde{d}_{1 n}}$ and $m_{\tilde{d}_{2n}}$), respectively. Note
that the squared sfermion masses are linear functions of the mSUGRA
parameters $M_0^2$ and $M_{1/2}^2$, see for example
Ref.~\cite{Drees:1995hj}.  For the calculation of Eq.~(\ref{llloop}) and
all following calculations, we have used the two-component spinor
formalism as described in Ref.~\cite{Dreiner:2008tw}.

For the first two sfermion generations, the sfermion mixing angles are
small and we approximate Eq.~(\ref{llloop}) by using the mass
insertion approximation (MIA) as described in
Ref.~\cite{Allanach:2007qc}. The slepton (and down-squark) mass
eigenstates are replaced by the respective left-- and right--handed
eigenstates with mass $m_{\tilde{\ell}_{L n}}$ and $m_{\tilde{\ell}_{R
 n}}$. The mixing angle can be approximated by
\beq 
\sin 2\tilde{\phi}_n^{\ell} = \frac{2
 (M^{LR}_{\tilde{\ell}})^{2}_{n} }{m^2_{\tilde{\ell}_{L n}} -
 m^2_{\tilde{\ell}_{R n}}} \, ,
\label{mia1} 
\eeq
where
\beq 
(M_{\tilde{\ell}}^{LR})^{2}_n = m_{{\ell}_n} \left[ \frac{(\textbf{h}_{E})_{nn}}{(\textbf{Y}_{E})_{nn}} - \mu \tan\beta \right] 
\label{mia2}
\eeq
denotes the left--right mixing matrix element of the charged sleptons
of generation $n$. $(\textbf{h}_{E})_{nn}$ is the trilinear
soft-breaking analog of the lepton Yukawa matrix element
$(\textbf{Y}_{E})_{nn}$ \cite{Allanach:2003eb}.

A similar formula is obtained for $\sin 2\tilde{\phi}_n^d$. One
only needs to replace in Eq.~(\ref{mia1}) and Eq.~(\ref{mia2}) $\ell
\leftrightarrow d$, $\tilde{\ell} \leftrightarrow \tilde{d}$,
$(\textbf{Y}_{E})_{nn} \leftrightarrow (\textbf{Y}_{D})_{nn}$ and
$(\textbf{h}_{E})_{nn} \leftrightarrow (\textbf{h}_{D})_{nn}$, where
$(\textbf{h}_{D})_{nn}$ is the soft-breaking analog of the down-quark
Yukawa matrix element $(\textbf{Y}_{D})_{nn}$.

\subsubsection{Contributions from Neutral Scalar--Neutralino--Loops}
\label{snusnu_loop}

As the final source of neutrino masses, we consider contributions
arising from loops with neutral scalars and neutralinos, \textit{cf.}
Refs.~\cite{Grossman:1997is,Grossman:1999hc,Grossman:2000ex}.  Most
important is the contribution from sneutrino--antisneutrino mixing, as
we will see in Eq.~(\ref{msnu_approx}).

If CP is conserved, sneutrinos $\tilde\nu_i$ and 
antisneutrinos $\tilde\nu_i^*$ mix to form CP--invariant mass eigenstates
\begin{eqnarray}
\tilde{\nu}_i^+&\equiv& \frac{1}{\sqrt{2}}(\tilde\nu_i+\tilde\nu_i^*) \, ,\\
\tilde{\nu}_i^-&\equiv& \frac{1}{i\sqrt{2}}(\tilde\nu_i-\tilde\nu_i^*) \, .
\end{eqnarray}

\noindent If lepton number is conserved, the $\tilde\nu_i^\pm$ masses
are degenerate and the CP-even (CPE) and CP-odd (CPO) contributions to
the neutrino mass from neutral scalar--neutralino--loops cancel,
\textit{cf.}  Fig.~\ref{snusnu}.

In contrast, if lepton number is violated, the $\tilde\nu_i^\pm
$ masses are in general different, so the cancellation is no longer
exact.  This is due to the fact that the CPE and CPO neutrinos mix
differently with the CPE and CPO Higgs fields, respectively.  The size
of this contribution to the neutrino masses is roughly proportional to
the mass splitting $\Delta m_{\tilde{\nu}_i}^2 = m_{\tilde{\nu}^+_i}^2
- m_{\tilde{\nu}_i^-}^2$, \textit{cf.}  Eq.~(\ref{msnu_approx}) and
Refs.~\cite{Grossman:1997is,Grossman:1999hc,Grossman:2000ex}.

The neutral scalar--neutralino-loops, shown in Fig.~\ref{snusnu}, lead
to the following contributions to the neutrino mass matrix
\cite{Allanach:2007qc} 
\barr (m_{\nu}^{\tilde\nu \tilde{\nu}} )_{ij} &
=& \frac{1}{32 \pi^2} \sum_{k=1}^4
\sum_{L=1}^5 m_{\tilde{\chi}^0_k} (g N_{1k} - g_2 N_{2k})^2 \nonumber\\
& & \times \biggl [Z^+_{(2+i)L} Z^+_{(2+j)L} \textrm{B}_0(0,m^2_{H^0_L},m_{\tilde{\chi}^0_k}^2) \nonumber\\
& & - Z^-_{(2+i)L} Z^-_{(2+j)L} \textrm{B}_0(0,m^2_{A^0_L},m_{\tilde{\chi}^0_k}^2) \biggr] \, ,\nonumber\\
\label{msnu}
\earr
where $m_{\tilde{\chi}^0_k}$ $(k=1 \dots 4)$ are the neutralino masses
and $N$ is the $4\times 4$ neutralino mixing matrix in the bino, wino,
Higgsino basis \cite{Gunion:1984yn}.  The two-point Passarino-Veltman
function is conventionally denoted $\textrm{B}_0$
\cite{Denner:1991kt}.  $m_{H^0_L}$ ($m_{A^0_L}$) with $L=1, \ldots, 5$
are the mass eigenvalues of the CPE (CPO) neutral Higgs bosons and CPE
(CPO) sneutrino fields. They can be obtained with the help of the
unitary matrix $Z^+$ ($Z^-$), which diagonalizes the mass
matrices of the CPE (CPO) neutral scalars, {\it i.e.}
\barr
(Z^+)^T \cM_{\textrm{CPE}} Z^+ &=& \textrm{diag}(m^2_{h^0}, 
m^2_{H^0},m^2_{\tilde{\nu}^+_1},m^2_{\tilde{\nu}^+_2}, 
m^2_{\tilde{\nu}^+_3}) \nonumber\\
&\equiv & \textrm{diag}(m^2_{H^0_L}) 
\earr 
and 
\barr
(Z^-)^T \cM_{\textrm{CPO}} Z^- &=& \textrm{diag}(m^2_{G^0}, m^2_{A^0},m^2_{\tilde{\nu}^-_1},m^2_{\tilde{\nu}^-_2}, m^2_{\tilde{\nu}^-_3}) \nonumber\\
&\equiv & \textrm{diag}(m^2_{A^0_L}) \;; \earr see
Ref.~\cite{Allanach:2007qc} for additional details.

In order to analyze the dependence of this contribution on the mSUGRA
parameters, we make use of the fact that in the $\text{B}_3$ mSUGRA
model, Eq.~(\ref{msnu}) can be approximated by \cite{Dedes:2006ni}
\barr
(m_{\nu}^{\tilde\nu \tilde{\nu}} )_{ij} & \approx & \frac{1}{32 \pi^2} \sum_{k=1}^4 m_{\tilde{\chi}^0_k}^3 (g N_{1k} - g_2 N_{2k})^2 \nonumber\\
& & \times \frac{\Delta m_{\tilde{\nu}_i}^2}{(m_{\tilde{\nu}_i}^2 -
m_{\tilde{\chi}^0_k}^2)^2} \ln \left(
\frac{m_{\tilde{\chi}^0_k}^2}{m_{\tilde{\nu}_i}^2} \right)
\delta_{ij}
\label{msnu_approx}
\earr
by expanding around $m^2_{H^0_{L>2}}$ and $m^2_{A^0_{L>2}}$.  The mass
splitting, $\Delta m_{\tilde{\nu}_i}^2$, in Eq.~(\ref{msnu_approx})
between CPE and CPO sneutrinos of generation $i$ is then given by
\cite{Grossman:2000ex}
\barr
\Delta m_{\tilde{\nu}_i}^2 &=& \frac{- 4 \tilde{B}^2 M_Z^2 
m_{\tilde{\nu}_i}^2 \sin^2\beta}{(m^2_{H^0} - m_{\tilde{\nu}_i}^2) 
(m^2_{h^0} - m_{\tilde{\nu}_i}^2)(m^2_{A^0} - m_{\tilde{\nu}_i}^2)} 
\nonumber \\
& & \times \frac{(\tilde{B} v_i - \tilde{D}_i v_d)^2}{(v_d^2 
+ v_i^2) (\tilde{B}^2 + \tilde{D}_i^2)}\, .
\label{mass_splitting}
\earr

\subsubsection{Numerical Implementation}
\label{numerical_numass}

The numerical calculation of the neutrino mass matrix was done in the
following way.  We first employed {\tt
SOFTSUSY-3.0.12}~\cite{Allanach:2001kg,Allanach:2009bv} to obtain
the low energy mass spectrum \footnote{We use as SM inputs for
{\tt SOFTSUSY} the following parameters: $M_Z=91.1876$ GeV ($m_t=
165.0$ GeV) for the pole mass of the $Z$ boson (top quark); $\alpha
^{-1}(M_Z)=127.925$ and $\alpha_s(M_Z)=0.1176$ for the gauge couplings
in the $\overline{MS}$ scheme; $m_b(m_b)=4.20$ GeV, $m_u(2\text{GeV})
=0.0024$ GeV, $m_d(2\text{GeV})=0.00475$ GeV, $m_s(2\text{GeV})=0.104$
GeV and $m_c(m_c)=1.27$ GeV for the light quark masses in the 
$\overline{MS}$ scheme.}.
We then used our own program to calculate the neutrino mass matrix.
The tree--level contribution was derived from
Eq.~(\ref{mnutree_matrix}). For the $\lam \lam$-- and $\lam'
\lam'$--loops, we employed Eq.~(\ref{llloop}), if third generation
sfermions were involved. However, for sfermions of the first two
generations we used the MIA as given in Eqs.~(\ref{mia1}) and
(\ref{mia2}). 


For the neutral scalar--neutralino--loops, we in principle employed
Eq.~(\ref{msnu}). However, instead of performing the large numerical
cancellation between CPE and CPO neutral scalars directly [square
bracket in Eq.~(\ref{msnu})], we used an MIA to calculate the deviation
from exact cancellation in the R-parity conserving (RPC) limit,
following Ref.~\cite{Allanach:2007qc}.  The resulting formula is quite
lengthy and we refer the interested reader to
Ref.~\cite{Allanach:2007qc} for details. We have cross checked our
program with the help of Eq.~(\ref{msnu_approx})
and Eq.~(\ref{mass_splitting}).  All our calculations are performed in the
CP-conserving limit.

\section{$\nu$--Masses: Dependence on mSUGRA Parameters}
\label{mSUGRA}

In the literature it has frequently been assumed that the tree--level
contribution to the neutrino mass, Eq.~(\ref{mnutree}), in the
$\text{B}_3$ mSUGRA model dominates over the loop contributions,
\textit{cf.} for example Refs.~\cite{Allanach:2003eb,Allanach:2007qc}.
However, as has been noted in Ref.~\cite{Carlos:1996du}, in certain
regions of $\text{B}_3$ mSUGRA parameter space, the tree--level
neutrino mass vanishes even when $\kappa_i \neq 0$.

We demonstrate this effect in Fig.~\ref{figa0}, where we display the
tree--level neutrino mass (solid red line) as a function of $A_0$. The
other $\text{B}_3$ mSUGRA parameters are given by Point I with
$\lam'_{233}|_{\rm GUT}=10^{-5}$, {\it cf.}
Sec.~\ref{example_points}. We see that the tree--level mass,
$m_{\nu}^{\textrm{tree}}$, vanishes around $A_0\approx 910$ GeV. In
the vicinity of this minimum, $m_{\nu}^{\textrm{tree}}$ drops by
several orders of magnitude over a wide range of $A_0$, and it is
therefore not a (large) fine--tuning effect. In this case the loop
contributions will dominate the neutrino mass matrix, resulting in
much weaker bounds on the involved $\bf{\Lam}$ coupling, \textit{cf.}
Sec.~\ref{bounds}. Thus the bound crucially depends on the choice of
$A_0$.

We emphasize that the range of $A_0$ for which weaker bounds
may be obtained is quite large.  In an interval of $\Delta A_0 \approx
100\,\mathrm{GeV}$ around the minimum, we obtain bounds on
$\lam'_{233}$ that are at least one order of magnitude smaller than
the bound derived at for example $A_0=0\gev$. Much weaker bounds can
therefore be obtained without a lot of fine tuning.

In this section, we aim to explain in detail the origin of this
cancellation, considering as an explicit example the case ${\bf \Lam}
\in \{\lam'_{ijk} \}$. We focus on the dependence of
$m_{\nu}^{\textrm {tree}}$ on the mSUGRA parameter $A_0$, because it
is always possible to find a value of $A_0$ [for a given set of
parameters $\tan\beta$, $M_{1/2}$, $M_0$, and sgn$(\mu)$] such that
the tree--level neutrino mass vanishes. All arguments can analogously
be applied to a $\lam_ {ijk}$ coupling, as discussed in
App.~\ref{lam_LLE}.  Note for the further discussion that we can
always obtain a positive ${\bf \Lam}$ by absorbing a possible sign of
${\bf \Lam}$ via a re--definition $L \rightarrow -L$ and $E
\rightarrow -E$ of the lepton doublet and lepton singlet superfields,
respectively.  We also note that the generated neutrino masses scale
roughly with ${\bf \Lam}^2$, {\it cf.}  the following discussion.

\subsection{$A_0$ Dependence of the Tree--Level Neutrino Mass}
\label{A0}
We now discuss the dependence of the tree--level neutrino mass at
$M_{\rm EW}$ as a function of $A_0$ at $M_{\rm GUT}$.  Recall from
Sec.~\ref{tree_level} that
\begin{equation}
m_{\nu}^{\textrm{tree}} \propto \Delta_i^2 = \left( v_i - v_d 
\frac{\kappa_i} { \mu} \right)^2 \, .
\label{mnutree_prop}
\end{equation}
From the RGE of $\kappa_i$, Eq.~(\ref{RGEkappa}), we obtain as the
dominant contribution
\begin{equation}
\kappa_i \propto \mu  \lam'_{ijk} (\textbf{Y}_D)_{jk} \equiv  \mu  
\lam'_{ijk} \frac{(m_d)_{jk}}{v_d}
\end{equation}
at all energy scales, where $(m_d)_{jk}$ denotes a matrix element of
the down quark mass matrix. Therefore,
\begin{equation}
v_d \frac{\kappa_i}{\mu} \propto  \lam'_{ijk} \cdot  (m_d)_{jk}\, ,
\end{equation}
without further dependence on mSUGRA parameters. 

\begin{figure}[t!]
\begin{center}
\epsfig{figure=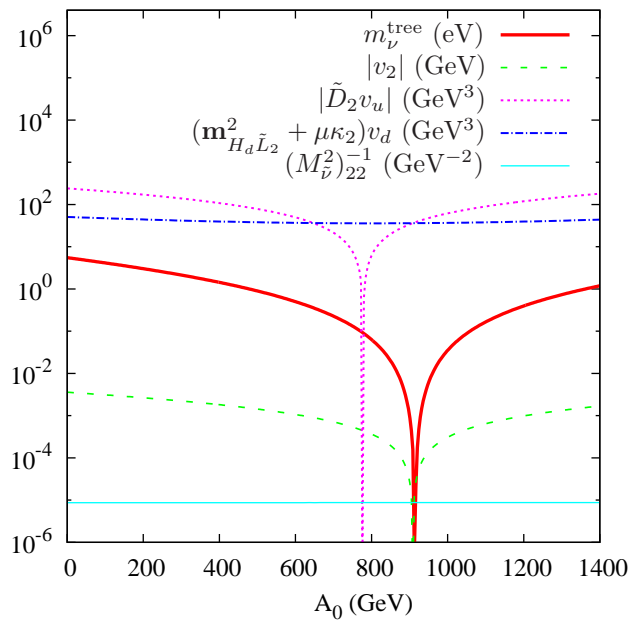}
\caption{$A_0$ dependence of $m_{\nu}^{\textrm{tree}}$ and the terms
determining the sneutrino vev $v_2$, Eq.~(\ref{vi}), at the REWSB
scale (used in {\tt SOFTSUSY} to calculate the sneutrino vev). Since
the scale affects the parameters only logarithmically, there are
only minor changes when running to $M_{\textrm{EW}}$. 
The other $\text{B}_3$ mSUGRA parameters are that of Point I with
$\lam'_{233}|_{\rm GUT}=10^{-5}$, Sec.~\ref{example_points}.
\label{figa0}}
\end{center}
\end{figure}

Thus, the dependence of the tree--level neutrino mass,
Eq.~(\ref{mnutree_prop}), on the mSUGRA parameters is solely
through the sneutrino vev $v_i$ \footnote{Note that there is one
exception, namely the direct proportionality
$m_{\nu}^{\textrm{tree}} \propto 1/M_{1/2}$, {\it cf.}
Eq.~(\ref{mnutree}).  However, compared to $v_i$, the impact of this
term on $m_{\nu}^{\textrm{tree}}$ and thus on the bounds of the
trilinear LNV couplings is much weaker.}. In
Fig.~\ref{figa0}, the dashed green line explicitly shows the
dependence of $|v_i|\,,\;i=2$ on $A_0$. It possesses a
clear minimum which is close to the minimum of
$m_{\nu}^{\textrm{tree}}$.

This behavior can be understood by taking a look at the (tree--level)
formula for the vev $v_i$, Eq.~(\ref{minVI}). For $\mathbf{\Lam} \in
\{\lam'_{ijk} \}$ it can be written as
\begin{eqnarray}
v_i = \frac{1}{(M_{\tilde{\nu}}^2)_{ii}} \biggl [ \tilde{D}_i v_u - 
(\mathbf{m}^2_{h_d \tilde{L}_i}+ \mu \kappa_i) v_d   \biggr ] \, ,
\label{vi}
\end{eqnarray}
with
\beq 
(M_{\tilde{\nu}}^2)_{ii} = (\ml)_{ii} \;+\; \frac{1}{2} M_Z^2 \cos2\beta \, .
\label{pre}
\eeq
Here, we have neglected terms proportional to $\kappa_i^2$ and
$v_i^2$, because they are much smaller than $(\ml)_{ii}$ and $M_Z^2$.
Note that we only obtain one non--zero sneutrino vev because
$\lam'_{ijk}$ violates only one lepton flavor. 

In many regions of parameter space the sneutrino vev in
Eq.~(\ref{mnutree_prop}) is at least two orders of magnitude larger
than the term $v_d \kappa_i / \mu$.  Thus the minimum of the neutrino
mass can only occur when the sneutrino vev is drastically reduced.
As we shall see, the sneutrino vev becomes very small, when
there is a cancellation between the two terms in Eq.~(\ref{vi}).

The second term of $v_i$ in Eq.~(\ref{vi}), $(\mathbf{m}^2_{h_d
\tilde{L}_i}+ \mu \kappa_i)v_d$, and the prefactor
$1/(M_{\tilde{\nu}}^2)_{ii}$ are always positive and depend only
weakly on $A_0$.  This can be seen in Fig.~\ref{figa0} for
$(\mathbf{m}^2_{h_d \tilde{L}_i}+ \mu \kappa_i)v_d$ (dotted--dashed blue
line) and also for $1/(M_{\tilde{\nu}}^2)_{ii}$ (solid turquoise line).  
This behavior can be easily understood:

The soft breaking parameter, $\mathbf{m}^2_{h_d \tilde{L}_i}$,
Eq.~(\ref{LNV_Lsoft}), is zero at $M_{\rm GUT}$ and is generated at
lower scales via~\cite{Allanach:2003eb}
\begin{equation}
16 \pi^2 \frac{d\mathbf{m}^2_{h_d \tilde{L}_i}}{dt} = -  \lam'_{ijk} 
(\mathbf{Y}_D)_{jk}
\mathcal{F}(\tilde{m}^2) - 6 h'_{ijk} (\mathbf{h}_D)_{jk} \, ,
\label{running_mHL}
\end{equation}
where $\mathcal{F}(\tilde{m}^2)$ is a linear function of the
soft-breaking scalar masses squared and of the down--type Higgs mass
parameter squared. $h'_{ijk}$ [$(\mathbf{h}_D)_{jk}$] is the
soft-breaking analog of $\lam'_{ijk}$ [$(\mathbf{Y}_D)_{jk}$] with $h'
_{ijk}=\lam'_{ijk}\times A_0$ [$(\mathbf{h}_D)_{jk}=(\mathbf{Y}_D)_
{jk} \times A_0$] at $M_{\rm GUT}$. The second term in
Eq.~(\ref{running_mHL}) thus depends on $A_0^2$. However, $\mathcal{F}
(\tilde{m}^2)$ is in general much larger than $A_0^2$ due to several
contributions from soft breaking masses \cite{Allanach:2003eb}.
Therefore, varying $A_0$ does not significantly change the magnitude
of $\mathbf{m}^2_{h_d \tilde{L}_i}$ as long as $A_0$ is not much
larger than the sfermion masses. 

Concerning the term $\mu \kappa_i$ in $(\mathbf{m}^2_{h_d
\tilde{L}_i}+ \mu \kappa_i)v_d$, we note from the RGE for
$\kappa_i$, Eq.~(\ref{RGEkappa}), that the only $A_0$ dependence of
$\kappa_i$ stems from its proportionality to $\mu$. $\mu$ at $M_{\rm
EW}$ can be approximated by~\cite{Drees:1995hj}
\begin{equation}
\mu^2 = c_1 M_0^2 + c_2 M_{1/2}^2 + c_3 A_0^2 + c_4 A_0 M_{1/2}- \frac{M_Z^2}{2} \;.
\label{mu2}
\end{equation}
Here $c_1$ and $c_2$ are numbers of $\cal{O}$(1) whereas $c_3$ and
$c_4$ are only of $\cal{O}$($10^{-1}-10^{-2}$) \footnote{All $c_i$
depend also weakly on tan$\beta$.  However, this becomes only
relevant for very small tan$\beta$ \cite{Drees:1995hj}.}.
Therefore, except for $A_0 \gg M_0, M_{1/2}$, the order of magnitude
of $\mu$ remains constant when varying $A_0$. 

We conclude that $(\mathbf{m}^2_{h_d \tilde{L}_i}+ \mu \kappa_i)v_d$
depends only weakly on $A_0$ and therefore, $\tilde{D}_i$ is decisive
for the $A_0$ dependence of the vev $v_i$ and thus of
$m_{\nu}^{\textrm{tree}}$. If the first term in Eq.~(\ref{vi}),
$\tilde{D}_i v_u$, is positive and only slightly larger than the
(nearly constant) second term, $(\mathbf{m}^2_{h_d \tilde{L}_i}+ \mu
\kappa_i)v_d$, $v_i$ can equal $v_d \kappa_i / \mu$ and we get
$m_{\nu}^{\textrm{tree}} =0$, {\it cf.} Eq.~(\ref{mnutree_prop}).

The strong $A_0$ dependence of the magnitude of $\tilde{D}_i v_u$ is
also displayed in Fig.~\ref{figa0} (dotted magenta line). We
observe that $|\tilde{D}_i v_u|$ is often larger than
$(\mathbf{m}^2_{h_d \tilde{L}_i}+ \mu \kappa_i)v_d$ (dotted--dashed blue
line).  However, near the tree--level neutrino mass minimum (solid red
line), it drops below $(\mathbf{m}^2_{h_d \tilde{L}_i}+ \mu
\kappa_i)v_d$ and $v_i$ can equal $v_d \frac{\kappa_i}{\mu}$.  In this case
$m_{\nu}^{\textrm{tree}}$, Eq.~(\ref{mnutree_prop}), vanishes.

In order to understand this behavior of $\tilde{D}_i$, we need to
understand how $\tilde{D}_i$ is generated via the RGEs. Recall that
$\tilde{D}_i=0$ at $M_{\rm GUT}$ within the $\text{B}_3$ mSUGRA model.
The generation of $\tilde{D}_i$ primarily depends on the running of
the trilinear soft breaking mass $h'_{ijk}$ \cite{Allanach:2003eb},
\begin{equation}
16 \pi^2 \frac{d\tilde{D}_i}{dt} = - 6 \mu (\textbf{Y}_D)_{jk}  h'_{ijk} + \dots\, .
\label{RGEDi2}
\end{equation}
We find the contribution in Eq.~(\ref{RGEDi}) proportional to
$\tilde{B}$ is typically much smaller \footnote{Only in parameter
 regions with small $\tan\beta$ and small $M_{1/2}$, a term
 proportional to $\tilde B$, Eq.~(\ref{RGEDi}), becomes equally
 important.  This is because $\tilde B$ increases with decreasing
 $\tan\beta$~\cite{Allanach:2003eb} whereas $\mu \times h'_{ijk}$
 decreases with decreasing $M_{1/2}$, {\it cf.} Eq.~(\ref{mu2}) and
 Eq.~(\ref{RGEh'}).  The term proportional to $\tilde B$ in
 Eq.~(\ref{RGEDi}) is then enhanced with respect to the term
 proportional to $h'_{ijk}$. However, in this parameter region
 $v_i$ will typically end up being negative because $\tilde D_i$ is
 further reduced than the other term in $v_i$, 
 such that the latter dominates.  Then there can
 be no cancellation in the tree--level neutrino mass,
 Eq.~(\ref{mnutree_prop}).}  and we here focus on the effects due
to $h'_{ijk}$.  The dominant terms of the corresponding RGE
are given by \cite{Allanach:2003eb,Bernhardt:2008jz}
\barr 
16 \pi^2 \frac{d
h'_{ijk}}{d t} = \frac{16}{3} g_3^2 \: (2 M_3 \lam'_{ijk} -
h'_{ijk}) + \dots \, ,
\label{RGEh'}
\earr
where $g_3$ ($M_3$) denotes the SU(3) gauge coupling (gaugino mass).
At $M_{\text{GUT}}$ this equation simplifies to 
\barr 
16 \pi^2 \frac{d
h'_{ijk}}{d t} = \frac{16}{3} g_{\rm GUT}^2 (2 M_{1/2} - A_0) \;
\lam'_{ijk} + \dots \, .
\label{RGEh'_GUT} 
\earr

\begin{figure}[t!]
\begin{center}
\epsfig{figure=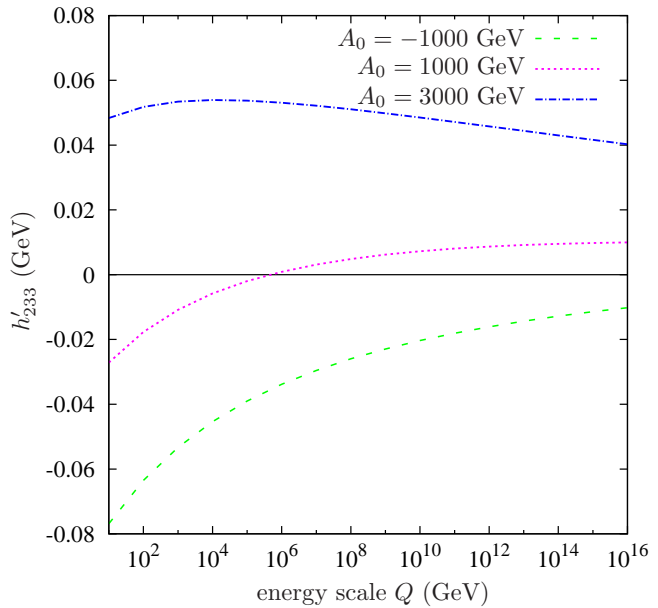,scale=0.93}
\caption{Running of $ h'_{233}$ for various values of $A_0$. The other
$\text{B}_3$ mSUGRA parameters are that of Point I, 
Sec.~\ref{example_points}, with ${\lam'_{233}}|_{\rm GUT}= 10^{-5}$
and $M_{1/2}=500$ GeV.
\label{hr_running}}
\end{center}
\end{figure}

\begin{figure}[t!]
\begin{center}
\epsfig{figure=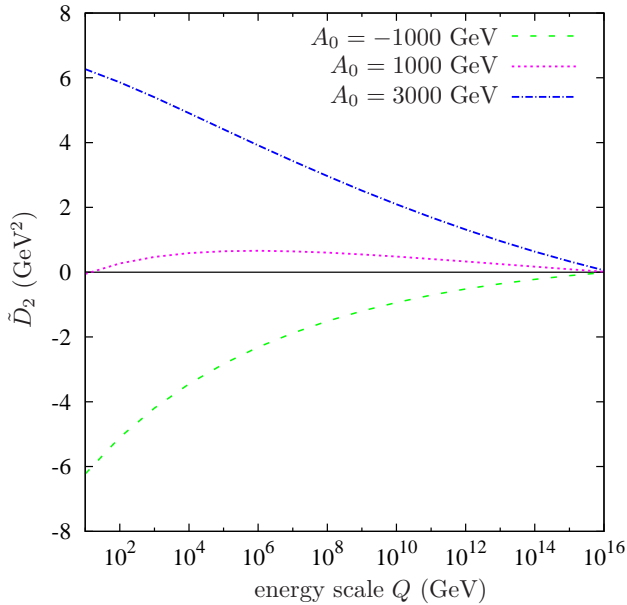,scale=0.93}
\caption{Running of the bilinear coupling $\tilde{D}_2$, Eq.~(\ref{RGEDi2}),
for the same parameter sets as those in Fig.~\ref{hr_running}.
\label{D2_running}}
\end{center}
\end{figure}

Keeping for now all parameters except $A_0$ fixed (with sgn$(\mu)=+1$
and $\lam'_{ijk}>0$ \footnote{From Eq.~(\ref{RGEkappa}) it is easy to
see that this implies sgn$(\kappa_i)= + 1$ below $M_{\rm GUT}$. }),
we can classify the running of $h'_{ijk}$, Eq.~(\ref{RGEh'}) and
Eq.~(\ref{RGEh'_GUT}), in the following way (see also
Ref.~\cite{Bernhardt:2008jz} for a detailed discussion):
\begin{itemize}
\item[(a)] $\bf A_0\ll2 M_{1/2}$ (including negative values of $A_0$):
Since the right hand side (RHS) of the RGE for $h'_{ijk}$,
Eq.~(\ref{RGEh'}), is always positive and large, $h'_{ijk}$ is
quickly reduced from its initial value of $A_0 \times \lam'_{ijk}$
and even becomes negative when running to lower energies. This
behavior is displayed in Fig.~\ref{hr_running} (dashed green line),
where the running of $h'_{233}$ is shown for different boundary
conditions at $M_{\rm GUT}$.

\item[(b)] $\bf A_0 \approx 2 M_{1/2}$: If the size of $A_0$ is
comparable to $2 M_{1/2}$, $h'_{ijk}$ will be fairly constant at
high energies, {\it cf.} the dotted magenta line in
Fig.~\ref{hr_running}.  However, when running to lower energies it
will still start decreasing, but more slowly than in case (a).
This is due to the fact that $M_3$ and $\lam'_{ijk}$ themselves
increase significantly (by factors of approx. 2.5 and 3,
respectively; see Ref.~\cite{Bernhardt:2008jz}) when running to
lower energies. Thus the term $2 M_3 \lam'_{ijk}$ eventually
dominates in Eq.~(\ref{RGEh'}) even if initially $A_0 \gtrsim 2
M_{1/2}$. This leads to a small, negative $h'_{ijk}$ at low
energies.

\item[(c)]$\bf A_0 \gg 2 M_{1/2}$: $h'_{ijk}$ is large at
$M_{\text{GUT}}$ and is further increased when running to lower
energies. This is due to the negative RHS of the RGE for $h'_{ijk}$,
Eq.~(\ref{RGEh'}); see also the dotted--dashed blue line in
Fig.~\ref{hr_running}. \\
\textbf{Caveat:} Since the term $2 M_3 \lam'_{ijk}$ in
Eq.~(\ref{RGEh'}) increases by a factor of approximately $8 \approx
3 \cdot 2.5$ when running from $M_{\text{GUT}}$ to $M_{\rm EW}$ [as
mentioned in (b)], $h'_{ijk}$ only strictly displays the behavior of
case (c) when $A_0 \gtrsim 20 M_{1/2}$. Otherwise, $h'_{ijk}$ will
decrease once the term $2 M_3 \lam'_{ijk}$ dominates.
\end{itemize}

Because  $\tilde{D}_i$ is zero at $M_{\rm GUT}$ and, according to Eq.~(\ref{RGEDi2}), also proportional 
to the integral of $h'_{ijk}$ over $\ln(Q)$, points (a) - (c) have the following consequences 
for $\tilde{D}_i$:
\begin{itemize}
\item[(a)] $\bf A_0 \ll 2 M_{1/2}$: Since $h'_{ijk}$ always becomes
negative below some energy scale close to $M_{\text{GUT}}$, the RHS
of Eq.~(\ref{RGEDi2}) is positive.  This leads to a large negative
$\tilde{D}_i$ at $M_Z$ as can be seen in Fig.~\ref{D2_running}
(dashed green line).  Consequently, all terms except $\tilde{D}_i v_u$
become negligible in $v_i$, Eq.~(\ref{vi}), and thus $|v_i|$ at
$M_{\rm EW}$ is large, dominating the tree--level neutrino mass,
Eq.~(\ref{mnutree_prop}).

\item[(b)] $\bf A_0 \approx 2 M_{1/2}$: Due to the initially negative
RHS of Eq.~(\ref{RGEDi2}) at energies close to $M_{\text{GUT}}$
(where $h'_{ijk}\approx A_0 \times \lam'_{ijk}$), $\tilde{D}_i$
first increases when running to lower energies but then starts
decreasing once $h'_{ijk}$ becomes negative, {\it cf.} the 
dotted magenta lines in Fig.~\ref{hr_running} and Fig.~\ref{D2_running}. At
some energy scale $Q$, $\tilde{D}_i$ becomes small such that $v_i$,
Eq.~(\ref{vi}), can equal $v_d \frac{\kappa_i}{\mu}$.  A
cancellation between these two terms in $m_{\nu}^{\textrm{tree}}$,
Eq.~(\ref{mnutree_prop}), at the scale $Q$ will then occur. This
corresponds to a vanishing tree--level neutrino mass if $Q = M_{\rm
 EW}$.

\item[(c)]$\bf A_0 \gg 2 M_{1/2}$: The RHS of Eq.~(\ref{RGEDi2}) is
always negative with a large magnitude such that we get a large
positive $\tilde{D}_i$ at the weak scale, \textit{cf.} the 
dotted--dashed blue line in Fig.~\ref{D2_running}. As in case (a), $\tilde{D}_i
v_u$ provides the main contribution to $|v_i|$, Eq.~(\ref{vi}).
Therefore, $|v_i|$ is large and dominates $m_{\nu}^{\textrm{tree}}$,
Eq.~(\ref{mnutree_prop}).
\end{itemize}

Summarizing, the tree--level neutrino mass has a minimum in the
parameter region where the size of $A_0$ is comparable to $2 M_{1/2}$.
This is mainly due to the running of the parameters $\tilde{D}_i$ and
$h'_{ijk}$ that affect the sneutrino vevs; in particular due to a
partial cancellation in Eq.~(\ref{RGEh'}). Note that in
Fig.~\ref{figa0} the tree--level neutrino mass vanishes at
$A_0\approx910\gev$, which is indeed close to $2M_{1/2}$.

As we see in the following section, the position of the minimum is
shifted towards higher values of $A_0$ for small $\tan\beta$.  In this
case, a change of the sign of the bilinear Higgs parameter $\mu$ also
has a significant impact.

\subsection{Dependence of the Tree--Level Neutrino Mass on the 
other mSUGRA Parameters}
\label{msugras}

\begin{figure*}[t!]
\setlength{\unitlength}{1in}
\subfigure[Tree--level neutrino mass as a function of $A_0$ and $M_{1/2}$.
\label{mnu_A0_M12}]{
\begin{picture}(3,2.3)
\put(-0.12,0){\epsfig{figure=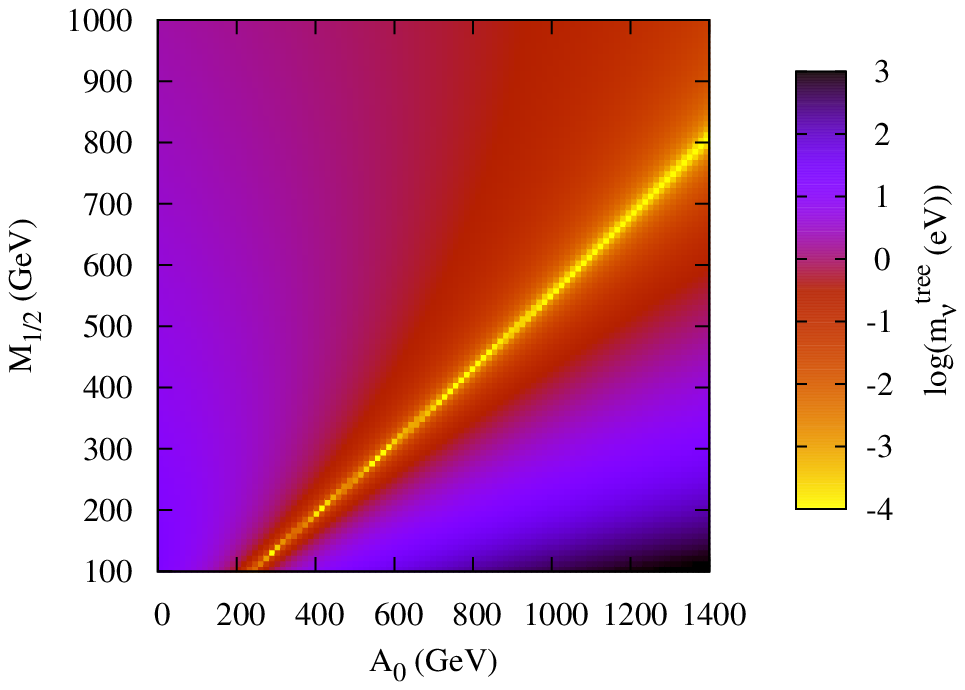,width=0.495\textwidth}}
\end{picture}}\hfill
\subfigure[Tree--level neutrino mass as a function of $A_0$ and $M_0$.
\label{mnu_A0_M0}]{
\begin{picture}(3,2.3)
\put(-0.12,0){\epsfig{figure=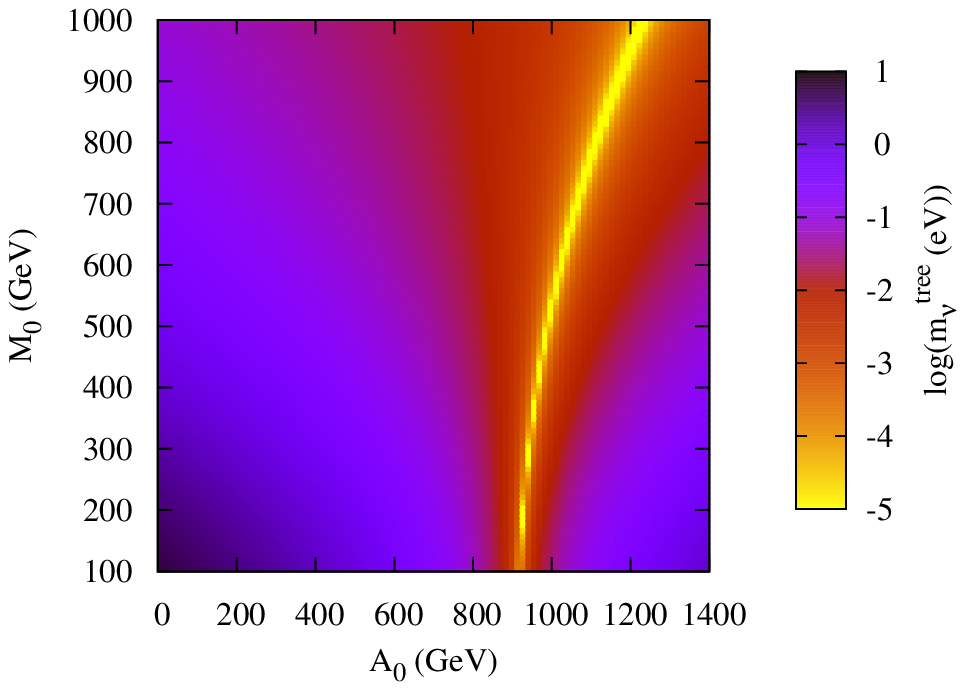,width=0.495\textwidth}}
\end{picture}}

\subfigure[Tree--level neutrino mass as a function of $A_0$ and $\tan\beta$ for $\text{sgn}(\mu)=+1$.
\label{mnu_A0_tanb_pmu}]{
\begin{picture}(3,2.3)
\put(0,0){\epsfig{figure=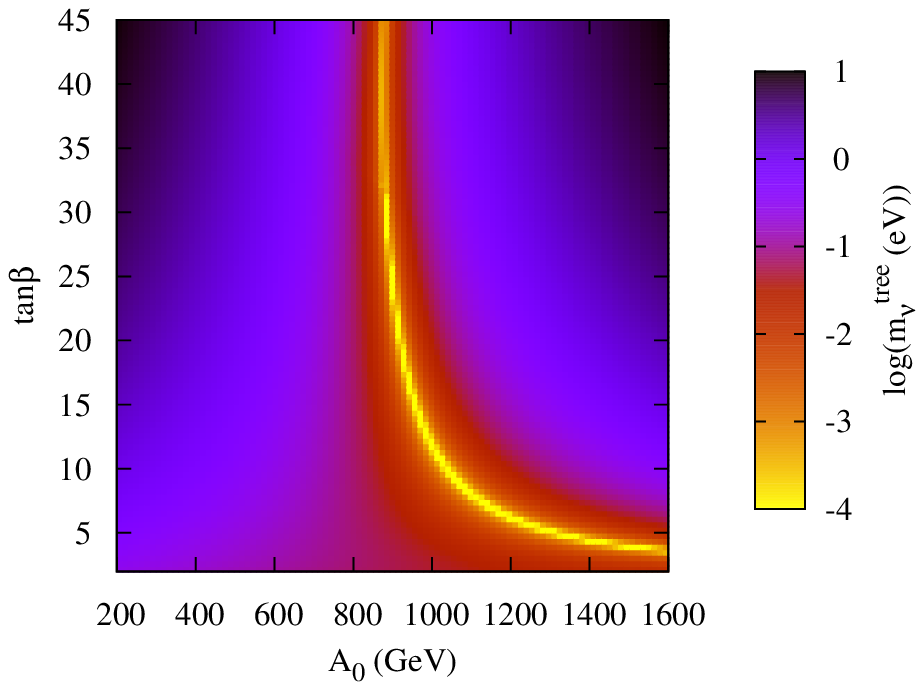,width=0.48\textwidth}}
\end{picture}}\hfill
\subfigure[Tree--level neutrino mass as a function of $A_0$ and $\tan\beta$ for $\text{sgn}(\mu)=-1$.
\label{mnu_A0_tanb_nmu}]{
\begin{picture}(3,2.3)
\put(0,0){\epsfig{figure=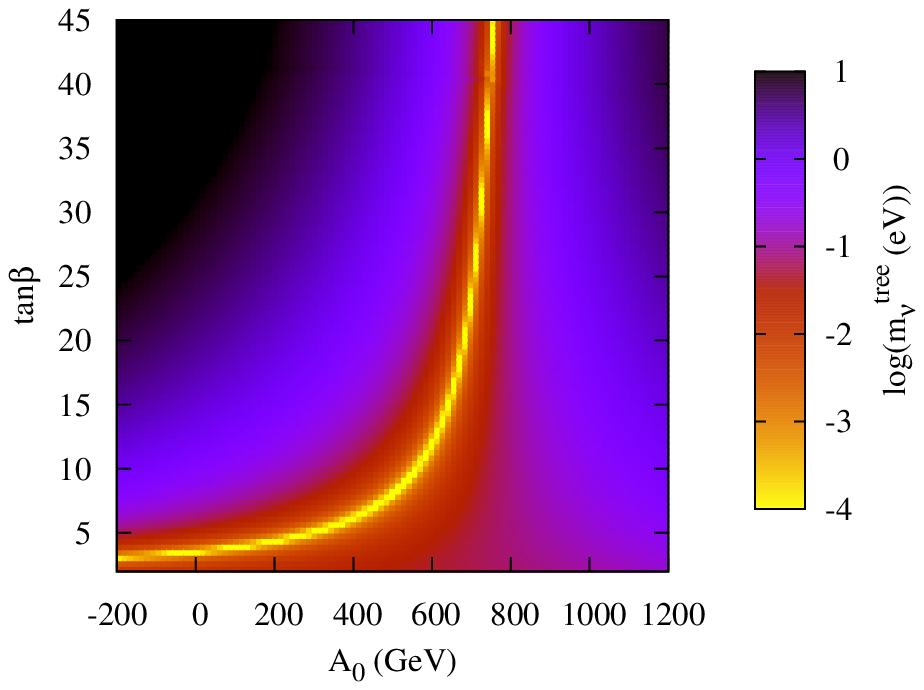,width=0.48\textwidth}}
\end{picture}}

\caption{Two dimensional plots of the tree--level neutrino mass.  The
plots are centered around parameter Point I,
Sec.~\ref{example_points}, with ${\lam'_{233}}|_{\rm GUT}=10^{-5}$.
The yellow regions signify parts of the parameter space where the
neutrino mass becomes smaller than $10^{-4}\text{eV}$,
Figs.~\ref{mnu_A0_M12}, \ref{mnu_A0_tanb_pmu}, \ref{mnu_A0_tanb_nmu},
or smaller than $10^{-5}\text{eV}$, Fig.~\ref{mnu_A0_M0} .
\label{fig2D_mnu}}
\end{figure*}

In App.~\ref{further_msugra}, we discuss in detail how the neutrino
mass matrix depends on the other mSUGRA parameter besides $A_0$.
Here we summarize the most important effects and
illustrate them in Fig.~\ref{fig2D_mnu}.

In Fig.~\ref{fig2D_mnu}, we show two dimensional mSUGRA parameter
scans of the tree--level neutrino mass. The other mSUGRA parameters
are those of Point I, Sec.~\ref{example_points}, with
${\lam'_{233}}|_ {\rm GUT}=10^{-5}$. One scan parameter is
always $A_0$ in order to show how the position of the minimum, which
was described in the last section, changes with the other mSUGRA
parameters.

Fig.~\ref{mnu_A0_M12} shows the $A_0$--$M_{1/2}$ plane. We can clearly
see that the position of the neutrino mass minimum is at $A_0 \approx
2 M_{1/2}$ as was concluded above. This illustrates that varying $M_{1/2}$
has a similar effect on the running of $h'_{ijk}$, Eq.~(\ref{RGEh'})
and Eq.~(\ref{RGEh'_GUT}), as varying $A_0$. This is clear from the
arguments (a)-(c) in Sec.~\ref{A0}. We could just rephrase the case
differentiation as
\begin{itemize}
\item[(a)] $\bf M_{1/2}  \gg A_0/2\,$.
\item[(b)] $\bf M_{1/2} \approx A_0/2\,$.
\item[(c)]$\bf M_{1/2} \ll A_0/2 \,$.
\end{itemize}

For $\mathbf \Lam \in \{\lam_{ijk}\}$ the relation is altered to $A_0
\approx M_{1/2}/2$. The change of the prefactor is due to the fact
that $\lam_{ijk}$ couples only leptonic fields to each other.
Consequently, only superfields carrying SU(2) and U(1) charges, but
not SU(3) charges, contribute to the relevant RGEs; see
App.~\ref{lam_LLE} for more details.

In Fig.~\ref{mnu_A0_M0} we present the tree--level neutrino mass as a
function of $A_0$ and $M_0$.  We observe that the position of the
neutrino mass minimum is fairly insensitive to $M_0$, compared to
$A_0$, $M_{1/2}$ and tan$\beta$ (see below). The minimum is shifted to
slightly higher values of $A_0$ for large $M_0$. However at large
$M_0$, the interval around the minimum in the $A_0$ direction where
the the tree--level neutrino mass is considerably reduced (and
therefore the bounds on $\lam'_{ijk}$ are substantially weakened) is
significantly broadened.

Finally we show in Fig.~\ref{mnu_A0_tanb_pmu} and
Fig.~\ref{mnu_A0_tanb_nmu} the $A_0$--$\tan\beta$ plane for $\text
{sgn}(\mu)=+1$ and $\text{sgn}(\mu)=-1$, respectively. In
Fig.~\ref{mnu_A0_tanb_pmu} we can see that for low tan$\beta$, the
neutrino mass minimum shifts to higher values of $A_0$. This is due to
a decrease of the down--type Yukawa coupling for low tan$\beta$ leading
to a decrease of the RHS of Eq.~(\ref{RGEDi2}). This decrease needs to
be balanced by increasing $A_0$; recall that $h'_{ijk}=\lam'_{ijk}
\times A_0$ at $M_{\rm GUT}$ in Eq.~(\ref{RGEDi2}).

A comparison of Fig.~\ref{mnu_A0_tanb_pmu} and
Fig.~\ref{mnu_A0_tanb_nmu} also shows that there is a {\it mirror
effect} around $A_0 = 800$ GeV ($\approx 2 M_{1/2}$) when we change the sign of the bilinear Higgs parameter
$\mu$. This happens because of a reversal of the sign of the RGE for
$\tilde D_i$, {\it cf.} App.~\ref{sgnmu}. Therefore the shift of the
minimum for low $\tan\beta$ now appears towards {\it lower} values of
$A_0$.

\subsection{The Dependence of the Loop Contributions to the Neutrino
Mass on the mSUGRA Parameters}
\label{msugra_abh_loops}

\begin{figure}[t!]
\begin{center}
\epsfig{figure=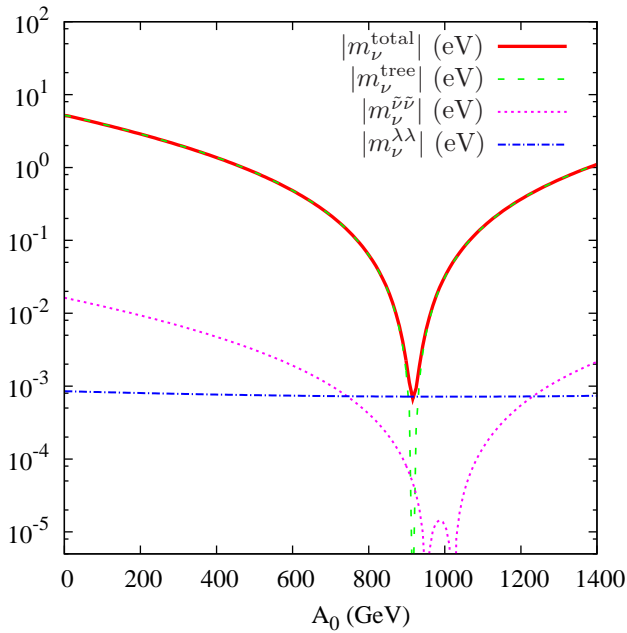}
\caption{$A_0$ dependence of the different contributions to the
neutrino mass at the REWSB scale for the $\text{B}_3$ mSUGRA Point I,
Sec.~\ref{example_points}, with $\lam'_{233}|_{\rm GUT}=
10^{-5}$. Note that only the absolute values of the contributions to
the neutrino mass are displayed. $m_{\nu}^{\textrm{tree}}$ and
$m_{\nu}^{\mathbf{\lam\lam}}$ are negative whereas $m_{\nu}^{\tilde\nu
\tilde{\nu}}$ is mostly positive. $m_{\nu}^{\tilde\nu
\tilde{\nu}}$ is only negative between the two minima of $|m_{\nu}^{\tilde\nu
\tilde{\nu}}|$; see Sec.~\ref{dep_loops} for details.
\label{figloops_LQD}}
\end{center}
\end{figure}

\begin{figure}[t!]
\begin{center}
\epsfig{figure=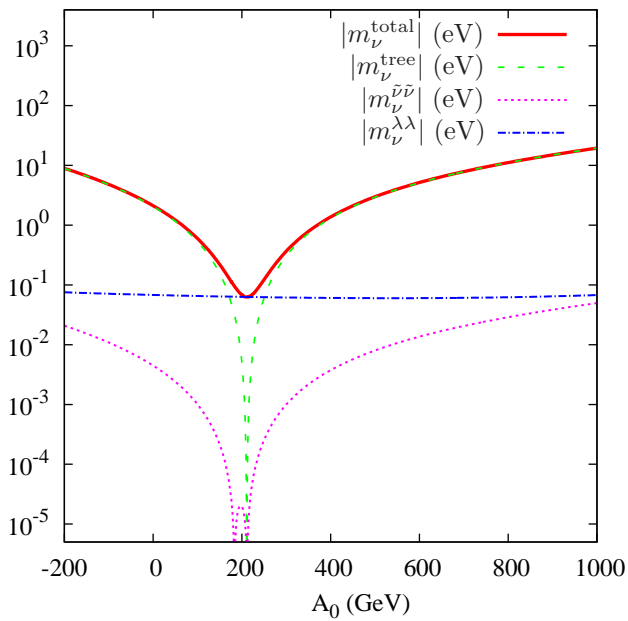}
\caption{Same as Fig.~\ref{figloops_LQD}, but for the $\text{B}_3$ mSUGRA Point II, 
Sec.~\ref{example_points}, with $\lam_{233}|_{\rm GUT}= 10^{-4}$.
\label{figloops_LLE}}
\end{center}
\end{figure}

The loop contributions to the neutrino mass matrix are usually several
orders of magnitude smaller than the tree--level contribution
\cite{Allanach:2003eb,Allanach:2007qc}. However, in the region around
the tree--level neutrino mass minimum, the loops dominate as shown in
Fig.~\ref{figloops_LQD} and Fig.~\ref{figloops_LLE}. Therefore, we now
briefly discuss the dependence of the loop contributions on the mSUGRA
parameters. 
\begin{itemize}
\item \textit{$\lam \lam$-- and $\lam' \lam'$--loops:} This contribution
to the neutrino mass, $m^{\lam \lam}_{\nu}$, depends only weakly on
the mSUGRA parameters, in particular it depends
 logarithmically on the relevant sfermion mass. For example,
varying $A_0$ from 0 to 1400 GeV ($-200$ GeV to $1000$ GeV) around Point
I (Point II) leaves the magnitude of $m^{\lam \lam}_{\nu}$ nearly
unchanged \footnote{In principle, there is an $A_0$ dependence that
 stems from left-right mixing of the sfermions inside the loop,
 {\it cf.} the first term in Eq.~(\ref{mia2}). However, in most
 regions of parameter space we have $\mu \tan \beta \gg A_0$. In
 this case only the second term in Eq.~(\ref{mia2}) plays a role.};
{\it cf.} the dotted--dashed blue line in Fig.~\ref{figloops_LQD}
(Fig.~\ref{figloops_LLE}). However, increasing $M_0$ or $M_{1/2}$
results in a decreasing $m^{\lam \lam}_{\nu}$: as the SUSY spectrum
gets heavier the sfermions in the loops decouple.

\item \textit{Neutral scalar--neutralino--loops:} This contribution to
the neutrino mass, $m^{\tilde{\nu}\tilde{\nu}}_{\nu}$, as a function
of $A_0$ possesses a minimum which lies in the vicinity of the
$m_\nu^{\textrm{tree}}$ minimum. However, there is no exact alignment.
This behavior can be understood by noting that the minima of
$m^{\tilde{\nu}\tilde{\nu}}_{\nu}$ arise due to the vanishing of
$\tilde{D}_i$, because roughly \beq m_\nu^{\tilde \nu \tilde \nu}
\propto \tilde D_i^2 \;,
\label{double_minimum}
\eeq 
{\it cf.} App.~\ref{dep_loops}.  This can be seen in
Fig.~\ref{figloops_LQD} as well as in Fig.~\ref{figloops_LLE} (dotted
magenta line).  Again, increasing $M_0$ or $M_{1/2}$ will in general
decrease $m^{\tilde{\nu}\tilde{\nu}}_{\nu}$, because the SUSY mass
spectrum gets heavier.

\item NLO corrections to the sneutrino vevs are typically at least one
order of magnitude smaller than the tree level quantities determining
the sneutrino vevs, $\mathbf{m}^2_{h_d \tilde L_i}\times
v_d/(M_{\tilde{\nu}})^2_{ii}$ and $\tilde{D}_i \times v_u
/(M_{\tilde{\nu}})^2_{ii} $, in Eq.~(\ref{minVI})~\cite{Chun:1999bq}.
For illustration, one could consider this as a $\mathcal{O}(10\%)$
correction to $\mathbf{m}^2_{h_d\tilde L_i}$. This shift upwards of
the dotted--dashed blue line in Fig.~\ref{figa0}  slightly changes the
position of the tree--level neutrino mass minimum, but does not alter
any of the conclusions drawn in this section.  Since the effects that
we investigate in this paper arise mainly from the contribution
$\tilde D_i v_u$ to the sneutrino vevs (see Sec.~\ref{A0}), these
corrections are not important for the qualitative analysis.

\end{itemize}

For parameter Points I and II, Sec.~\ref{example_points}, the $A_0$
interval, $\Delta A_0$, where the loops dominate is relatively small,
{\it cf.} Fig.~\ref{figloops_LQD} and Fig.~\ref{figloops_LLE}.
However, there are other parameter regions where the loops dominate in
intervals of $\Delta A_0 = \mathcal{O}(100 \, \text{GeV})$! This is
for example the case if one varies $A_0$ around the benchmark point
SPS1a \cite{Allanach:2002nj}.  We investigate the resulting bounds on
the LNV trilinear couplings in the following section.


\section{Bounds on Trilinear $\text{B}_3$ Couplings from $\nu$--Masses}
\label{bounds}

In this section, we calculate upper bounds on all trilinear LNV
couplings $\mathbf{\Lam} \in \{\lam_{ijk}, \lam'_{ijk}\}$ at $M_{\rm
GUT}$ from the cosmological upper bound on the sum of neutrino masses
as given in Eq.~(\ref{bound_numass}).  Note that in good approximation
\beq 
m_{\nu}|_{\rm EW} \propto \mathbf{\Lambda}^2|_{\rm GUT} \;,
\label{mnu_lam}
\eeq 
as explained in Sec.~\ref{A0} \footnote{
This is clear since the LNV parameters $\tilde{D}_i$, $\kappa_i$
and $\mathbf{m}^2_{h_d \tilde{L}_i}$ that determine the sneutrino vev
are generated proportional to $\mathbf{\Lam}$ at $M_{\text{GUT}}$,
{\it cf.} Sec.~\ref{A0}. From $m_{\nu}^{\textrm{tree}} \propto
v_i^2$ we then obtain the relationship in Eq.~(\ref{mnu_lam}).
}, Eq.~(\ref{llloop}) and App.~\ref{dep_loops}. 
Based on this approximation we employ
an iterative procedure to account for effects beyond
Eq.~(\ref{mnu_lam}).

In Sec.~\ref{compare}, we first compare our bounds with those given in
Ref.~\cite{Allanach:2003eb}, where the mSUGRA parameters of the
benchmark point SPS1a \cite{Allanach:2002nj} (in addition to
$\mathbf{\Lambda}$) were used. We choose the same mSUGRA
parameters beside $A_0$ in order to show how the bounds change in the
vicinity of the tree--level neutrino mass minimum, {\it cf.}
Sec.~\ref{A0}. We then perform in Sec.~\ref{2Dscan} two dimensional
parameter scans around the benchmark scenarios Point I and Point II
({\it cf.} Sec.~\ref{example_points}) to show more generally how the
bounds depend on the $\text{B}_3$ mSUGRA parameters.

In our parameter scans we exclude parameter regions where a tachyon
occurs \cite{Allanach:2003eb} or where the LEP2 exclusion bound on the
light SSM Higgs mass is not fulfilled
\cite{Schael:2006cr,Barate:2003sz}.  However, we reduce the LEP2 bound
by 3 GeV in order to account for numerical uncertainties of {\tt
SOFTSUSY} \cite{Allanach:2003jw,Degrassi:2002fi,Allanach:2004rh}.  For
instance, in the decoupling limit (where the light Higgs, $h^0$, is
SM-like) a lower bound of
\beq 
m_{h^0} > 111.4\, \textrm{GeV}
\label{higgs_bound}
\eeq 
is imposed. In the figures, we also show contour lines for the
$2\sigma$ window of the SUSY contribution to the anomalous magnetic
moment of the muon
\cite{Prades:2009qp,Bennett:2006fi,Jegerlehner:2009ry,Miller:2007kk}
\begin{equation}
8.6 \times 10^{-10} < \delta a_{\mu}^{\rm SUSY} < 40.6 \times 10^{-10} \, .
\label{amu}
\end{equation} 
For more details see Ref.~\cite{Bernhardt:2008jz} and references
therein.

We also note that the complete parameter space which we investigate in
the following (having rejected parameter regions which contain
tachyons or violate the LEP2 Higgs bound) is consistent with the
experimental upper bound on the branching ratio of $B_s \rightarrow
\mu^+ \mu^-$ \cite{Barberio:2008fa}, {\it i.e.}
\begin{equation}
\text{BR}(B_s \rightarrow \mu^+ \mu^-) < 4.7 \times 10^{-8} \, ,
\end{equation} 
and with the $2 \sigma$ window for the branching ratio of
$b\rightarrow s \gamma$, \cite{Barberio:2008fa,Buras:2002tp},
\begin{equation} 
2.74 \times 10^{-4} < \text{BR}(b \rightarrow  s \gamma) < 4.30 \times 10^{-4} \;.
\end{equation}  
We employed {\tt micrOMEGAs2.2} \cite{Belanger:2006is}, for the
evaluation of $\delta a^{\rm SUSY}_\mu$, $\text{BR}(B_s \rightarrow
\mu^+ \mu^-)$, and $\text{BR}(b \rightarrow s \gamma)$. Note that
there is a significant correlation in mSUGRA models between the muon
anomalous magnetic moment and $B_s \rightarrow \mu^+ \mu^-$
\cite{Dedes:2001fv}. Furthermore, we are well above the
standard supersymmetric mass bounds, as for example on the charginos.

\subsection{Comparison with Previous Results}
\label{compare}

\begin{table*}[t]
\begin{center}
\begin{tabular}{|c|c|c|c|c|c|c|}
\hline
&\multicolumn{3}{c|}{Up mixing}
&\multicolumn{3}{c|}{Down mixing} \\ \hline
$A_0$ (GeV)  & -100 &   500 & 550  &   -100 &   500 & 550 \\ \hline
$\;\lam'_{111}\;$ 		& \ 2.0$\times  {10^{-3}}\;$ & \ 2.7$\times 10^{-2}\;$ & \ 8.3$\times 10^{-2}\;$ 
		& \ 9.7$\times {10^{-4}}\;$ & \ 1.3$\times 10^{-2}\;$ & \ 5.3$\times 10^{-2}\;$ \\ 
$\lam'_{211}$ 		& 2.0$ \times {10^{-3}}$ & 2.7$\times 10^{-2}$ & \ 8.3$\times 10^{-2}\;$ 
		& 9.7$\times {10^{-4}}$ & 1.4$\times 10^{-2}$ & \ 5.3$\times 10^{-2}\;$\\  
$\lam'_{311}$		& 2.0$ \times {10^{-3}}$ & 2.7$\times 10^{-2}$  & \ 8.3$\times 10^{-2}\;$ 
		& 9.6$\times {10^{-4}}$& 1.3$\times 10^{-2}$ & \ 5.3$\times 10^{-2}\;$\\
		
$\;\lam'_{121},\; \lam'_{112}$ 		& \ (1.3$\times {10^{-1}})^t\;$ & \ (1.7$\times {10^{-1}})^t\; $& \ (1.7$\times {10^{-1}})^t\; $  
		& \ 4.3$\times {10^{-4}}\;$ & \ 6.0$\times 10^{-3}\;$ & \ 2.7$\times 10^{-2}\;$ \\  
$\;\lam'_{221},\; \lam'_{212}$	& \ (1.3$\times {10^{-1}})^t\;$ & \ (1.7$\times {10^{-1}})^t\;$ & \ (1.7$\times {10^{-1}})^t\;$   
		& \ 4.3$\times {10^{-4}}\;$ & \ 6.0$\times 10^{-3}\;$ & \ 2.7$\times 10^{-2}\;$ \\
$\;\lam'_{321},\; \lam'_{312}$	& \ (1.3$\times {10^{-1}})^t\;$ & \ (1.7$\times {10^{-1}})^t\;$ & \ (1.7$\times {10^{-1}})^t\;   $
		& \ 4.3$\times {10^{-4}}\;$ & \ 5.9$\times 10^{-3}\;$ & \ 2.6$\times 10^{-2}\;$ \\
		
$\;\lam'_{131}\;$ 		& \ (1.4$\times {10^{-1}})^t\;$ & \ (1.9$\times {10^{-1}})^t\; $& \ (1.9$\times {10^{-1}})^t\;  $ 
		& \ 6.9$\times {10^{-4}}\;$ & \ 9.5$\times 10^{-3}\;$ & \ 4.2$\times 10^{-2}\;$ \\
$\lam'_{231}$ 		& \ (1.4$\times {10^{-1}})^t\; $ & \ (1.9$\times {10^{-1}})^t\; $ & \ (1.9$\times {10^{-1}})^t\;$   
		& \ 6.9$\times {10^{-4}}\;$ & \ 9.5$\times 10^{-3}\;$ & \ 4.3$\times 10^{-2}\;$ \\
$\lam'_{331}$		& \ (1.4$\times {10^{-1}})^t\;$ & \ (1.9$\times {10^{-1}})^t\;$ & \ (1.9$\times {10^{-1}})^t\; $  
	& \ 6.8$\times {10^{-4}}\;$ & \ 9.3$\times 10^{-3}\;$ & \ 4.2$\times 10^{-2}\;$ \\

$\lam'_{122}$ & 9.1$\times  {10^{-5}}$&  1.3$\times 10^{-3}$ &  5.3$\times 10^{-3}$ 
		& 8.9$\times {10^{-5}}$& 1.2$\times 10^{-3}$ &  5.2$\times 10^{-3}$ \\ 			
$\lam'_{222}$ 	& 9.1$\times {10^{-5}}$&  1.3$\times 10^{-3}$ &  5.3$\times 10^{-3}$ 
		& 8.9$\times {10^{-5}}$& 1.2$\times 10^{-3}$ &  5.2$\times 10^{-3}$\\ 
$\lam'_{322}$ & 9.0$\times {10^{-5}}$&  1.3$\times 10^{-3}$ &  5.3$\times 10^{-3}$ 
		& 8.8$\times {10^{-5}}$& 1.2$\times 10^{-3}$&  5.2$\times 10^{-3}$\\

$\lam'_{132}$ & 2.4$\times  {10^{-2}}$&  \ (1.9$\times {10^{-1}})^t\;$ &  \ (1.9$\times {10^{-1}})^t\;$  
		& 5.8$\times {10^{-5}}$& 8.0$\times 10^{-4}$ &  3.9$\times 10^{-3}$ \\
$\lam'_{232}$ 	& 2.4$\times  {10^{-2}}$&   \ (1.9$\times {10^{-1}})^t\;$ &  \ (1.9$\times {10^{-1}})^t\; $
		& 5.8$\times {10^{-5}}$& 8.0$\times 10^{-4}$ &  3.9$\times 10^{-3}$ \\  
$\lam'_{332}$ & 2.4$\times  {10^{-2}}$&   \ (1.9$\times {10^{-1}})^t\;$ &  \ (1.9$\times {10^{-1}})^t\;$ 
		& 5.8$\times {10^{-5}}$& 7.9$\times 10^{-4}$ &  3.8$\times 10^{-3}$ \\

$\lam'_{113}$ & 4.2$\times  {10^{-3}}$&  5.5$\times 10^{-2}$ &   1.9$\times 10^{-1}$
		& 6.3$\times {10^{-4}}$& 8.7$\times 10^{-3}$ &  3.8$\times 10^{-2}$ \\
$\lam'_{213}$ 	& 4.2$\times  {10^{-3}}$&  5.5$\times 10^{-2}$ &   1.9$\times 10^{-1}$
		& 6.3$\times {10^{-4}}$& 8.7$\times 10^{-3}$ &  3.8$\times 10^{-2}$ \\  
$\lam'_{313}$  & 4.2$\times  {10^{-3}}$&  5.4$\times 10^{-2}$ &   1.7$\times 10^{-1}$ 
		& 6.2$\times {10^{-4}}$& 8.6$\times 10^{-3}$ &  3.7$\times 10^{-2}$ \\

$\lam'_{123}$ & 5.9$\times  {10^{-4}}$&  8.7$\times 10^{-3}$ &   2.4$\times 10^{-2}$
		& 5.3$\times {10^{-5}}$& 7.4$\times 10^{-4}$ &  3.4$\times 10^{-3}$ \\
$\lam'_{223}$ 	& 5.9$\times  {10^{-4}}$&  8.7$\times 10^{-3}$ &   2.4$\times 10^{-2}$
		& 5.3$\times {10^{-5}}$& 7.4$\times 10^{-4}$ &  3.4$\times 10^{-3}$ \\  
$\lam'_{323}$  & 5.8$\times  {10^{-4}}$&  8.5$\times 10^{-3}$ &   2.4$\times 10^{-2}$
		& 5.3$\times {10^{-5}}$& 7.2$\times 10^{-4}$ &  3.4$\times 10^{-3}$ \\

$\lam'_{133}$ & 2.3$ \times {10^{-6}}$& 3.2$\times 10^{-5}$ &  1.3$\times 10^{-4}$  
		&2.3$\times {10^{-6}}$& 3.2$\times 10^{-5}$ &  1.3$\times 10^{-4}$\\ 
$\lam'_{233}$ 	& 2.3$\times {10^{-6}}$& 3.2$\times 10^{-5}$ &  1.3$\times 10^{-4}$
		&2.3$\times {10^{-6}}$& 3.2$\times 10^{-5}$ &  1.3$\times 10^{-4}$\\  
$\lam'_{333}$ & 2.3$\times{10^{-6}}$& 3.1$\times 10^{-5}$ &  1.3$\times 10^{-4}$
		& 2.3$\times {10^{-6}}$& 3.1$\times 10^{-5}$ &  1.3$\times 10^{-4}$\\  
\hline
\end{tabular} 
\caption{Upper bounds on the trilinear couplings $\lam'_{ijk}$,
Eq.~(\ref{LNV_superpot}), at $M_{\text{GUT}}$ for several
values of $A_0$ (second row).  The other mSUGRA parameters are
those of SPS1a \cite{Allanach:2002nj}. We assume up-mixing
(down-mixing) in column 2-4 (5-7), {\it cf.} Sec.~\ref{quark_mixing}. 
Bounds arising from the absence of tachyons are in 
parentheses and marked by a superscript $t$: $()^t$.}
\label{tab:bounds}
\end{center}
\end{table*}

\begin{table*}[t]
\begin{center}
\begin{tabular}{|c|c|c|c|}
\hline
$A_0$ (GeV)  & -100  &   200 & 120 \\ \hline
$\;\lam_{211}\;$ 		& \ 1.1$\times  {10^{-1}}\;$ & \ \ 2.7$\times 10^{-1}\;$ & \ (7.1$\times  {10^{-1}})^t\;$  \\ 
$\;\lam_{311}\;$ 		& \ 1.1$\times  {10^{-1}}\;$ & \ 2.7$\times 10^{-1}\;$  & \ (7.1$\times  {10^{-1}})^t\;$  \\		
$\;\lam_{231}\;$ 		& \ (5.5$\times  {10^{-1}})^t\;$ & \ (6.7$\times 10^{-1})^t\;$ & \ (7.1$\times  {10^{-1}})^t\;$  \\		
$\;\lam_{122}\;$ 		& \ 4.7$\times  {10^{-4}}\;$ & \ 1.7$\times 10^{-3}\;$ & \ 4.9$\times  {10^{-3}}\;$  \\	
$\;\lam_{322}\;$ 		& \ 4.7$\times  {10^{-4}}\;$ & \ 1.7$\times  {10^{-3}}\;$  &  \ 4.9$\times  {10^{-3}}\;$  \\	
$\;\lam_{132}\;$ 		& \ (5.5$\times  {10^{-1}})^t\;$ & \ (6.7$\times 10^{-1})^t\;$ & \ (7.1$\times  {10^{-1}})^t\;$   \\	
$\;\lam_{123}\;$ 		& \ (5.1$\times  {10^{-1}})^t\;$ & \ (6.3$\times 10^{-1})^t\;$ & \ (6.7$\times  {10^{-1}})^t\;$  \\	
$\;\lam_{133}\;$ 		& \ 2.7$\times  {10^{-5}}\;$ & \ 1.0$\times  {10^{-4}}\;$ & \ 2.8$\times  {10^{-4}}\;$  \\	
$\;\lam_{233}\;$ 		& \ 2.7$\times  {10^{-5}}\;$ & \ 1.0$\times  {10^{-4}}\;$ & \ 2.8$\times  {10^{-4}}\;$ \\	
		\hline
\end{tabular}
\caption{Upper bounds on the trilinear couplings $\lam_{ijk}$, at
$M_{\text{GUT}}$ for different values of $A_0$ (first row). The other
mSUGRA parameters are those of SPS1a \cite{Allanach:2002nj}.  Bounds
arising 
from the absence of tachyons are marked by $()^t$.
\label{tab:bounds2}}
\end{center}
\end{table*}
		
\normalsize In Ref.~\cite{Allanach:2003eb}, bounds on single
couplings $\mathbf \Lambda$ at $M_{\rm GUT}$ in the $\text{B}_3$
mSUGRA model were determined for the mSUGRA parameters of
SPS1a, in particular $A_0=-100$ GeV.  However, the possibility
of obtaining much weaker bounds on the coupling
$\mathbf{\Lam}$ in the region of the tree--level neutrino mass minimum
was not exploited.  Note that the bounds in 
Ref.~\cite{Allanach:2003eb} were also obtained for a less
restrictive cosmological bound of $ \sum m_{\nu_i} < 0.71 \,
\textrm{eV}$ \footnote{Ref.~\cite{Allanach:2003eb} did not calculate 
the dominant loop contributions to the neutrino mass matrix. However,
these are negligible for SPS1a, because SPS1a lies far away from the
tree-level mass minimum.}.  We present here an update of these results
by using Eq.~(\ref{bound_numass}). We then explore the mSUGRA
parameter dependence of the bounds.

In Tab.~\ref{tab:bounds} and Tab.~\ref{tab:bounds2} ($\mathbf \Lam \in
\{\lam'_{ijk}\}$ and $\mathbf \Lam \in \{\lam_{ijk}\}$, respectively),
we compare the previous results with bounds (at $M_{\rm GUT}$) that we
obtain for identical $\text{B}_3$ mSUGRA parameter points, where only
the choice of $A_0$ differs.  In order to obtain corresponding bounds
at $M_{\rm EW}$ one needs to take into account the RGE evolution of
the couplings.  Quantitatively this results in multiplying the bounds
in Tab.~\ref{tab:bounds} (Tab.~\ref{tab:bounds2}) by roughly a factor
of 3.5 (1.5), {\it cf.}
Ref.~\cite{Carlos:1996du,Allanach:1999ic,Allanach:2003eb,Dreiner:2008rv,Dreiner:2008ca}.

In addition to $A_0 = -100$ GeV (SPS1a), we choose two parameter
points which lie $\Delta A_0 \approx 10$ GeV and $\Delta A_0 \approx
60-70$ GeV, away from the neutrino mass minimum.  In
Tab.~\ref{tab:bounds} ($\mathbf \Lam \in \{\lam'_{ijk}\}$), we choose $A_0
= 500$ GeV (column 3 and 6) and $A_0 = 550$ GeV (column 4 and 7). In
Tab.~\ref{tab:bounds2} ($\mathbf \Lam \in \{\lam_{ijk}\}$), we choose $A_0
= 200$ GeV (column 3) and $A_0 = 120$ GeV (column 4). This enables us
to examine the dependence of the bounds on $A_0$ around the
tree--level mass minimum.

Note that at SPS1a and when varying $A_0$, the neutrino mass
minimum for $\lam'_{ijk} \not = 0$ lies at $A_0 = 563$~GeV. This value
is mostly independent of the choice of the indices $i,j,k$. This is
clear because the condition for the minimum to occur, $A_0 \approx 2
M_{1/2}$, does not depend on $i,j,k$, {\it cf.} Sec.~\ref{mSUGRA}.
Similarly, for $\lam_{ijk}|_{\rm GUT} \not = 0$ the minimum 
is expected at $A_0 \approx M_{1/2}/2$. For the SPS1a parameters we 
thus obtain $A_0 \approx 127$~GeV
\footnote{This value is smaller than would be expected by estimating $A_0
    \approx M_{1/2}/2 = 250$ GeV, because we are considering a
    parameter point with relatively low $\tan\beta$ ($\tan\beta = 10$
    for SPS1a). As discussed in Sec.~\ref{msugras}, this leads to a
    shift of the tree--level neutrino mass minimum towards lower
    values of $A_0$, {\it cf.} Fig.~\ref{mnu_A0_tanb_pmu}.}.

We first concentrate on Tab.~\ref{tab:bounds}. Comparing the columns
for $A_0=-100$ GeV and then for $A_0=500$ GeV, {\it i.e.}  approaching
the minimum up to $\Delta A_0 = 63$ GeV, the bounds from too large
neutrino masses are weakened by a factor of 13--15. When we go even
closer, {\it i.e.} $A_0 = 550$ GeV and $\Delta A_0 = 13$ GeV, the
bounds are weakened by a factor of 40--64 compared to
$A_0=-100$~GeV. As we discuss below, in the case of up-mixing, some
couplings in Tab.~\ref{tab:bounds} (column 2-4) can not be restricted
at all by too large neutrino masses. In this case we show the
bounds at $M_{\rm GUT}$ [marked by $()^t$], that one obtains from the
absence of tachyons; see also Ref.~\cite{Allanach:2003eb}.

We differentiate in Tab.~\ref{tab:bounds} between up-- and
down--type quark mixing, {\it cf.} Sec.~\ref{quark_mixing}. Different
quark mixing has important consequences for the bounds on the
couplings $\lam'_{ijk}$ if $j\neq k$. As is clear from
Sec.~(\ref{A0}), the tree--level neutrino mass is generated
proportional to $\lam'_{ijk}\times (\textbf{Y}_{D})_{jk}$.  Thus, no
tree--level mass is generated at this level when we consider $j \neq
k$ and up--type mixing (which implies a diagonal $\textbf{Y}_{D}$).
But, an additional $\lam'_{ikk}$ coupling will be generated via RGE
running at lower scales, {\it cf.}  Ref.~\cite{Allanach:2003eb}. This
coupling will still generate a tree--level neutrino mass, which is
however suppressed by the additional one--loop effect
\footnote{Note that also the loop contributions are strongly
suppressed, because the $\lam' \lam'$--loops are proportional to
$\lam'_{ijk} \times \lam'_{ikj}$, Eq.~(\ref{llloop}), and the neutral
scalar loops are aligned with the tree-level mass, {\it cf.}
Sec.~\ref{dep_loops}.}.

This effect can be seen in Tab.~\ref{tab:bounds}, if we compare for
example the upper bounds on $\lam'_{223}$ and $\lam'_{233}$ for up--
and down--type quark mixing.  The ratio between these bounds is
roughly 200 in the case of up--type mixing whereas there is only one
order of magnitude difference for down--type mixing.

In the latter case, the ratio between the $\lam'_{223}$ and
$\lam'_{233}$ bounds originates mainly from the ratio
\beq
\frac{(\textbf{Y}_{D})_{23}}{(\textbf{Y}_{D})_{33}} = \frac{({\bf
V}_{CKM})_{23}}{({\bf V}_{CKM})_{33}} \, ,
\label{Yukawa_hierarchy}
\eeq
since the tree--level mass is generated via $\lam'_{223} \times
(\textbf{Y}_{D})_{23}$ and $\lam'_{233} \times (\textbf{Y}_{D})_{33}$,
respectively.

To conclude, the bounds from the generation of neutrino masses (at
least in the case of down--type mixing) are usually the strongest
bounds on the couplings $\lam'_{ijk}$ at $M_{\rm GUT}$. As considered
in Ref.~\cite{Allanach:2003eb}, they range from $\mathcal{O}(10^{-4})$
to $\mathcal{O}(10^{-6})$ for the parameter point SPS1a (column 5 in
Tab.~\ref{tab:bounds}).  However, there is a large window around the
tree-level neutrino mass minimum, where bounds may be obtained that
are between one and two orders of magnitude weaker than those in
Ref.~\cite{Allanach:2003eb}. Around the minimum, the couplings are
only bounded from above by $\mathcal{O}(10^{-2})$ to $\mathcal{O}
(10^{-4})$ ({\it cf.}  column 7 in Tab.~\ref{tab:bounds}). Thus, other
low energy bounds become competitive
\cite{Allanach:1999ic,Barbier:2004ez,other}.

We now discuss in Tab.~\ref{tab:bounds2} the case of a non-vanishing
coupling $\lambda_{ijk}$ at $M_{\rm GUT}$. Contrary to
Tab.~\ref{tab:bounds}, in the case considered in
Tab.~\ref{tab:bounds2} the quark mixing assumption does not affect the
bounds since $\lam_{ijk}$ couples only to lepton superfields. Due to
the antisymmetry $\lam_{ijk} = - \lam_{jik}$ there are only 9
independent couplings.

We observe in Tab.~\ref{tab:bounds2} that if $i\not=j\not=k
\not=i$ there are no bounds from too large neutrino
masses. The only bound we obtain stems from the absence of
tachyons. This is because we assume a diagonal lepton Yukawa matrix
$\textbf{Y}_{E}$ as stated in Sec.~\ref{quark_mixing} and therefore,
only couplings of the form $\lam_{ikk}$ can generate a neutrino mass
\footnote{This would change drastically if the $\textbf{Y}_{E}$ were 
strongly mixed~\cite{Allanach:2003eb}.}.  

For these couplings, the bounds at $M_{\rm GUT}$ for $A_0=-100$ GeV
(column 2) range from $1.1 \times 10^{-1}$ ($\lam_{211}$ and $\lam_
{311}$) to $2.7 \times 10^{-5}$ ($\lam_{133}$ and $\lam_{233}$). If we
approach the tree--level mass minimum, {\it i.e.} going from column 2
to column 4 with $A_0=120$ GeV, the bound is weaker than the tachyon
bound ($\lam_{211}$ and $\lam_{311}$) or it is weakened to $2.8 \times
10^{-4}$ ($\lam_{133}$ and $\lam_{233}$). The bounds from neutrino
masses are thus decreased by roughly a factor of 10.

Comparing the bounds on $\lam_{ikk}$ at $M_{\rm GUT}$, one can see
nicely how the choice of $k$ influences the strength of the bound. The
bounds resemble the hierarchy between the lepton Yukawa couplings
$(\textbf{Y}_{E})_{kk}$ analogously to 
Eq.~(\ref{Yukawa_hierarchy}). Therefore, the bounds are strongest for
$k=3$.

In contrast to Tab.~\ref{tab:bounds}, the bounds are only reduced by
one order of magnitude when we approach the tree-level mass
minimum. This is because the loop contributions play an important role
for the bounds in Tab.~\ref{tab:bounds2}, as we discuss in the
following section.

\subsection{Influence of Loop Contributions}
\label{importance_of_loops}

We now shortly discuss the influence of the neutrino mass loop
contributions on the bounds.  Typically, one expects that the
closer we approach the tree--level neutrino mass minimum the
more important the loop contributions become. This is because the
loops are not aligned to the tree--level mass, {\it cf.}
Sec.~\ref{msugra_abh_loops}.

However, in the case of the neutral scalar loops there is still
partial alignment, because both the tree--level mass minimum and the
minima of the neutral scalar loops crucially depend on the vanishing
of the bilinear LNV parameter $\tilde D_i$, {\it cf.} Sect.~\ref{msugra_abh_loops}.
Therefore, it is the $\lam' \lam'$--loops and $\lam \lam$--loops,
Sec.~\ref{lamlam_loop}, that are relevant whenever the loop
contributions become dominant over the tree--level contributions.

We now give a few examples. For $\mathbf \Lam \in \{\lam_{ijk}\}$,
Tab.~\ref{tab:bounds2}, the loop contributions dominate over the
tree--level mass in a range of $\Delta A_0 \approx \pm 50$ GeV around
the tree--level mass minimum at $A_0=127$ GeV. Therefore, the bounds
in this region are much more restrictive ({\it i.e.} the value of the
bounds {\it decreases}) when taking into account the loop
contributions.  For example,
\beq
\frac{\lam_{233}^{\textrm{tot}}}{\lam_{233}^{\textrm{tree}}} \approx
0.3 \, ,
\label{loop_influence}
\eeq for $A_0=120$ GeV; column 4 in Tab.~\ref{tab:bounds2}. Here,
$\lam_{233}^{\textrm{tot}}$ is the bound on $\lam_{233}$ at $M_{\rm
GUT}$ if we take into account both tree--level and
loop--contributions to the neutrino mass. In contrast,
$\lam_{233}^{\textrm{tree}}$ would be the bound if we only employ the
tree-level mass.

Further away from the minimum, the influence of the loop contributions
is weaker.  The bounds are strengthened by approximately $5\%$ for
$A_0=200$ GeV (column 3 of Tab.~\ref{tab:bounds2}) and $<1\%$ for
$A_0=-100$ GeV (column 2 of Tab.~\ref{tab:bounds2}).

The loop contributions are less
important for the bounds in Tab.~\ref{tab:bounds}, {\it i.e.}
$\mathbf \Lam \in \{\lam'_{ijk}\}$. For example, even near the
tree-level mass minimum (column 4 and 7 with $A_0=550$ GeV), the
bounds become only stronger by up to $20\%$ if we take the loop
induced neutrino masses in addition to the tree--level mass into
account. 

\subsection{Dependence of Bounds on $\text{B}_3$ mSUGRA Parameters}
\label{2Dscan}

\begin{figure*}[t!]
\setlength{\unitlength}{1in}
\subfigure[Upper bounds on $\lam'_{233}$ at $M_{\rm GUT}$ 
in the $A_0$--$M_{1/2}$ plane. The parameter space below 
the green line is disfavored by $\delta^{\rm SUSY}_{\mu}$; 
see Eq.~(\ref{amu}).
\label{fig2D.a}]{
\begin{picture}(3,2.3)
\put(0,0){\epsfig{figure=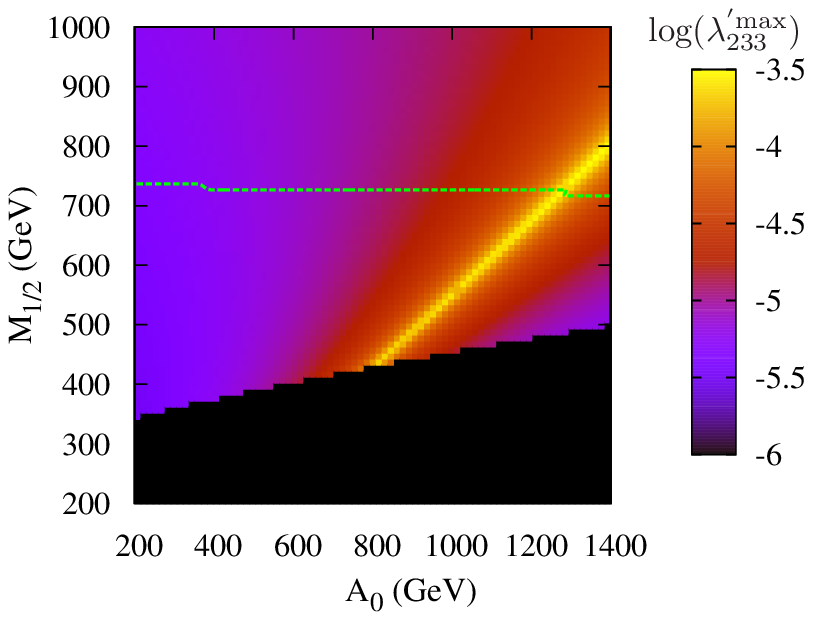,width=0.45\textwidth}}
\end{picture}}\hfill
\subfigure[Upper bounds on $\lam'_{233}$ at $M_{\rm GUT}$ 
in the $A_0$--$\tan \beta$ plane. The parameter space below 
the green line is disfavored by $\delta^{\rm SUSY}_{\mu}$; 
see Eq.~(\ref{amu}).
\label{fig2D.b}]{
\begin{picture}(3,2.3)
\put(0,0){\epsfig{figure=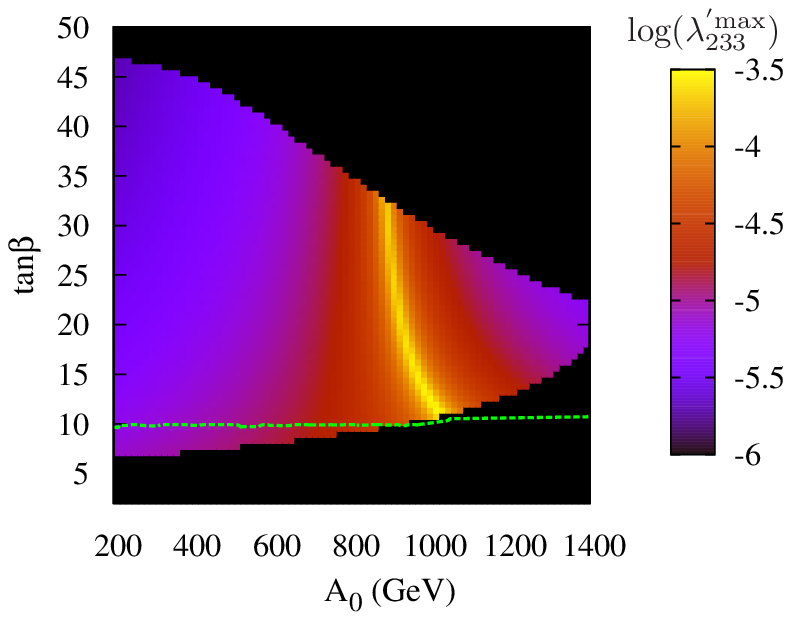,width=0.45\textwidth}}
\end{picture}}
\subfigure[Upper bounds on $\lam'_{233}$ at $M_{\rm GUT}$ 
in the $A_0$--$M_{0}$ plane. The parameter space above 
the green line is disfavored by $\delta^{\rm SUSY}_{\mu}$; 
see Eq.~(\ref{amu}). 
\label{fig2D.c}]{
\begin{picture}(3,2.3)
\put(0,0){\epsfig{figure=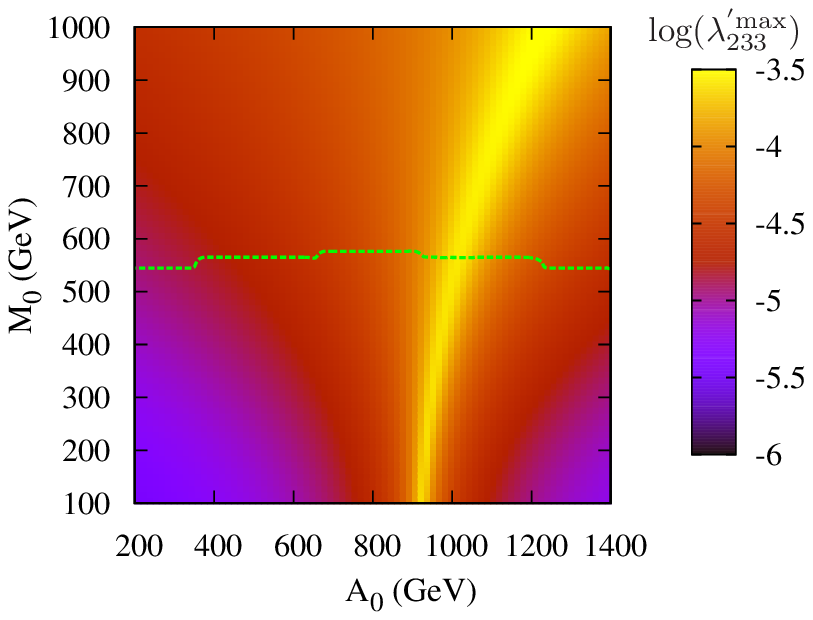,width=0.45\textwidth}}
\end{picture}}\hfill
\caption{Upper bounds on $\lam'_{233}$ at $M_{\rm GUT}$ from the
cosmological bound on the sum of neutrino masses,
Eq.~(\ref{bound_numass}), as a function of mSUGRA parameters. The
parameter scans are centered around the benchmark Point~I,
{\it cf.} Sec.~\ref{example_points}. The blackened-out region
denotes parameter points where tachyons occur or where the
LEP2 Higgs bound is violated.
\label{fig2D}}
\end{figure*}

\begin{figure*}[t!]
\setlength{\unitlength}{1in}
\subfigure[Same as Fig.~\ref{fig2D.a}, but for $\lam_{233}$.
\label{fig2D_LE.a}]{
\begin{picture}(3,2.3)
\put(0,0){\epsfig{figure=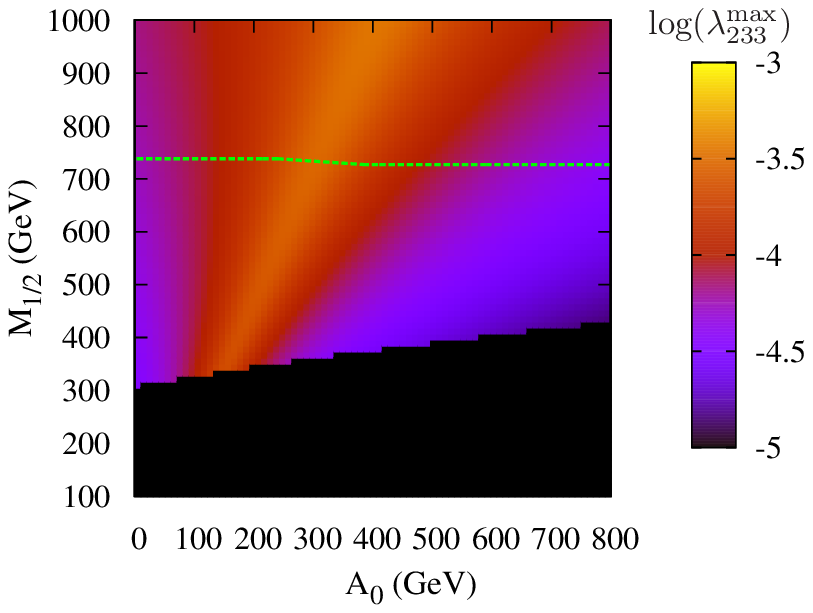,width=0.45\textwidth}}
\end{picture}}\hfill
\subfigure[Same as Fig.~\ref{fig2D.b}, but for $\lam_{233}$.
\label{fig2D_LE.b}]{
\begin{picture}(3,2.3)
\put(0,0){\epsfig{figure=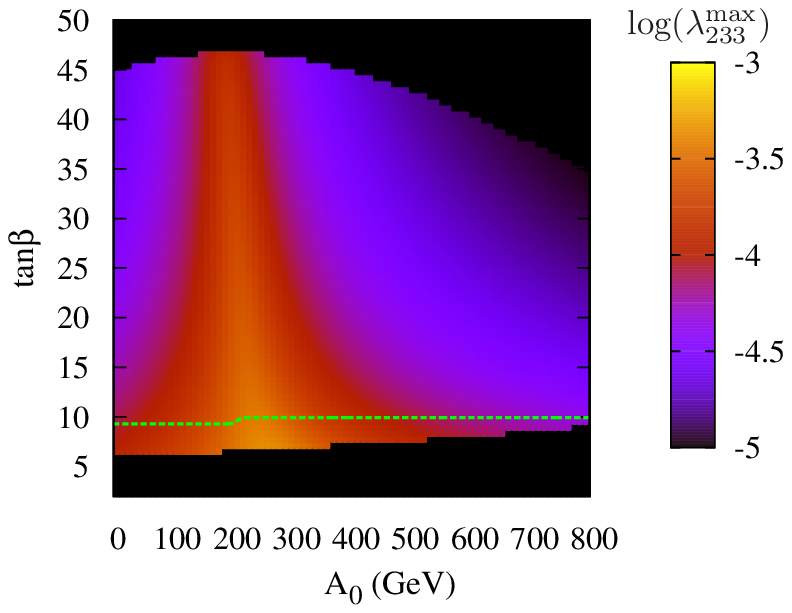,width=0.45\textwidth}}
\end{picture}}
\subfigure[Same as Fig.~\ref{fig2D.c}, but for $\lam_{233}$.
\label{fig2D_LE.c}]{
\begin{picture}(3,2.3)
\put(0,0){\epsfig{figure=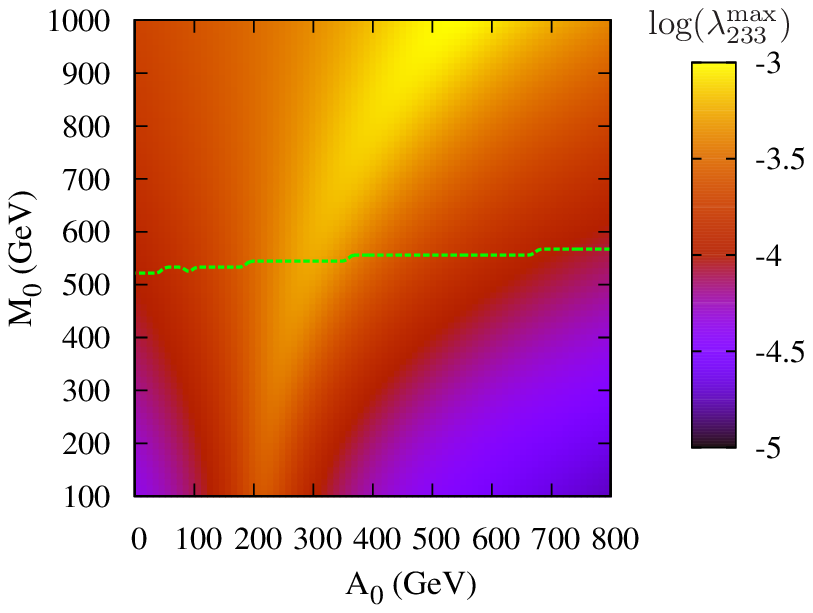,width=0.45\textwidth}}
\end{picture}}\hfill
\caption{Same as Fig.~\ref{fig2D}, but for $\lam_{233}$ at $M_{\rm
GUT}$ and for the benchmark scenario Point~II, {\it cf.}
Sec.~\ref{example_points}.
\label{fig2D_LE}}
\end{figure*}

In this section, we discuss the dependence of the bounds on $\Lam \in
\{\lam_{ijk}, \lam'_{ijk} \}$ at $M_{\rm GUT}$ on the $\text{B}_3$
mSUGRA parameters.  For that purpose we perform two-dimensional
parameter scans around the benchmark scenarios, Point~I and Point~II,
of Sec.~\ref{example_points}.  For the calculation of the bounds all
contributions to the neutrino mass considered in Sec.~\ref{chapM} are
included.  We will focus here on the couplings $\lam'_{233}$ and
$\lam_{233}$, because these couplings have the strongest
constraints from neutrino masses, {\it cf.} Tab.~\ref{tab:bounds} and
Tab.~\ref{tab:bounds2}.

We have analyzed in Sec.~\ref{mSUGRA} how the neutrino mass changes
with the mSUGRA parameters.  Due to its approximate proportionality to
$\mathbf \Lam^2$, {\it cf.} Eq.~(\ref{mnu_lam}), the analysis in
Sec.~\ref{mSUGRA} is directly transferable to the mSUGRA dependence of
bounds on the LNV trilinear couplings. Therefore, the parameter scans
presented in this section, {\it i.e.} Fig.~\ref{fig2D} and
Fig.~\ref{fig2D_LE}, resemble closely those in Fig.~\ref{fig2D_mnu},
Sec.~\ref{mSUGRA}.

We show in Fig.~\ref{fig2D} [Fig.~\ref{fig2D_LE}] how the bounds on
$\lam'_{233}$ [$\lam_{233}$] at $M_{\rm GUT}$ vary with mSUGRA
parameters. We present in Figs.~\ref{fig2D.a}--\ref{fig2D.c}
[Figs.~\ref{fig2D_LE.a}--\ref{fig2D_LE.c}] the $A_0$--$M_{1/2}$,
$A_0$--$\tan \beta$, and $A_0$--$M_0$ planes, respectively. The bounds
are shown on a logarithmic scale.  The blackened out regions designate
areas of parameter space which are rejected due to tachyons in the
model or violation of the LEP2 bound on the lightest Higgs
mass, {\it cf.}  Eq.~(\ref{higgs_bound}).  Furthermore, we
include contour lines of the $2 \sigma$ window for the SUSY
contribution to the anomalous magnetic moment of the muon,
Eq.~(\ref{amu}). Imposing Eq.~(\ref{amu}) disfavors the parameter
space below [above] the green contour line in Figs.~\ref{fig2D.a},
\ref{fig2D.b}, \ref{fig2D_LE.a}, and~\ref{fig2D_LE.b}
[Fig.~\ref{fig2D.c} and Fig.~\ref{fig2D.c}].

We observe in Fig.~\ref{fig2D} that the strictest bounds on
$\lam'_{233}$ from too large neutrino masses are of $\mathcal{O}(
10^{-6})$ . However, there are sizable regions of parameter space
where the bounds are considerably weakened.  For example, in the
$A_0$--$M_{1/2}$ plane, Fig.~\ref{fig2D.a}, the bounds are of
$\mathcal{O}(10^{-6})$ only in approximately half of the parameter
space whereas in the other half, the bounds are $\mathcal{O}(10^{-5})$
or weaker.  In roughly $10\%$ of the allowed region in
Fig.~\ref{fig2D}, the bounds even lie at or above
$\mathcal{O}(10^{-4})$! In this region, the loop contributions to the
heaviest neutrino mass are essential for determining the bounds since
the corresponding tree--level neutrino mass vanishes, {\it cf.} also
the discussion in Sec.~\ref{importance_of_loops}.

We can see in Fig.~\ref{fig2D_LE} a similar behavior for the
parameter dependence of the bounds on $\lam_{233}$. Here, the
strongest bounds are now of $\mathcal{O}(10^{-5})$.  
However, for
example in the $A_0$--$M_0$ plane, Fig.~\ref{fig2D_LE.c}, the
bounds are as strong as $\mathcal{O}(10^{-5})$ in only
about $25\%$ of
the parameter plane.  The remaining $75\%$ have bounds of
$\mathcal{O}(10^{-4})$ ($50\%$) or even $\mathcal{O}(10^{-3})$
($25\%$)!

Up to now, we have analyzed how the bounds on the trilinear
LNV couplings $\lam'_{233}$ and $\lam_{233}$ vary with the mSUGRA
parameters. However, from the analysis in Sec.~\ref{compare},
we can easily deduce how most of these bounds change for
different couplings $\lam'_{ijk}$ and $\lam_{ijk}$, {\it i.e} for
different indices $i,j,k$. For $\lam'_{ijk}$ the index $i$ does not
significantly influence the bound, because the employed Yukawa
coupling, $(\mathbf{Y}_D)_{jk}$, via which the tree--level mass
is generated, does not depend on $i$.  But, the situation is totally
different when we change the indices $j,k$. In general, for
$\lam'_{ijk}$ (and down--mixing) the bounds will display the hierarchy
of the down--type Yukawa couplings.  Therefore, bounds for couplings
$\lam'_{i11}$ are about three orders of magnitude weaker than bounds
for the couplings $\lam'_{i33}$ as long as the other $\text{B}_3$
mSUGRA parameter are the same.  We also observe a similar behavior
for $\lam'_{ijk}$ with up--mixing and for $\lam_{ijk}$ [using
$(\mathbf{Y}_E)_{jk}$ instead of $(\mathbf{Y}_D)_{jk}$], if $j=k$; {\it
cf.} the discussion in Sec.~\ref{compare}.

To conclude, one can use the Yukawa matrix $\mathbf{Y}_D$
($\mathbf{Y}_E$) to easily translate the bounds in Fig.~\ref{fig2D}
(Fig.~\ref{fig2D_LE}) to bounds on couplings other than $\lam'_{233}$
($\lam_{233}$).

\section{Summary and Conclusion}
\label{conclusion}

We have calculated upper bounds on all trilinear lepton number
violating couplings at the grand unification scale within the
$\text{B}_3$ ({\it i.e.} lepton number violating) minimal supergravity
(mSUGRA) model, which result from the cosmological bound on the sum of
neutrino masses. We have shown that these bounds on the
couplings can be weaker by one to two orders of magnitude compared to
the ones which were previously presented in the literature for the
benchmark scenario SPS1a; {\it cf.}  Sec.~\ref{bounds}. In general,
the bounds can be as weak as $\mathcal{O}(10^{-1})$. Thus other low
energy bounds become competitive.

The reason for these large effects is that the tree--level
neutrino mass depends strongly on the trilinear soft-breaking
$A_0$--parameter (and also similarly on the gaugino masses). We
concluded in Sec.~\ref{mSUGRA}, that in regions of parameter space
with $A_0\approx 2 M_{1/2}$ ($A_0\approx M_{1/2}/2$) for
$\lambda'_{ijk}|_{\rm GUT} \not = 0$ ($\lambda_{ijk}|_{\rm GUT} \not =
0$), a cancellation between the different contributions to the
tree--level mass can occur. We have explained this effect in detail
and have shown that such a cancellation is significant in large
regions of the mSUGRA parameter space.  For example, the bounds can be
weakened by one order of magnitude in $A_0$ intervals of up to
$\mathcal{O}(100$ GeV), see Figs.~\ref{fig2D} and
\ref{fig2D_LE}.  Therefore, much weaker bounds (compared to previous
ones) can be obtained without significant fine--tuning.

In order to obtain the correct bounds in the vicinity of the
tree--level neutrino mass minimum, we included the main loop
contributions to the neutrino mass matrix; {\it cf.}
Sec.~\ref{chapM}.  We also described in Sec.~\ref{mSUGRA} and
App.~\ref{further_msugra} for the first time the dependence of the
tree--level {\it and} loop induced neutrino mass on all mSUGRA
parameters. Although we concentrated in this work on the $\text{B}_3$
mSUGRA model, the mechanisms described will also work in more general
R--parity violating models.
	
Our work can help to find new supersymmetric scenarios that are
consistent with the observed neutrino masses and mixings. We have
shown in this publication how the (typically large) hierarchy between the tree--level and
1--loop neutrino masses can systematically be reduced. Together with
at least one additional lepton number violating coupling, one can use
this mechanism to match the ratio between tree--level and
  1--loop induced masses to the observed neutrino mass hierarchy, both
for hierarchical neutrino masses and for a degenerate spectrum. 

We also note, as described in the introduction, that large lepton number
violating couplings can lead to distinct collider signatures. We will
address these topics in future publications.

\begin{acknowledgments}
We thank Ben Allanach, Howie Haber, Jong
Soo Kim and Steve C. H. Kom for discussions. The work of H. Dreiner was partially financed by the SFB--TR 33
`The Dark Universe' and partially by DOE grant DE-FG02-04ER41286.
S. Grab's work was financed by the DOE grant DE-FG02-04ER41286. The
work of M. Hanussek was funded by the Konrad--Adenauer--Stiftung,
the Bonn Cologne Graduate School and the Deutsche Telekom Stiftung.

\end{acknowledgments}

\appendix 

\section{$\nu$-Masses: Dependence on Further $\rm B_3$ mSUGRA Parameters}
\label{further_msugra}

In Sec.~\ref{A0}, we described in detail the dependence of the
tree--level neutrino mass, Eq.~(\ref{mnutree}), on the $\text{B}_3$
mSUGRA parameter $A_0$. We also reviewed some further effects in
Sec.~\ref{msugras}. In this appendix, we explain now in more detail
the dependence of the tree--level neutrino mass and the loop induced
masses on the remaining $\text{B}_3$ mSUGRA parameters.

\subsection{$M_{1/2}$ Dependence}\label{m12}

\begin{figure}[t!]
\begin{center}
\epsfig{figure=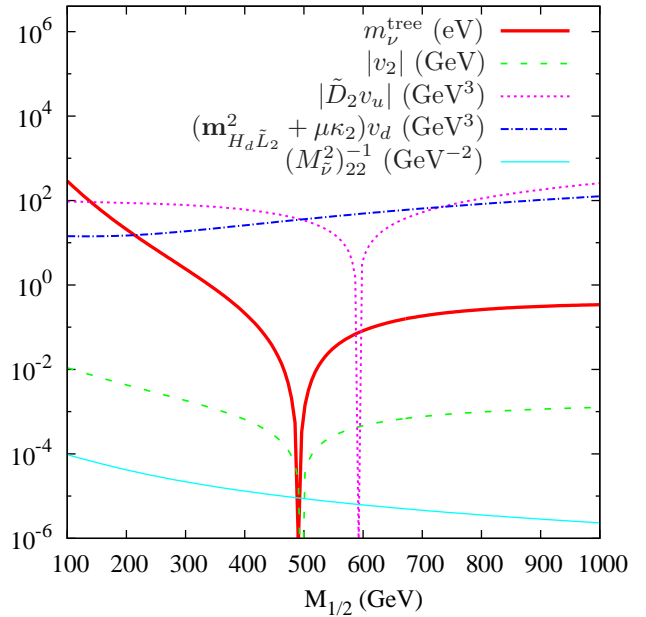}
\caption{Same as Fig.~\ref{figa0}, but now for the mSUGRA parameter 
$M_{1/2}$ instead of $A_0$. 
\label{m12dep}}
\end{center}
\end{figure}

The tree--level neutrino mass minimum can be explained equivalently in
terms of its dependence on $M_{1/2}$ instead of its dependence on
$A_0$. This is because varying $M_{1/2}$ has a similar effect on the
sneutrino vev $v_i$, Eq.~(\ref{vi}), as varying $A_0$, {\it cf.}
Sec.~\ref{A0} and Sec.~\ref{msugras}. However, when varying $M_{1/2}$
there are {\it additional effects} coming on the one hand from the
dependence of $\mu^2$, $(M_{\tilde{\nu}}^2)_{ii}$ and
$\mathbf{m}^2_{H_d \tilde{L}_i}$ on $M_{1/2}$. These quantities are
linear functions of $M_{1/2}^2$. For $\mu^2$ this can bee seen from
Eq.~(\ref{mu2}). For $(M_{\tilde{\nu}}^2)_{ii}$ and $\mathbf{m}^2_{H_d
\tilde{L}_i}$ this follows because the respective RGEs are functions
of the squared sfermion masses \cite{Allanach:2003eb}. One obtains for
example \cite{Drees:1995hj}
\begin{equation}
(M_{\tilde{\nu}}^2)_{ii} \approx M_0^2 + 0.52 M_{1/2}^2 + 
\frac{1}{2} M_Z^2 \cos2\beta \,. 
\label{Mnuii} 
\end{equation}
On the other hand, there is also a direct proportionality of
$m_{\nu}^{\textrm{tree}}$ to $M_{1/2}^{-1}$, \textit{cf.}
Eq.~(\ref{mnutree}). All these additional effects do not significantly
influence the position of the tree--level neutrino mass minimum, {\it
i.e.} $A_0 \approx 2 M_{1/2}$ still holds for $\mathbf{\Lam} \in
 \{\lam'_{ijk} \} $; see Sec.~\ref{mSUGRA}.  However, the effects add a
global slope to the terms (as a function of $M_{1/2}$), which
contribute to the tree level mass.  This behavior can be seen in
Fig.~\ref{m12dep}.

We show in Fig.~\ref{m12dep} the same contributions as in
Fig.~\ref{figa0}, but now as a function of $M_{1/2}$ instead of
$A_0$. Here $A_0$ has been fixed to 900 GeV. On the one hand, we
observe that the quantities $\tilde{D}_i v_u$ (dotted
magenta line) and $(\mathbf{m}^2_{h_d \tilde{L}_i}+ \mu \kappa_i) v_d$ 
(dotted-dashed blue line) are nearly constant for low values of $M_{1/2}$, but
they have a positive slope for large values of $M_{1/2}$. This is
mainly due to their dependence on $\mu$; {\it cf.}  Eq.~(\ref{RGEDi})
[Eq.~(\ref{RGEkappa})] for $\tilde{D}_i$ [$\kappa_i$].  On the other
hand $(M_{\tilde{\nu}}^{-2})_{ii}$ (solid turquoise line) has a
negative slope for all values of $M_{1/2}$ because of
Eq.~(\ref{Mnuii}). Overall this leads to a steep decrease of the
tree--level neutrino mass (solid red line) in the region of low
$M_{1/2}$, whereas in the region of large $M_{1/2}$, the various
contributions' dependence on $M_{1/2}$ roughly cancels, see
Fig.~\ref{m12dep}.

Going beyond the plot, for $M_{1/2}\rightarrow \infty$ the
tree--level mass scales with $M_{1/2}^{-1}$, as follows from the
different contributions to $m_{\nu}^{\textrm{tree}}$ in
Eq.~(\ref{mnutree}).  Such a behavior is expected, because SUSY
decouples from the SM sector in the limit $M_{1/2}\rightarrow \infty$.

\subsection{$\tan \beta$ Dependence}
\label{tanb}

\begin{figure}[t!]
\begin{center}
\epsfig{figure=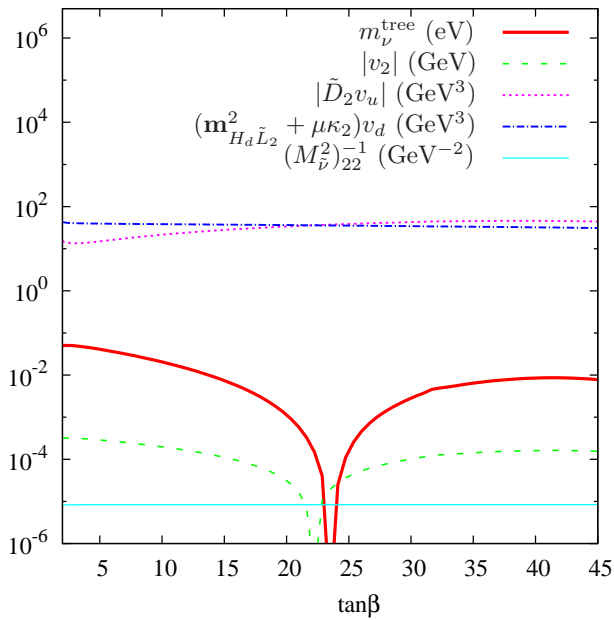,scale=0.96}
\caption{Same as Fig.~\ref{figa0}, but now for the mSUGRA parameter
$\tan\beta$ instead of $A_0$. \label{tanbdep}}
\end{center}
\end{figure}

Varying $\tan\beta$ most importantly affects the tree--level neutrino
mass via the term $\tilde{D}_i v_u$ in Eq.~(\ref{vi}). The RGE for
$\tilde{D}_i$, Eq.~(\ref{RGEDi2}), is proportional to the down--type
Yukawa coupling $(\textbf{Y}_D)_{jk} \equiv (\textbf{m}_d)_{jk} /
v_d$. Therefore, 
\beq \tilde{D}_i v_u \propto c_1 + c_2
\frac{v_u}{v_d} \equiv c_1 + c_2 \tan\beta \; , 
\eeq at $M_{\rm EW}$.
The factors $c_1$ and $c_2$ depend on the other mSUGRA parameters but
their magnitude is approximately independent of $\tan\beta$.  However, there is a
dependence of $\text{sgn}(c_2)$ on $\tan\beta$ via the RGE of $h'_{ijk}$.  Especially in case~(b) of
Sec.~\ref{A0}, {\it i.e.}  in the region around the tree--level
neutrino mass minimum, this becomes relevant
\footnote{In case~(a), $c_2$ remains always negative and in case~(c),
$c_2$ is positive.}.  

This (weak) $\tan \beta$ dependence of
$|\tilde{D}_i v_u|$ is illustrated in Fig.~\ref{tanbdep} for our
$\text{B}_3$ mSUGRA parameter set Point I; see
Sec.~\ref{example_points}.  One observes that the dotted magenta line
($|\tilde{D}_i v_u|$) increases between $\tan\beta = 2$ and $\tan\beta
\approx 40$. Here, $\text{sgn}(c_2) > 0$.  Above $\tan\beta \approx 40$,
$|\tilde{D}_i v_u|$ starts decreasing, {\it i.e.} $\text{sgn}(c_2) < 0$.
This is due to the enhancement of the down--type Yukawa coupling when
increasing $\tan\beta$, since this reduces $h'_{ijk}$ further and
further until it becomes negative.  This decrease of $|\tilde{D}_i v_u|$ is only partially
visible in Fig.~\ref{tanbdep} since the parameter region with high
$\tan\beta$ is excluded due to tachyons.

One can also see in Fig.~\ref{tanbdep} that the other term determining
the sneutrino vev, $(\mathbf{m}^2_{h_d \tilde{L}_i}+ \mu
\kappa_i)v_d$, which is displayed as a dotted-dashed blue
line, is fairly constant regarding tan$\beta$. This contribution to
the sneutrino vev is subtracted from the first term, $\tilde{D}_i v_u$
(dotted magenta line), so that the sneutrino vev becomes zero when the
two lines intersect; see Eq.~(\ref{vi}).

We observe this intersection in Fig.~\ref{tanbdep} at $\tan\beta
\approx 22$, thus yielding the tree--level neutrino mass minimum in
this region.  In theory, there could even arise {\it two} minima
because above $\tan\beta \approx 40$ $\tilde{D}_i v_u$ starts
decreasing again, leading to another intersection with
$(\mathbf{m}^2_{h_d \tilde{L}_i}+ \mu \kappa_i)v_d$.  However, as
mentioned before, this usually happens in an excluded region of
parameter space.

As is also illustrated in Fig.~\ref{tanbdep}, there is quite a
sizable difference between the two terms which determine the
sneutrino vev, {\it i.e.}  $(\mathbf{m}^2_{h_d \tilde{L}_i}+ \mu
\kappa_i)v_d$ (dotted--dashed blue line) and $\tilde{D}_i v_u$ (dotted
magenta line) in the region of low $\tan\beta$.  If we are looking for
a neutrino mass minimum in this region of parameter space, we need to
adjust $A_0$ towards higher values, which will increase $h'_{ijk}$
[{\it cf.} Eq.~(\ref{RGEh'})]. Therefore, increasing $A_0$ will shift
the dotted magenta line upwards until it intersects with the
dotted-dashed blue line at the desired low $\tan\beta$ value. This
shift of the tree--level neutrino mass minimum to higher $A_0$ is
clearly visible in Fig.~\ref{mnu_A0_tanb_pmu}.  For $\tan\beta=20$,
the minimum lies at $A_0 \approx 900$ GeV whereas for $\tan\beta=5$,
it has shifted to $A_0 \approx 1300$ GeV.

\subsection{sgn($\mu$) Dependence}
\label{sgnmu}
A change of sgn($\mu$) notably affects the tree--level neutrino mass via the 
RGE running of $\tilde{D}_i$ [Eq.~(\ref{RGEDi2})], in which the overall 
sign of the RGE is changed. Therefore, the sign of $\tilde D_i$ itself is 
reversed at any energy scale but its magnitude is mostly unaffected. 
Consequently, the $A_0$ value where $\tilde D_i=0$ is still mostly the same 
after a sign change. 

However, at the position of the tree--level neutrino mass minimum, $\tilde D_i$ 
needs to be slightly larger than zero in order to cancel the other terms contributing 
to the tree--level mass, {\it cf.} Sec.~\ref{A0} and Sec.~\ref{tanb}. When we are 
at a parameter point where the tree--level neutrino mass minimum occurs for positive $\mu$
({\it i.e.} $\tilde D_i$ is small and positive), a sign change to $\text{sgn}(\mu)=-1$ will yield 
a $\tilde D_i$ which is small and negative. The other contributing terms undergo no 
overall sign change. If we would like to obtain a neutrino mass minimum now, $\tilde D_i$ 
needs to be increased in order to become slightly larger than zero again. 
This can be achieved by {\it decreasing} $A_0$, Sec.~\ref{A0}, (or, equivalently, {\it increasing} 
$M_{1/2}$, Sec.~\ref{m12}) since this increases $\tilde D_i$ via $h'_{ijk}$ in its RGE, Eq.~(\ref{RGEDi2}), 
when $\mu$ is negative. Therefore, the tree--level minimum will occur at smaller values of $A_0$
(or equivalently larger values of $M_{1/2}$) when we change $\text{sgn}(\mu)=+1$ to $\text{sgn}(\mu)=-1$.

This effect becomes more important when we go to regions of low
$\tan\beta$.  Here the influence of $h'_{ijk}$ on $\tilde D_i$,
Eq.~(\ref{RGEDi2}), becomes weaker due to the decrease of the
down--type Yukawa coupling, as we discussed in Sec.~\ref{tanb}.  In
order to still obtain a positive $\tilde D_i$ after reversing
sgn($\mu$), $h'_{ijk}$ has to decrease in a more substantial fashion
than for large $\tan\beta$. Therefore, the parameter point where the
tree--level neutrino mass minimum is located will shift to smaller
$A_0$ when changing $\text{sgn}(\mu)=+1$ to $\text{sgn}(\mu)=-1$,
especially for $\tan\beta \lesssim 10$.

Overall, this leads to a ``mirroring" of the tree--level mass minimum
curve in the $A_0$--$\tan\beta$ plane around $A_0 = 800\textrm{
GeV}(\approx 2 M_{1/2})$.  This can be seen in
Fig.~\ref{mnu_A0_tanb_pmu} and Fig.~\ref{mnu_A0_tanb_nmu}: for
$\text{sgn}(\mu)=+1$ the minimum shifts to higher values of $A_0$ with
decreasing $\tan\beta$, whereas for $\text{sgn}(\mu)=-1$ the minimum
shifts to lower values of $A_0$.

\subsection{$M_0$ Dependence}
\label{m0}

\begin{figure}[t!]
\begin{center}
\epsfig{figure=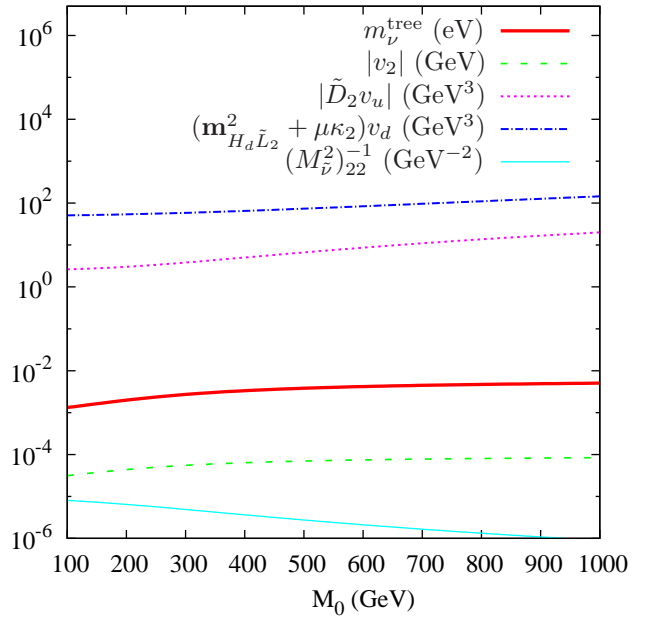}
\caption{Same as Fig.~\ref{figa0}, but now for the mSUGRA parameter
$M_0$ instead of $A_0$.
\label{m0dep}}
\end{center}
\end{figure}

Varying $M_0$ does not greatly affect the tree--level neutrino
mass. However, similar effects as those described in Sec.~\ref{m12}
as {\it additional effects}, arise due to the dependence of
several parameters on $M_0^2$, {\it cf.} for example Eq.~(\ref{mu2})
and Eq.~(\ref{Mnuii}).  This can be seen in Fig.~\ref{m0dep}, where we
again show the terms, which enter the tree--level neutrino
mass formula, Eq.~(\ref{mnutree}). We can see that most of the
quantities depend only weakly on $M_0$.  This results in a nearly
constant tree--level neutrino mass, {\it cf.} solid red line in
Fig.~\ref{m0dep}.  

However, the above mentioned $M_0^2$ dependences lead to a moderate
shift of the tree--level neutrino mass minimum towards higher values
of $A_0$ when increasing $M_0$. Explaining this in detail is fairly
lengthy because the $M_0$ dependence of the parameters determining the
tree--level neutrino mass is not as straightforward as the dependence
on other mSUGRA parameters. However, the effect is shown numerically
in Fig.~\ref{mnu_A0_M0}.

It should be noted that there is a similar {\it mirror effect}
when changing sgn($\mu$) as for $\tan\beta$.  For
sgn($\mu)=-1$, the minimum shifts towards {\it lower} values of $A_0$
when increasing $M_0$.

\subsection{Changes for $\mathbf \Lam \in \lam_{ijk}$}
\label{lam_LLE}
\begin{figure}[t!]
\begin{center}
\epsfig{figure=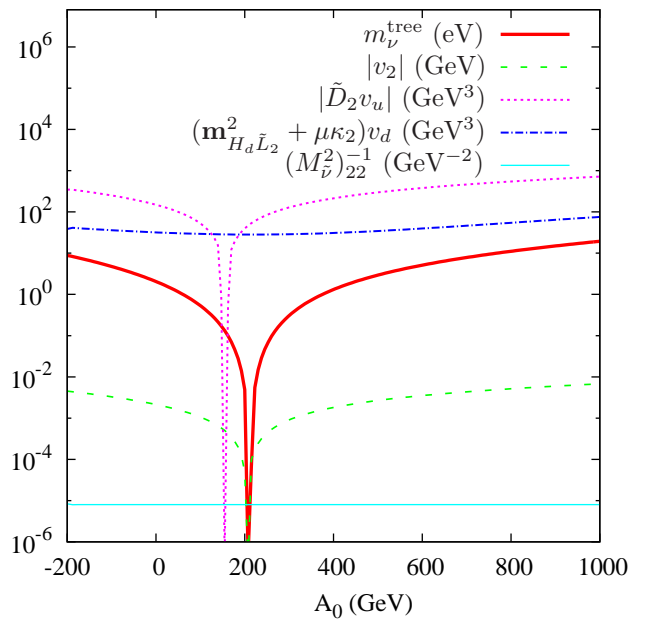}
\caption{Same as Fig.~\ref{figa0}, but now for the $\text{B}_3$ mSUGRA
Point II, Sec.~\ref{example_points}, with ${\lam_{233}}|_{\rm
GUT}=10^{-4}$.
\label{figa0_loop} }
\end{center}
\end{figure}

We now consider the case of $\mathbf \Lam \in \{ \lam_{ijk}
\}$ instead of $\mathbf \Lam \in \{\lam'_{ijk}\}$.  Since
$\lam_{ijk}$ only couples lepton superfields to each other (as opposed
to the $\lam'_{ijk}$ operator which also involves quark superfields),
the RGEs in Sec.~\ref{A0} are reduced by a (color) factor of 3
\cite{Carlos:1996du,Allanach:2003eb}. In addition, the down quark
Yukawa matrix elements, $(\mathbf{Y}_D)_{jk}$, need to be replaced by
the respective lepton Yukawa matrix elements, $(\mathbf{Y}_E)_{jk}$.
Otherwise, the structure of the RGEs remains the same.

The only RGE where there are more extensive relevant changes
is that for $h_{ijk}$ (which replaces $h'_{ijk}$); {\it cf.}
Eq.~(\ref{LNV_Lsoft}). Eq.~(\ref{RGEh'}) must be replaced by
\cite{Allanach:2003eb}
\begin{eqnarray}
16 \pi^2 \frac{d h_{ijk}}{d t} &=& \frac{9}{5} g_1^2 (2 M_1 \lam_{ijk} 
- h_{ijk}) \nonumber \\
&  & +  3  g_2^2 (2 M_2 \lam_{ijk} - h_{ijk})   + \dots \, ,
\label{RGEh}
\end{eqnarray}
with $h_{ijk}=A_0 \times \lam_{ijk}$ at $M_{\rm GUT}$.  This looks
exactly the same as the RGE for $h'_{ijk}$, Eq.~(\ref{RGEh'}), only
with $g_3$ and $M_3$ replaced by $g_\alpha$ and $M_\alpha$
($\alpha=1,2$). However, it is important to realize that the running
of $g_\alpha$ and $M_\alpha$ is different from the running of $g_3$
and $M_3$. As was mentioned in Sec.~\ref{A0}, the latter quantities
\textit{increase} when running to lower energy scales whereas the
former \textit{decrease} \cite{Martin:1997ns}.

This has important consequences for the position of the tree--level
neutrino mass minimum. The terms $g_\alpha^2 M_\alpha \lam_{ijk}$ of
Eq.~(\ref{RGEh}) now decrease [as opposed to $g_3^2 M_3 \lam'_{ijk}$
in Eq.~(\ref{RGEh'})].  It is thus necessary to choose $A_0$ smaller
in order to have a smaller $h_{ijk}$ at $M_{\rm GUT}$ and at lower
scales to compensate for this. Quantitatively, we checked numerically
that we now need $A_0 \approx M_{1/2}/2$ ($\mathbf \Lam \in
\{\lam_{ijk}\}$) to achieve a vanishing tree--level neutrino mass rather
than $A_0 \approx 2 M_{1/2}$ ($\mathbf \Lam \in\{ \lam'_{ijk}\}$) as
was the case in Sec.~\ref{A0}.

For illustrative purpose, we show in Fig.~\ref{figa0_loop} the $A_0$
dependence of the tree--level neutrino mass (solid red line) and of
the terms determining the sneutrino vev $v_2$ for a non-vanishing
coupling $\lam_{233}$ at $M_{\rm GUT}$.  Fig.~\ref{figa0_loop} is
equivalent to Fig.~\ref{figa0} beside the fact that we now employ the
parameter Point II with ${\lam_{233}}|_{\rm GUT}=10^{-4}$ instead of
the parameter Point I with ${\lam'_{233}}|_{\rm GUT}=10^{-5}$, {\it
cf.}  Sec.~\ref{example_points}. The qualitative behavior of all terms
is the same in both figures. However, in Fig.~\ref{figa0_loop}
the minima are shifted to lower values of $A_0$ compared to
Fig.~\ref{figa0}.

We conclude that the line of argument explaining the minimum of the
tree--level neutrino mass in the case of $\mathbf \Lam \in
\{\lam'_{ijk}\}$ still holds for $\mathbf \Lam \in \{\lam_{ijk}\}$.
However, the position of the minimum now shifts to $A_0 \approx
M_{1/2}/2$.

\subsection{$A_0$ Dependence of the Neutral Scalar--Neutralino--Loops}
\label{dep_loops}

According to Eqs.~(\ref{msnu_approx}) and
(\ref{mass_splitting}), the dominant loop contribution from neutral
scalar--neutralino--loops to the neutrino mass matrix,
$(m^{\tilde{\nu}\tilde{\nu}}_{\nu})_{ii}$, is proportional to 
\barr
(m_{\nu}^{\tilde\nu \tilde{\nu}} )_{ii} &\propto & \; (\tilde D_i v_d
- \tilde B v_i)^2 \nonumber \\ & & \times f(m^2_{\tilde{\chi}^0_k},
m^2_{\tilde{\nu}_i}, m^2_{H_0}, m^2_{A_0}, m^2_{h_0}) \, ,
\label{mnu_snu_approx}
\earr
where $f$ is a function of the neutralino, sneutrino and Higgs masses
squared, respectively.

The $A_0$ dependence of Eq.~(\ref{mnu_snu_approx}) is mainly
determined by $\tilde D_i$, since the
$A_0$ dependence of $v_i$ is governed by $\tilde D_i (A_0)$, 
\barr 
v_i(A_0) \propto \tilde D_i(A_0) + c\;, 
\earr 
where the term $c$ depends mainly on the other mSUGRA parameters but
barely on $A_0$, as discussed in Sec.~\ref{A0}.
Therefore $(m_{\nu}^{\tilde\nu \tilde{\nu}} )_{ii} $
is roughly proportional to $\tilde D_i^2$. The behavior of $\tilde D_i$
has been discussed in detail in Sec.~\ref{A0} in the context of the
tree--level neutrino mass. We have shown that there is always a value
of $A_0$ where $\tilde D_i$ becomes zero.  Thus the neutral
scalar--neutralino loops display a similar minimum as the tree--level
neutrino mass. The position of the minimum is close to the tree--level
one, but not exactly aligned. This can be seen by comparing the
dotted magenta line and dashed green line in Fig.~\ref{figloops_LQD} and
Fig.~\ref{figloops_LLE}.  However, since Eq.~(\ref{mnu_snu_approx}) is
only an approximate formula [for the exact formula, {\it cf.}
Eq.~(\ref{msnu})], the real curve is slightly shifted downwards such
that its minimum reaches negative values. Therefore
$|(m_{\nu}^{\tilde\nu \tilde{\nu}} )_{ii} |$ in
Fig.~\ref{figloops_LQD} and Fig.~\ref{figloops_LLE} appears to have
two minima.

It is also immediately obvious from Eq.~(\ref{mnu_snu_approx}) that
the scalar--neutralino--loops are roughly proportional to $[\mathbf
\Lam \times (\mathbf{Y}_D)_{jk}]^2$ like the tree--level mass.


\begin{thebibliography}{99}

\bibitem{Fukuda:1998ah}
Y.~Fukuda {\it et al.}  [Super-Kamiokande Collaboration],
Phys.\ Rev.\ Lett.\  {\bf 82} (1999) 2644
[arXiv:hep-ex/9812014].

\bibitem{Fukuda:1998fd}
Y.~Fukuda {\it et al.}  [Super-Kamiokande Collaboration],
Phys.\ Rev.\ Lett.\  {\bf 81} (1998) 1158
[Erratum-ibid.\  {\bf 81} (1998) 4279]
[arXiv:hep-ex/9805021].

\bibitem{Ahmad:2002jz}
Q.~R.~Ahmad {\it et al.}  [SNO Collaboration],
Phys.\ Rev.\ Lett.\  {\bf 89} (2002) 011301
[arXiv:nucl-ex/0204008].

\bibitem{Aharmim:2005gt}
B.~Aharmim {\it et al.}  [SNO Collaboration],
Phys.\ Rev.\  C {\bf 72} (2005) 055502
[arXiv:nucl-ex/0502021].

\bibitem{Apollonio:2002gd}
M.~Apollonio {\it et al.}  [CHOOZ Collaboration],
Eur.\ Phys.\ J.\  C {\bf 27} (2003) 331
[arXiv:hep-ex/0301017].

\bibitem{Cleveland:1998nv}
B.~T.~Cleveland {\it et al.},
Astrophys.\ J.\  {\bf 496} (1998) 505.

\bibitem{GonzalezGarcia:2007ib}
M.~C.~Gonzalez-Garcia and M.~Maltoni,
Phys.\ Rept.\  {\bf 460} (2008) 1
[arXiv:0704.1800 [hep-ph]];
M.~C.~Gonzalez-Garcia, M.~Maltoni and J.~Salvado,
JHEP {\bf 1004} (2010) 056
[arXiv:1001.4524 [hep-ph]].


\bibitem{Barate:1997zg}
R.~Barate {\it et al.}  [ALEPH Collaboration],
Eur.\ Phys.\ J.\  C {\bf 2} (1998) 395.

\bibitem{Assamagan:1995wb}
K.~Assamagan {\it et al.},
Phys.\ Rev.\  D {\bf 53} (1996) 6065.

\bibitem{Bonn:2001tw}
J.~Bonn {\it et al.},
Nucl.\ Phys.\ Proc.\ Suppl.\  {\bf 91} (2001) 273.

\bibitem{Lobashev:2001uu}
V.~M.~Lobashev {\it et al.},
Nucl.\ Phys.\ Proc.\ Suppl.\  {\bf 91} (2001) 280.

\bibitem{Amsler:2008zzb}
C.~Amsler {\it et al.}  [Particle Data Group],
Phys.\ Lett.\  B {\bf 667} (2008) 1.

\bibitem{Cirelli:2006kt}
M.~Cirelli and A.~Strumia,
JCAP {\bf 0612} (2006) 013
[arXiv:astro-ph/0607086].

\bibitem{Goobar:2006xz}
A.~Goobar, S.~Hannestad, E.~Mortsell and H.~Tu,
JCAP {\bf 0606} (2006) 019
[arXiv:astro-ph/0602155].

\bibitem{Wendell:2010md}
R.~Wendell {\it et al.}  [Kamiokande Collaboration],
arXiv:1002.3471 [hep-ex].


\bibitem{Minkowski:1977sc}
P.~Minkowski,
Phys.\ Lett.\  B {\bf 67} (1977) 421.

\bibitem{Mohapatra:1979ia}
R.~N.~Mohapatra and G.~Senjanovic,
Phys.\ Rev.\ Lett.\  {\bf 44} (1980) 912.

\bibitem{Yanagida}
T.~Yanagida,
proc. of the Workshop:
{\it Baryon Number of the Universe and Unified Theories}, 
Tsukuba, Japan, 1979.

\bibitem{GellMann:1980vs}
M.~Gell-Mann, P.~Ramond and R.~Slansky,
in the proc. of the {\it Supergravity Stony Brook Workshop}, 
ed. by P. van Nieuwenhuizen and D.Z. Freedman (North Holland Publ. Co.), 1979.

\bibitem{Mohapatra:1980yp}
R.~N.~Mohapatra and G.~Senjanovic,
Phys.\ Rev.\  D {\bf 23} (1981) 165.

\bibitem{Jezabek:1998du}
M.~Jezabek and Y.~Sumino,
Phys.\ Lett.\  B {\bf 440} (1998) 327
[arXiv:hep-ph/9807310].

\bibitem{Haber:1984rc}
H.~E.~Haber and G.~L.~Kane,
Phys.\ Rept.\  {\bf 117} (1985) 75.

\bibitem{Martin:1997ns}
S.~P.~Martin,
arXiv:hep-ph/9709356.

\bibitem{Coleman:1967ad}
S.~R.~Coleman and J.~Mandula,
Phys.\ Rev.\  {\bf 159} (1967) 1251.

\bibitem{Haag:1974qh}
R.~Haag, J.~T.~Lopuszanski and M.~Sohnius,
Nucl.\ Phys.\  B {\bf 88} (1975) 257.

\bibitem{Drees:1996ca}
M.~Drees,
arXiv:hep-ph/9611409.

\bibitem{Gildener:1976ai}
E.~Gildener,
Phys.\ Rev.\  D {\bf 14} (1976) 1667.

\bibitem{Veltman:1980mj}
M.~J.~G.~Veltman,
Acta Phys.\ Polon.\  B {\bf 12}, 437 (1981).

\bibitem{Sakai:1981gr}
N.~Sakai,
Z.\ Phys.\  C {\bf 11} (1981) 153.

\bibitem{Witten:1981nf}
E.~Witten,
Nucl.\ Phys.\  B {\bf 188} (1981) 513.

\bibitem{Hall:1983id}
L.~J.~Hall and M.~Suzuki,
Nucl.\ Phys.\  B {\bf 231} (1984) 419.

\bibitem{Joshipura:1994ib}
A.~S.~Joshipura and M.~Nowakowski,
Phys.\ Rev.\  D {\bf 51} (1995) 2421
[arXiv:hep-ph/9408224].

\bibitem{Nowakowski:1995dx}
M.~Nowakowski and A.~Pilaftsis,
Nucl.\ Phys.\  B {\bf 461} (1996) 19
[arXiv:hep-ph/9508271].

\bibitem{Grossman:1997is}
Y.~Grossman and H.~E.~Haber,
Phys.\ Rev.\ Lett.\  {\bf 78} (1997) 3438
[arXiv:hep-ph/9702421].

\bibitem{Grossman:1999hc}
Y.~Grossman and H.~E.~Haber,
arXiv:hep-ph/9906310.

\bibitem{Grossman:2000ex}
Y.~Grossman and H.~E.~Haber,
Phys.\ Rev.\  D {\bf 63} (2001) 075011
[arXiv:hep-ph/0005276].

\bibitem{Nardi:1996iy}
E.~Nardi,
Phys.\ Rev.\  D {\bf 55} (1997) 5772
[arXiv:hep-ph/9610540].


\bibitem{Davidson:2000ne}
S.~Davidson and M.~Losada,
Phys.\ Rev.\  D {\bf 65} (2002) 075025
[arXiv:hep-ph/0010325].

\bibitem{Dedes:2006ni}
A.~Dedes, S.~Rimmer and J.~Rosiek,
JHEP {\bf 0608}, 005 (2006)
[arXiv:hep-ph/0603225].

\bibitem{Dreiner:2007uj}
H.~K.~Dreiner, J.~Soo Kim and M.~Thormeier,
arXiv:0711.4315 [hep-ph].

\bibitem{Allanach:2007qc}
B.~C.~Allanach and C.~H.~Kom,
JHEP {\bf 0804} (2008) 081
[arXiv:0712.0852 [hep-ph]].

\bibitem{Sakai:1981pk}
N.~Sakai and T.~Yanagida,
Nucl.\ Phys.\  B {\bf 197} (1982) 533.

\bibitem{Weinberg:1981wj}
S.~Weinberg,
Phys.\ Rev.\  D {\bf 26} (1982) 287.

\bibitem{Allanach:1999ic}
B.~C.~Allanach, A.~Dedes and H.~K.~Dreiner,
Phys.\ Rev.\  D {\bf 60} (1999) 075014
[arXiv:hep-ph/9906209].

\bibitem{Ibanez:1991hv}
L.~E.~Ibanez and G.~G.~Ross,
Phys.\ Lett.\  B {\bf 260} (1991) 291.

\bibitem{Ibanez:1991pr}
L.~E.~Ibanez and G.~G.~Ross,
Nucl.\ Phys.\  B {\bf 368} (1992) 3.

\bibitem{Dreiner:2005rd}
H.~K.~Dreiner, C.~Luhn and M.~Thormeier,
Phys.\ Rev.\  D {\bf 73} (2006) 075007
[arXiv:hep-ph/0512163].

\bibitem{Banks:1991xj}
T.~Banks and M.~Dine,
Phys.\ Rev.\  D {\bf 45} (1992) 1424
[arXiv:hep-th/9109045].

\bibitem{Bhattacharyya:1996nj}
G.~Bhattacharyya,
Nucl.\ Phys.\ Proc.\ Suppl.\  {\bf 52A} (1997) 83
[arXiv:hep-ph/9608415].


\bibitem{Dreiner:1997uz}
H.~K.~Dreiner,
arXiv:hep-ph/9707435.

\bibitem{Barbier:2004ez}
R.~Barbier {\it et al.},
Phys.\ Rept.\  {\bf 420} (2005) 1
[arXiv:hep-ph/0406039].


\bibitem{Allanach:2003eb}
B.~C.~Allanach, A.~Dedes and H.~K.~Dreiner,
Phys.\ Rev.\  D {\bf 69} (2004) 115002
[Erratum-ibid.\  D {\bf 72} (2005) 079902]
[arXiv:hep-ph/0309196].

\bibitem{Martin:1993zk}
S.~P.~Martin and M.~T.~Vaughn,
Phys.\ Rev.\  D {\bf 50} (1994) 2282
[Erratum-ibid.\  D {\bf 78} (2008) 039903]
[arXiv:hep-ph/9311340].

\bibitem{Carlos:1996du}
B.~de Carlos and P.~L.~White,
Phys.\ Rev.\  D {\bf 54} (1996) 3427
[arXiv:hep-ph/9602381].

\bibitem{Besmer:2000rj}
T.~Besmer and A.~Steffen,
Phys.\ Rev.\  D {\bf 63} (2001) 055007
[arXiv:hep-ph/0004067].


\bibitem{Dreiner:1991pe}
H.~K.~Dreiner and G.~G.~Ross,
Nucl.\ Phys.\  B {\bf 365} (1991) 597.

\bibitem{Dreiner:2000vf}
H.~K.~Dreiner, P.~Richardson and M.~H.~Seymour,
Phys.\ Rev.\  D {\bf 63} (2001) 055008
[arXiv:hep-ph/0007228];
H.~K.~Dreiner, P.~Richardson and M.~H.~Seymour,
JHEP {\bf 0004} (2000) 008
[arXiv:hep-ph/9912407].


\bibitem{Dreiner:2008rv}
H.~K.~Dreiner, S.~Grab and M.~K.~Trenkel,
Phys.\ Rev.\  D {\bf 79} (2009) 016002
[Erratum-ibid.\  {\bf 79} (2009) 019902]
[arXiv:0808.3079 [hep-ph]].

\bibitem{Bernhardt:2008mz}
M.~A.~Bernhardt, H.~K.~Dreiner, S.~Grab and P.~Richardson,
Phys.\ Rev.\  D {\bf 78} (2008) 015016
[arXiv:0802.1482 [hep-ph]].

\bibitem{singleslep}
S.~Dimopoulos and L.~J.~Hall,
Phys.\ Lett.\  B {\bf 207} (1988) 210;
G.~Moreau, E.~Perez and G.~Polesello,
Nucl.\ Phys.\  B {\bf 604} (2001) 3
[arXiv:hep-ph/0003012];
D.~Choudhury, S.~Majhi and V.~Ravindran,
Nucl.\ Phys.\  B {\bf 660} (2003) 343
[arXiv:hep-ph/0207247];
L.~L.~Yang, C.~S.~Li, J.~J.~Liu and Q.~Li,
Phys.\ Rev.\  D {\bf 72} (2005) 074026
[arXiv:hep-ph/0507331];
Y.~Q.~Chen, T.~Han and Z.~G.~Si,
JHEP {\bf 0705} (2007) 068
[arXiv:hep-ph/0612076].

\bibitem{Allanach:1997sa}
B.~C.~Allanach, H.~K.~Dreiner, P.~Morawitz and M.~D.~Williams,
Phys.\ Lett.\  B {\bf 420} (1998) 307
[arXiv:hep-ph/9708495].

\bibitem{Arai:2010ci}
M.~Arai, K.~Huitu, S.~K.~Rai and K.~Rao,
arXiv:1003.4708 [hep-ph].


\bibitem{Dreiner:2006sv}
H.~K.~Dreiner, S.~Grab, M.~Kr{\" a}mer and M.~K.~Trenkel,
Phys.\ Rev.\  D {\bf 75} (2007) 035003
[arXiv:hep-ph/0611195].

\bibitem{Allanach:2009iv}
B.~C.~Allanach, C.~H.~Kom and H.~Pas,
Phys.\ Rev.\ Lett.\  {\bf 103} (2009) 091801
[arXiv:0902.4697 [hep-ph]].


\bibitem{Allanach:2006st}
B.~C.~Allanach, M.~A.~Bernhardt, H.~K.~Dreiner, C.~H.~Kom and P.~Richardson,
Phys.\ Rev.\  D {\bf 75} (2007) 035002
[arXiv:hep-ph/0609263].

\bibitem{Bernhardt:2008jz}
M.~A.~Bernhardt, S.~P.~Das, H.~K.~Dreiner and S.~Grab,
Phys.\ Rev.\  D {\bf 79} (2009) 035003
[arXiv:0810.3423 [hep-ph]].

\bibitem{Dreiner:2008ca}
H.~K.~Dreiner and S.~Grab,
Phys.\ Lett.\  B {\bf 679} (2009) 45
[arXiv:0811.0200 [hep-ph]].

\bibitem{Jack:2005id}
I.~Jack, D.~R.~T.~Jones and A.~F.~Kord,
Phys.\ Lett.\  B {\bf 632} (2006) 703
[arXiv:hep-ph/0505238].

\bibitem{Dreiner:2009fi}
H.~K.~Dreiner and S.~Grab,
AIP Conf.\ Proc.\  {\bf 1200} (2010) 358
[arXiv:0909.5407 [hep-ph]].



\bibitem{Hempfling:1995wj}
R.~Hempfling,
Nucl.\ Phys.\  B {\bf 478} (1996) 3
[arXiv:hep-ph/9511288].

\bibitem{Kaplan:1999ds}
D.~E.~Kaplan and A.~E.~Nelson,
JHEP {\bf 0001} (2000) 033
[arXiv:hep-ph/9901254].

\bibitem{Mira:2000gg}
J.~M.~Mira, E.~Nardi, D.~A.~Restrepo and J.~W.~F.~Valle,
Phys.\ Lett.\  B {\bf 492} (2000) 81
[arXiv:hep-ph/0007266].

\bibitem{Hirsch:2002ys}
M.~Hirsch, W.~Porod, J.~C.~Romao and J.~W.~F.~Valle,
Phys.\ Rev.\  D {\bf 66} (2002) 095006
[arXiv:hep-ph/0207334].

\bibitem{Hirsch:2000ef}
M.~Hirsch, M.~A.~Diaz, W.~Porod, J.~C.~Romao and J.~W.~F.~Valle,
Phys.\ Rev.\  D {\bf 62} (2000) 113008
[Erratum-ibid.\  D {\bf 65} (2002) 119901]
[arXiv:hep-ph/0004115].

\bibitem{Diaz:2003as}
M.~A.~Diaz, M.~Hirsch, W.~Porod, J.~C.~Romao and J.~W.~F.~Valle,
Phys.\ Rev.\  D {\bf 68} (2003) 013009
[Erratum-ibid.\  D {\bf 71} (2005) 059904]
[arXiv:hep-ph/0302021].

\bibitem{Bartl:2003uq}
A.~Bartl, M.~Hirsch, T.~Kernreiter, W.~Porod and J.~W.~F.~Valle,
JHEP {\bf 0311} (2003) 005
[arXiv:hep-ph/0306071].

\bibitem{Hirsch:2003fe}
M.~Hirsch and W.~Porod,
Phys.\ Rev.\  D {\bf 68} (2003) 115007
[arXiv:hep-ph/0307364].

\bibitem{Hirsch:2004he}
M.~Hirsch and J.~W.~F.~Valle,
New J.\ Phys.\  {\bf 6} (2004) 76
[arXiv:hep-ph/0405015].

\bibitem{Hirsch:2005ag}
M.~Hirsch, W.~Porod and D.~Restrepo,
JHEP {\bf 0503} (2005) 062
[arXiv:hep-ph/0503059].

\bibitem{deCampos:2007bn}
F.~de Campos, O.~J.~P.~Eboli, M.~B.~Magro, W.~Porod, D.~Restrepo, M.~Hirsch and J.~W.~F.~Valle,
JHEP {\bf 0805} (2008) 048
[arXiv:0712.2156 [hep-ph]].

\bibitem{deCampos:2008av}
F.~de Campos, M.~A.~Diaz, O.~J.~P.~Eboli, M.~B.~Magro, W.~Porod and S.~Skadhauge,
Phys.\ Rev.\  D {\bf 77} (2008) 115025
[arXiv:0803.4405 [hep-ph]].

\bibitem{Haber:1997if}
H.~E.~Haber,
Nucl.\ Phys.\ Proc.\ Suppl.\  {\bf 62} (1998) 469
[arXiv:hep-ph/9709450].

\bibitem{Ibanez:1982fr}
L.~E.~Ibanez and G.~G.~Ross,
Phys.\ Lett.\  B {\bf 110} (1982) 215.

\bibitem{msugramodel} 
A.~H.~Chamseddine, R.~Arnowitt and P.~Nath,
Phys.\ Rev.\ Lett.\  {\bf 49} (1982) 970;
L.~Alvarez-Gaume, M.~Claudson and M.~Wise,
Nucl.\ Phys.\ B {\bf 207} (1982) 96;
L.~Ibanez,
Phys.\ Lett.\ B {\bf 118} (1982) 73;
S.~K.~Soni and H.~A.~Weldon,
Phys.\ Lett.\ B {\bf 126} (1983) 215;
L.~J.~Hall, J.~D.~Lykken and S.~Weinberg,
Phys.\ Rev.\ D {\bf 27} (1983) 2359;  
R. Barbieri, S. Ferrara and C. A. Savoy, 
Phys.\ Lett.\ B  {\bf 119} (1982) 343.

\bibitem{Dreiner:2003hw}
H.~K.~Dreiner and M.~Thormeier,
Phys.\ Rev.\  D {\bf 69} (2004) 053002
[arXiv:hep-ph/0305270].

\bibitem{Allanach:2001kg}
B.~C.~Allanach,
Comput.\ Phys.\ Commun.\  {\bf 143} (2002) 305
[arXiv:hep-ph/0104145].

\bibitem{Allanach:2009bv}
B.~C.~Allanach and M.~A.~Bernhardt,
Comput.\ Phys.\ Commun.\  {\bf 181} (2010) 232
[arXiv:0903.1805 [hep-ph]].

\bibitem{Allanach:2007vi}
B.~C.~Allanach, M.~A.~Bernhardt, H.~K.~Dreiner, S.~Grab, C.~H.~Kom and P.~Richardson,
arXiv:0710.2034 [hep-ph].

\bibitem{Akeroyd:1997iq}
A.~G.~Akeroyd, M.~A.~Diaz, J.~Ferrandis, M.~A.~Garcia-Jareno and 
J.~W.~F.~Valle,
Nucl.\ Phys.\  B {\bf 529} (1998) 3
[arXiv:hep-ph/9707395].

\bibitem{Akeroyd:2001pm}
A.~G.~Akeroyd, C.~Liu and J.~H.~Song,
Phys.\ Rev.\  D {\bf 65} (2002) 015008
[arXiv:hep-ph/0107218].

\bibitem{Ellis:1983ew}
J.~R.~Ellis, J.~S.~Hagelin, D.~V.~Nanopoulos, K.~A.~Olive and M.~Srednicki,
Nucl.\ Phys.\  B {\bf 238} (1984) 453.

\bibitem{Schael:2006cr}
S.~Schael {\it et al.}  [ALEPH Collaboration and DELPHI Collaboration and
               L3 Collaboration and ],
Eur.\ Phys.\ J.\  C {\bf 47} (2006) 547
[arXiv:hep-ex/0602042].

\bibitem{Barate:2003sz}
R.~Barate {\it et al.}  [LEP Working Group for Higgs boson searches and
               ALEPH Collaboration and  and],
Phys.\ Lett.\  B {\bf 565} (2003) 61
[arXiv:hep-ex/0306033].

\bibitem{Bennett:2006fi}
G.~W.~Bennett {\it et al.}  [Muon G-2 Collaboration],
Phys.\ Rev.\  D {\bf 73} (2006) 072003
[arXiv:hep-ex/0602035].

\bibitem{Barberio:2008fa}
E.~Barberio {\it et al.}  [Heavy Flavor Averaging Group],
arXiv:0808.1297 [hep-ex].

\bibitem{Chun:1999bq}
E.~J.~Chun and S.~K.~Kang,
Phys.\ Rev.\  D {\bf 61} (2000) 075012
[arXiv:hep-ph/9909429].

\bibitem{Davidson:1999mc}
S.~Davidson, M.~Losada and N.~Rius,
Nucl.\ Phys.\  B {\bf 587} (2000) 118
[arXiv:hep-ph/9911317].

\bibitem{Dedes:2002dy}
A.~Dedes and P.~Slavich,
Nucl.\ Phys.\  B {\bf 657} (2003) 333
[arXiv:hep-ph/0212132].

\bibitem{Agashe:1995qm}
K.~Agashe and M.~Graesser,
Phys.\ Rev.\  D {\bf 54} (1996) 4445
[arXiv:hep-ph/9510439].

\bibitem{Drees:1995hj}
M.~Drees and S.~P.~Martin,
arXiv:hep-ph/9504324.

\bibitem{Dreiner:2008tw}
H.~K.~Dreiner, H.~E.~Haber and S.~P.~Martin,
arXiv:0812.1594 [hep-ph].

\bibitem{Gunion:1984yn}
J.~F.~Gunion and H.~E.~Haber,
Nucl.\ Phys.\  B {\bf 272} (1986) 1
[Erratum-ibid.\  B {\bf 402} (1993) 567].

\bibitem{Denner:1991kt}
A.~Denner,
Fortsch.\ Phys.\  {\bf 41} (1993) 307
[arXiv:0709.1075 [hep-ph]].

\bibitem{Allanach:2002nj}
B.~C.~Allanach {\it et al.},
in {\it Proc. of the APS/DPF/DPB Summer Study on the Future of 
Particle Physics (Snowmass 2001) } ed. N.~Graf,
Eur.\ Phys.\ J.\  C {\bf 25} (2002) 113
[arXiv:hep-ph/0202233].

\bibitem{Allanach:2003jw}
B.~C.~Allanach, S.~Kraml and W.~Porod,
JHEP {\bf 0303} (2003) 016
[arXiv:hep-ph/0302102].

\bibitem{Degrassi:2002fi}
G.~Degrassi, S.~Heinemeyer, W.~Hollik, P.~Slavich and G.~Weiglein,
Eur.\ Phys.\ J.\  C {\bf 28} (2003) 133
[arXiv:hep-ph/0212020].

\bibitem{Allanach:2004rh}
B.~C.~Allanach, A.~Djouadi, J.~L.~Kneur, W.~Porod and P.~Slavich,
JHEP {\bf 0409} (2004) 044
[arXiv:hep-ph/0406166].

\bibitem{Prades:2009qp}
J.~Prades,
arXiv:0909.2546 [hep-ph].

\bibitem{Jegerlehner:2009ry}
F.~Jegerlehner and A.~Nyffeler,
Phys.\ Rept.\  {\bf 477} (2009) 1
[arXiv:0902.3360 [hep-ph]].

\bibitem{Miller:2007kk}
J.~P.~Miller, E.~de Rafael and B.~L.~Roberts,
Rept.\ Prog.\ Phys.\  {\bf 70} (2007) 795
[arXiv:hep-ph/0703049].

\bibitem{Buras:2002tp}
A.~J.~Buras, A.~Czarnecki, M.~Misiak and J.~Urban,
Nucl.\ Phys.\  B {\bf 631} (2002) 219
[arXiv:hep-ph/0203135].

\bibitem{Belanger:2006is}
G.~Belanger, F.~Boudjema, A.~Pukhov and A.~Semenov,
Comput.\ Phys.\ Commun.\  {\bf 176} (2007) 367
[arXiv:hep-ph/0607059].

\bibitem{Dedes:2001fv}
A.~Dedes, H.~K.~Dreiner and U.~Nierste,
Phys.\ Rev.\ Lett.\  {\bf 87} (2001) 251804
[arXiv:hep-ph/0108037].


\bibitem{other}
H.~K.~Dreiner, M.~Kramer and B.~O'Leary,
Phys.\ Rev.\  D {\bf 75} (2007) 114016
[arXiv:hep-ph/0612278];
H.~K.~Dreiner, G.~Polesello and M.~Thormeier,
Phys.\ Rev.\  D {\bf 65} (2002) 115006
[arXiv:hep-ph/0112228];
D.~K.~Ghosh, S.~Raychaudhuri and K.~Sridhar,
Phys.\ Lett.\  B {\bf 396} (1997) 177
[arXiv:hep-ph/9608352];
G.~Bhattacharyya and D.~Choudhury,
Mod.\ Phys.\ Lett.\  A {\bf 10} (1995) 1699
[arXiv:hep-ph/9503263];
Y.~Kao and T.~Takeuchi,
arXiv:0910.4980 [Unknown].


\bibitem{Maltoni:2008ka}
M.~Maltoni and T.~Schwetz,
arXiv:0812.3161 [hep-ph].


\end{thebibliography}
\end{document}